\documentstyle[latexsym,amssymb,epsfig,aps,floats,eqsecnum,preprint,amsfonts]{revtex}
\def\laq{\raise 0.4ex\hbox{$<$}\kern -0.8em\lower 0.62
ex\hbox{$\sim$}}
\def\gaq{\raise 0.4ex\hbox{$>$}\kern -0.7em\lower 0.62ex\hbox{$\sim$}}

\font\tenbb=msbm10
\font\sevenbb=msbm7
\font\fivebb=msbm5
\newfam\bbfam
\textfont\bbfam=\tenbb \scriptfont\bbfam=\sevenbb
\scriptscriptfont\bbfam=\fivebb

\newcommand{\pa}{\partial} 

\newcommand{\vphi}{\varphi}

\newcommand{\beq}{\begin{equation}} 
\newcommand{\eeq}{\end{equation}}
\newcommand{\bea}{\begin{eqnarray}} 
\newcommand{\eea}{\end{eqnarray}}
\newcommand{\beam}{\begin{mathletters}} 
\newcommand{\eeam}{\end{mathletters}}
 
\newcommand{\wt}{\hat{t}} 
\newcommand{\wH}{\widehat{H}} 
\newcommand{\wF}{\widehat{\cal F}} 
\newcommand{\ww}{\widehat{\omega}} 

\begin{document}
\draft
\preprint{\vbox{\baselineskip=12pt
\rightline{IHES/P/99/90} 
\rightline{GRP/99/521} 
\vskip 0.2truecm
\rightline{gr-qc/0001013}}}

\title{\Large\bf Transition from inspiral to plunge 
in binary black hole coalescences}
\author{ Alessandra Buonanno${}^{a,b}$ and Thibault Damour${}^{b,c}$}
\address{$^a$ {\it Physics, Mathematics and Astronomy Department, \\
California Institute of Technology,
Pasadena, CA 91125 }\\
{$^b$ {\it Institut des Hautes Etudes Scientifiques, 91440
Bures-sur-Yvette, France} \\ 
{$^c$ {\it DARC, CNRS-Observatoire de Paris, 92195 Meudon, France}}}}
\maketitle
\begin{abstract}
Combining recent techniques giving non-perturbative 
re-summed estimates of the damping and conservative parts of the 
two-body dynamics, we describe 
the transition between the adiabatic phase and the plunge, 
in coalescing binary black holes with comparable masses 
moving on quasi-circular orbits. 
We give initial dynamical data for numerical relativity investigations, 
with a fraction of an orbit left, and provide, for data analysis 
purposes, an estimate of
the gravitational wave-form emitted throughout the inspiral, plunge and 
coalescence phases.
\end{abstract}
\newpage
\section{Introduction}\label{sec1}
The most promising candidate sources for ground based interferometric 
gravitational-wave (GW) detectors such as LIGO and VIRGO are binary 
systems made of massive (stellar) black holes \cite{LPP97,FH98,BCT98,PP99}. 
Such binary black holes (with individual 
masses in the range, say, $3-15 M_{\odot}$) pose special problems 
\cite{BCT98}, \cite{DIS2}. 

Let us recall that gravitational radiation damping
is efficient at circularizing such binary systems, and then
 drives, for a long 
time, a slow inspiraling quasi-circular motion of the binary system.
This quasi-circular ``adiabatic 
inspiral phase'' is expected to terminate abruptly, and to change to some 
type of ``plunge phase'' (leading to final coalescence) when the binary 
orbit shrinks down to the Last Stable (circular) Orbit (LSO) defined by 
the conservative part of the nonlinear relativistic force law between 
two bodies. [In the test-mass limit, the full nonlinear relativistic 
force law corresponds to geodesic motion in a Schwarzschild spacetime, 
and exhibits, as is well known, an LSO located at $R = 6 {GM}$. One 
expects that a comparable-mass system will still exhibit such an LSO;
see below.] 
Now, the signal to noise ratio (in an initial LIGO detector)
 for inspiral signals from comparable-mass 
black hole binaries reaches a {\it maximum} for $M \simeq 28 M_{\odot}$, 
which corresponds to a GW frequency for the waves emitted at the Last 
Stable (circular) Orbit (LSO) equal to $f_{\rm GW}^{\rm LSO} \simeq 170 
{\rm Hz}$, a value which is (not accidentally) very close to the location 
$f_{\rm det} \simeq 167 {\rm Hz}$ (for initial LIGO) of the minimum of 
the characteristic detector noise amplitude $h_n (f) \equiv \sqrt{f \, 
S_n (f)}$ (see Fig.~\ref{Fig1} of \cite{DIS2}). Therefore the first 
detections will probably concern massive systems with $M \sim 30 
M_{\odot}$. Moreover, Ref.~\cite{DIS2} has shown that when the total mass 
$M \equiv m_1 + m_2$ lies in the range $5-40 M_{\odot}$ the proximity 
(within a factor $\sim 2$) between the observationally most important 
frequencies\footnote{\label{f1} We neglect here the very small difference 
between 
the optimal frequency $f_{\rm det}$ for generic broad-band bursts, and 
the optimal frequency $f_p$ for inspiral signals (see \cite{DIS2}).} 
$f_{\rm det}$ and the GW frequency at the LSO, $f_{\rm GW}^{\rm LSO}$, was 
calling both for an especially careful treatment of the Fourier transform 
of the emitted waveform, and for an improved knowledge of the transition 
between the inspiral phase and the plunge phase.

The present paper will attempt to improve our knowledge of the transition 
between inspiral and plunge by combining two, recently proposed, 
{\it non-perturbative} techniques: Refs.~\cite{DIS} and \cite{BD99}. Let us 
first 
recall that, a few years ago, Will and collaborators \cite{LW90}, 
\cite{KWW} tried to attack the problem of the late-time evolution of 
compact binaries (including the transition from inspiral to plunge) by a 
direct use of the Damour-Deruelle \cite{DD81a,D82,D83} 
equations of motion. These equations of motion are given in the form of a 
{\it perturbative} expansion in powers of a small parameter $\varepsilon 
= v/c$ (``post-Newtonian'', or, in short, PN, expansion). In 
Ref.~\cite{LW90} a direct integration of these perturbative equations of 
motion (using the method of osculating elements) was used, while, in 
Ref.~\cite{KWW} it was proposed to improve the straightforward 
perturbative approach by using ``hybrid'' equations of motion. The 
``hybrid'' approach is a {\it partial} re-summation approach in which the 
perturbative terms in the equations of motion which survive in the test 
mass limit ($\nu \equiv m_1 \, m_2 / (m_1 + m_2)^2 \rightarrow 0$) are 
replaced by the known, exact ``Schwarzschild terms'', while the 
$\nu$-dependent terms are left as a perturbative expansion. Both the 
robustness \cite{WS}, \cite{SWmoriond} and the consistency \cite{DIS} of 
the hybrid approach of \cite{KWW} have been questioned. [In particular, 
it was pointed out in Ref.~\cite{DIS} that, in this approach, the 
supposedly small ``$\nu$-corrections'' represent, in several cases, a 
very large (larger than 100\%) modification of the corresponding 
$\nu$-independent terms.] Another sign of the unreliability of the hybrid 
approach is the fact that the recent study \cite{DJS1,DJS2} of 
the location of the LSO at the third post-Newtonian (3PN) accuracy has 
qualitatively
confirmed the 2PN-level results of the non-perturbative techniques to be 
discussed below (namely that the LSO is ``lower than 6GM''), thereby 
casting doubt on the most striking prediction of the hybrid approach (an 
LSO ``higher than 6GM'', i.e. with a lower orbital frequency).

By contrast with the perturbative approach of \cite{LW90} and the 
partially re-summed approach of \cite{KWW}, the present paper will rely on 
the systematic use of non-perturbative re-summation techniques. The basic 
philosophy underlying our approach is the following. We are interested in 
understanding, in quantitative detail, the combined influence on the 
inspiral $\rightarrow$ plunge transition of radiation reaction and of 
non-linear effects in the force law for {\it comparable-mass} binary 
systems (i.e. for systems for which $\nu \equiv m_1 \, m_2 / (m_1 + 
m_2)^2$ is around\footnote{\label{f2} Note that because $\nu$,
considered as a function of the ratio $ m_1 / m_2$, reaches 
its {\it maximum} $\nu_{\rm max} = 1/4$ for $m_1 / m_2 = 1$ it stays 
numerically near $1/4$ even for mass ratios quite different from 1. E.g., 
even for $m_1 / m_2 = 3$, $4 \nu  = 0.75$.} $1/4$). 

At present there exists no method for deriving, from first principles, 
non-perturbative expressions for the two-body equations of motion, 
especially in the case of interest where $4\nu$ is {\it not} a small 
parameter. As a substitute we shall combine two different re-summation 
techniques that have been recently introduced to deal with two separate 
aspects of the problem we wish to tackle.

The first re-summation technique, introduced in \cite{DIS}, allows one to 
get a non-perturbative, $\nu$-dependent, estimate of the rate of loss of 
angular momentum (under gravitational damping) in quasi-circular, 
comparable-mass binaries. The idea of \cite{DIS} is three-pronged: (i) to 
work with an invariant function of an invariant argument, $F(v)$, (ii) to 
inject some plausible information about the meromorphic structure of this 
function, and, finally, (iii) to use Pad\'e approximants to estimate 
$F(v)$ from the first few known terms in the perturbative (PN) expansion 
of $F(v)$. The second re-summation technique, introduced in \cite{BD99}, 
allows one to derive a non-perturbative, $\nu$-dependent, estimate of the 
(conservative part of the) nonlinear force law determining the motion of 
comparable binaries. The idea of \cite{BD99} is to map the real two-body 
problem on a simpler effective one-body problem, i.e. on the problem of 
the motion of a particle of mass $\mu \equiv m_1 \, m_2 / (m_1 + m_2)$ in 
some ``effective'' background metric $g_{\mu \nu}^{\rm eff} 
(x^{\lambda})$. The possibility (and uniqueness, given some natural 
requirement) of such a mapping, real $\rightarrow$ effective, was proven
at the 2PN level in \cite{BD99}. The extension of this approach at the 
3PN level has been recently discussed \cite{DJS1,DJS2} on the 
basis of the 3PN dynamics recently derived by Jaranowski and Sch\"afer 
\cite{JS98}. At the 2PN level the $\nu$-dependent terms in the effective 
metric were found to be numerically so small (around the LSO) that the 
need for a further (Pad\'e-type) re-summing of the effective metric 
coefficients did not arise. [However, note that Ref.~\cite{DJS2} has 
introduced, at 2PN and 3PN, the further idea of a specific, 
Pad\'e-improvement of $g_{\mu \nu}^{\rm eff} (x^{\lambda})$.] In this 
paper we shall show how one can combine the methods of \cite{DIS} and 
\cite{BD99} to derive a full force law (including radiation reaction) 
describing the quasi-circular motion of comparable-mass binaries. Our 
approach is intended to apply to any value of $\nu$, but is restricted to 
considering {\it quasi-circular} motions, where the radial velocity $\dot 
R$ is much smaller than the circular one $R \, \dot{\varphi}$. As we shall 
see, we shall consistently check that the condition $\dot R \ll R \, 
\dot{\varphi}$ holds true not only during the adiabatic inspiral, but 
also during the transition to the plunge, and even during most of the 
plunge.

We  apply our method, in this paper, to deriving two sorts of 
results which are of direct interest to the ongoing effort to detect 
gravitational waves. First, we shall give initial dynamical data (i.e. 
initial positions and momenta) for binary black holes that have just 
started their plunge motion. The idea here is that numerical relativity 
will probably not be able, before quite a few years, to accurately evolve 
binary systems over many (or even $\sim 10$) orbits. This is why we 
propose a method for computing accurate initial dynamical data at a 
moment so late in the evolution that there remains  (when $4 \nu \sim 1$)
{\it less} than one orbit to evolve.
[In the equal-mass case, $\nu = 1/4$, we shall compute data $\simeq 0.6$ orbit
before ``coalescence''.]
 Our contention (whose robustness we shall try to 
establish) is that suitably re-summed versions of {\it analytic} (PN) 
results allow one to push the evolution that far. [We shall use here 
2.5PN-accurate information for angular momentum loss and 2PN-accurate 
information for the conservative force law. However, as shown in 
\cite{DIS} and \cite{DJS1,DJS2} our method can be pushed to 
higher accuracy when the correspondingly needed PN results become 
unambiguously known.] Note that this attitude is opposite to the one 
taken in \cite{BCT98} in which it was assumed that ``there is little hope, 
via PN Pad\'e approximants, to evolve'' a binary system up to the moment 
where it can provide initial data for the final coalescence. Let us, 
however, immediately add that the present paper is still incomplete, in 
that we give only dynamical data $(\mbox{\boldmath$q$}_1 , 
\mbox{\boldmath$q$}_2 , \mbox{\boldmath$p$}_1 , \mbox{\boldmath$p$}_2)$ 
but we do not solve the remaining problem of constructing the initial 
gravitational data $(g_{ij} (x) , K_{ij} (x))$ determined (in principle) 
by $(\mbox{\boldmath$q$}_a , \mbox{\boldmath$p$}_a)$ (given, say, some 
no-incoming-radiation condition). We shall leave this (important) issue 
to future work.

The second aim of this work is to provide, for data analysis purposes, 
some estimate of the complete waveform emitted by the coalescence of two 
black holes (with negligible spins). We do not claim that this part of 
the work will be as accurate as the first one. The idea here is to 
provide a (hopefully $\sim$~10\% accurate) guess of the complete 
waveform, with its transition from an inspiral phase to a plunge one, 
followed by a coalescence ending in a stationary final state. In view of the 
recent realization \cite{DIS2} of the crucial importance of the details 
of the transition to the plunge for the construction of faithful GW 
templates (for massive binaries with $5 M_{\odot} \, \laq \, M \, \laq \, 40 
M_{\odot}$) even an approximate knowledge of the complete waveform will 
be a valuable information for data analysis (e.g. to test the accuracy of 
present templates, and/or to propose more accurate or, at least, more 
robust, templates).

While preparing this work for publication, we learned of the existence of 
an independent work of Ori and Thorne \cite{OT99} which deals
with the transition between the inspiral and the plunge in the test mass
 limit ( $\nu \rightarrow 0$).

\section{Conservative part of the two-body force law}
\label{sec2}
In this section, we recall the non-perturbative construction of the 
(conservative) two-body force law given in Ref.~\cite{BD99}. There it was 
shown that the conservative part (i.e. without radiation damping) of the 
dynamics of a binary system, represented in ADM phase-space coordinates 
$(\mbox{\boldmath$q$}_1^{\rm ADM} , \mbox{\boldmath$q$}_2^{\rm ADM} , 
\mbox{\boldmath$p$}_1^{\rm ADM} , \mbox{\boldmath$p$}_2^{\rm ADM})$, 
could be mapped (at the 2PN level), via the combination of an energy map, 
${\cal E}_{\rm eff} = f \, ({\cal E}_{\rm real})$, and a canonical 
transformation, $(\mbox{\boldmath$q$}_a^{\rm ADM} , 
\mbox{\boldmath$p$}_a^{\rm ADM}) \rightarrow (\mbox{\boldmath$q$}_a , 
\mbox{\boldmath$p$}_a)$, $a=1,2$, into the simpler dynamics of the 
geodesic motion of a particle of mass $\mu = m_1 \, m_2 / (m_1 + m_2)$ in 
some effective background geometry $g_{\mu \nu}^{\rm eff} (x)$:
\beq
\label{2.0}
ds_{\rm eff}^2 = g_{\mu \nu}^{\rm eff} (x^{\lambda}) \, dx^{\mu} \, 
dx^{\nu} = -A(R) \, c^2 \, dt^2 + B(R) \, dR^2 + C(R) \, R^2 \, 
(d\theta^2 + \sin^2 \theta \, d\varphi^2) \, .
\eeq
[See \cite{DJS2} for the generalization of this approach to the 3PN 
level.] Here the coordinates $(R,\theta,\varphi)$ are polar coordinates 
in the {\it effective} problem (describing the relative motion). They are 
related in the standard way ($Q^x = R \, \sin \theta \, \cos \varphi$, 
$Q^y = R \, \sin \theta \, \sin \varphi$, $Q^z = R \, \cos \theta$) to 
the (relative) effective Cartesian coordinates $\mbox{\boldmath$Q$} = 
\mbox{\boldmath$q$}_1 - \mbox{\boldmath$q$}_2$, where 
$\mbox{\boldmath$q$}_1$ and $\mbox{\boldmath$q$}_2$ are the effective 
coordinates of each body. One works in the center-of-mass frame of the 
binary system, i.e. $\mbox{\boldmath$p$}_1 + \mbox{\boldmath$p$}_2 = 
0 = \mbox{\boldmath$p$}_1^{\rm ADM} + 
\mbox{\boldmath$p$}_2^{\rm ADM}$. The canonical conjugate of the relative 
position $\mbox{\boldmath$Q$}$ is the relative momentum 
$\mbox{\boldmath$P$} = \mbox{\boldmath$p$}_1 = - \mbox{\boldmath$p$}_2$. 
In most of this paper we shall work with the effective phase-space 
coordinates $(\mbox{\boldmath$Q$} , \mbox{\boldmath$P$})$ [or rather with 
scaled versions of their polar~\footnote{\label{f3} Note that we have the usual 
relations, such as, $P_R = n^i \, P_i$ with $n^i = Q^i / R$, and 
$P_{\varphi} = Q^x \, P_y - Q^y \, P_x$.} counterparts $(R , \theta , 
\varphi ; P_R , P_{\theta} , P_{\varphi})$]. We shall only discuss at the 
end how to construct the more physically relevant ADM phase space 
coordinates $(\mbox{\boldmath$q$}_a^{\rm ADM} , 
\mbox{\boldmath$p$}_a^{\rm ADM})$ from $(\mbox{\boldmath$Q$} , 
\mbox{\boldmath$P$})$.

In absence of damping (to be added later), the evolution (with respect to 
the real ADM time coordinate $t_{\rm real}$) of $(\mbox{\boldmath$Q$} , 
\mbox{\boldmath$P$})$ is given by Hamilton's equations
\bea
\label{2.1}
&& \frac{dQ^i}{dt_{\rm real}} - \frac{\partial H_{\rm real}^{\rm 
improved}(
\mbox{\boldmath$Q$},\mbox{\boldmath$P$})}
{\partial P_i} = 0 \,,\\
\label{2.2}
&& \frac{dP_i}{dt_{\rm real}} + \frac{\partial H_{\rm real}^{\rm 
improved}(
\mbox{\boldmath$Q$},\mbox{\boldmath$P$})}
{\partial Q^i} = 0 \,,
\eea
where the {\it real} (i.e. giving the $t_{\rm real}$-evolution, and the 
real two-body energy) {\it improved} (i.e. representing a non-perturbative 
re-summed estimate of the real PN Hamiltonian) Hamiltonian reads
\beq 
\label{2.3}
H_{\rm real}^{\rm improved}(\mbox{\boldmath$Q$},\mbox{\boldmath$P$}) = 
M\,c^2\,\sqrt{1 + 2\nu\,\left (\frac{H_{\rm 
eff}(\mbox{\boldmath$Q$},\mbox{\boldmath$P$}) - 
\mu\,c^2}{\mu\,c^2} \right )}\,,
\eeq
and 
\beq
\label{2.4}
H_{\rm eff} (\mbox{\boldmath$Q$} , \mbox{\boldmath$P$}) = \mu \, c^2 \, 
\sqrt{A(Q) \left[ 1 + \frac{(\mbox{\boldmath$n$} \cdot 
\mbox{\boldmath$P$})^2}{\mu^2 \, c^2 \, B(Q)} + 
\frac{(\mbox{\boldmath$n$} 
\times \mbox{\boldmath$P$})^2}{\mu^2 \, c^2\,C(Q)} \right]} \, . 
\eeq
Here $Q \equiv \sqrt{\delta_{ij} \, Q^i \, Q^j} = R$,  
$n^i = Q^i / Q$ is the unit vector in the radial 
direction,  
and the scalar and vector products are performed as in Euclidean space. 
Henceforth, we shall pose $t \equiv t_{\rm real}$, $H \equiv H_{\rm 
real}^{\rm 
improved}$ 
and use the following notation:
\beq
\label{2.5}
M \equiv m_1 + m_2 \,, \quad \quad \mu \equiv \frac{m_1\,m_2}{M} \,, 
\quad \quad  \nu \equiv \frac{\mu}{M} 
\equiv \frac{m_1\, m_2}{(m_1 + m_2)^2} \, . 
\eeq

In polar coordinates, restricting ourselves to planar 
motion in the equatorial plane $\theta = \pi/2$ and to the Schwarzschild 
gauge ($C(Q) = 1$), we get the equations of motion
\bea
\label{2.6}
&&\frac{dR}{d t} = \frac{\pa H}{\pa P_R}
(R,P_R,P_\vphi)\,, \\
\label{2.7}
&& \frac{d \vphi}{d t} = \frac{\pa H}{\pa P_\vphi}
(R,P_R,P_\vphi)\,, \\
\label{2.8}
&& \frac{d P_R}{d t} + \frac{\pa H}{\pa R}
(R,P_R,P_\vphi)=0\,, \\
\label{2.9}
&& \frac{d P_\vphi}{d t} =0\,,
\eea
with
\beq
\label{2.10}
H(R,P_R,P_\vphi) = M\,c^2\,\sqrt{1 + 2\nu\,\left [
\sqrt{A(R)\,\left (1 + \frac{P_R^2}{\mu^2\,c^2\,B(R)} + 
\frac{P_\vphi^2}{\mu^2\,c^2\,R^2} \right )} -1 \right ]}\,.
\eeq
Like in any (non-degenerate) Hamiltonian system, this conservative dynamics 
is equivalent to a Lagrangian dynamics
\beq
\label{2.11}
L_{\rm real}^{\rm 
improved}(\mbox{\boldmath$Q$},\dot{\mbox{\boldmath$Q$}}) = 
P_i\,\dot{Q}^i - 
H_{\rm real}^{\rm improved}(\mbox{\boldmath$Q$},\mbox{\boldmath$P$})\,,
\eeq
with $P_i(\dot{Q})$ obtained by solving $\dot{Q}^i = \pa H/\pa P_i$. 
The Lagrangian equations of motion read:
\beq
\frac{d}{d t}\,\frac{\pa L_{\rm real}^{\rm improved}}{\pa \dot{Q}^i} - 
\frac{\pa L_{\rm real}^{\rm improved}}{\pa Q^i} =0\,.
\eeq
To ease the notation we denote $L \equiv L_{\rm real}^{\rm improved}$.

Finally, the 2PN-accurate metric coefficients $A(R)$, $B(R)$, 
Eq.~(\ref{2.0}), (in the Schwarzschild gauge where $C(R) \equiv 1$) read
\beq
\label{2.13}
A(R) = 1 - \frac{2 \, {GM}}{c^2 \, R} + 2\nu \left( 
\frac{{GM}}{c^2 \, R} \right)^3 \, ,
\eeq
\beq
\label{2.14}
B(R) \equiv D(R) / A(R) \, ,
\eeq
with 
\beq
\label{2.15}
D(R) = 1 - 6\nu \left( \frac{{GM}}{c^2 \, R} \right)^2 \, .
\eeq
Note that it was recently suggested \cite{DJS2} (because of the slow 
convergence of the 3PN contributions) to replace the straightforward 
expression (\ref{2.13}) by a suitably Padeed version, namely (at 2PN): 
$A_{P_2} (R) = 1-2u (1+\nu u^2)^{-1}$, where $u \equiv {GM} / c^2 \, 
R$. However, we have checked that this refinement has only a very minor 
effect on the results to be discussed below.

The re-summed (conservative) dynamics defined by the Hamiltonian 
(\ref{2.10}) contains a Last Stable (circular) Orbit (LSO) which is a 
$\nu$-deformed version of the well known Schwarzschild LSO. Let us recall 
that the radius of the LSO is obtained by imposing the existence of an
inflection point in the effective  potential
 $H(R, P_R = 0, {\cal J})$ for the radial motion,
\beq
\label{2.16}
\frac{\pa H}{\pa R} \, (R, P_R = 0, {\cal J}) = 0 = \frac{\pa^2 H}{\pa 
R^2} \, (R, P_R = 0, {\cal J}) \, ,
\eeq
where the total angular momentum ${\cal J} \equiv P_{\varphi}$ stays 
fixed. Eq.~(\ref{2.16}) has a solution in $R$ (for each value of $\nu$) 
only for some specific value of ${\cal J} = {\cal J}^{\rm LSO} (\nu)$. In 
terms of the rescaled variables $r \equiv c^2 R / {GM}$, $j \equiv c {\cal 
J} / (\mu \, {GM})$, $\widehat{\omega} \equiv {GM} \, 
\dot{\varphi}/c^3$, the LSO quantities defined, in the equal-mass case $\nu = 
1/4$, by the Hamiltonian (\ref{2.10}), take the following values 
\cite{BD99}
\bea
&&r_{\rm LSO} \left( {1}/{4} \right) = 5.718\,, \quad \quad \quad j_{\rm LSO} 
\left( {\small {1}/{4}} \right) = 3.404 \,, \nonumber \\
&&\widehat{\omega}_{\rm LSO} \left({1}/{4} \right) = 0.07340 \,, 
\quad \quad  \frac{{\cal E}_{\rm real}^{\rm LSO} \left({1}/{4} \right) - 
Mc^2}{Mc^2} = - 0.01501 \, . \label{2.17}
\eea
Note that the comparable-mass LSO is slightly more inwards (both in terms 
of the coordinate $R$ and in the sense of having a higher orbital 
frequency) than its corresponding rescaled test-mass limit: $r_{\rm LSO} 
(0) = 6$, $j_{\rm LSO} (0) = \sqrt{12} = 3.4641$, $\widehat{\omega}_{\rm 
LSO} (0) = 6^{-3/2} = 0.068041$.

As we shall need in the following to refer to the numerical value of 
$\widehat{\omega}_{\rm LSO} (\nu)$ for arbitrary values of $\nu$, we have fitted 
the 
result obtained by the (rather intricate) method of Ref.~\cite{BD99} to a simple 
polynomial in $\nu$. We find
\bea
\label{6.12}
&& \ww_{\rm LSO}(\nu) \simeq \omega_0\,[1 + \omega_1\,(4\nu) + 
\omega_2\,(4\nu)^2]\,,\\
\label{6.13}
&& \omega_0 = 0.0680414\,,
\quad \quad \omega_1 = 0.0693305\,, \quad \quad \omega_2 = 0.00935142\,.
\eea

\section{Incorporating radiation reaction effects}
\label{sec3}
We wish to augment the conservative dynamics described in 
the previous section by adding, as accurately as possible, radiation 
reaction effects. If we were doing it in the Lagrangian formalism we 
would write (in any coordinate system)
\beq
\label{3.1}
\frac{d}{d t}\,\frac{\pa L}{\pa \dot{Q}^i} - 
\frac{\pa L}{\pa Q^i} = {\cal F}_i^{\rm Lag}(Q,\dot{Q})\,.
\eeq
This would define the additional damping force ${\cal F}_i^{\rm Lag} 
(Q,\dot Q)$ needed in the Lagrangian formalism. In particular, in polar 
coordinates we would write (for planar motion $\theta = \pi / 2$):
\bea
\label{3.2}
&& \frac{d}{d t}\,\frac{\pa L}{\pa \dot{R}} - 
\frac{\pa L}{\pa R} = {\cal F}_R^{\rm 
Lag}(R,\varphi,\dot{R},\dot{\vphi})\,,\\
\label{3.3}
&& \frac{d}{d t}\,\frac{\pa L}{\pa \dot{\vphi}} = 
{\cal F}_\vphi^{\rm Lag}(R,\varphi,\dot{R},\dot{\vphi})\,.
\eea
We want to work in the Hamiltonian framework, hence
coming back to the coordinates $R$, $P_R$, $\vphi$ and $P_\vphi$ and 
imposing the constraint that the usual definition $P_i = \pa L/\pa 
\dot{Q}^i$ holds without corrections (which implies that the other usual 
relations $\dot{Q}^i = \pa H/\pa P_i$, $\pa H/\pa Q^i = -\pa L/\pa Q^i$ 
and Eq.~(\ref{2.11}) hold too) we get
\bea
\label{3.4}
&&\frac{dR}{d t} - \frac{\pa H}{\pa P_R}
(R,P_R,P_\vphi) = 0\,, \\
\label{3.5}
&& \frac{d \vphi}{d t} - \frac{\pa H}{\pa P_\vphi}
(R,P_R,P_\vphi) = 0 \,, \\
\label{3.6}
&& \frac{d P_R}{d t} + \frac{\pa H}{\pa R}
(R,P_R,P_\vphi)= {\cal F}_R^{\rm Ham}(R,\varphi,P_R,P_\vphi)\,, \\
\label{3.7}
&& \frac{d P_\vphi}{d t} = {\cal F}_\vphi^{\rm 
Ham}(R,\varphi,P_R,P_\vphi) \,,
\eea
where the Hamiltonian damping force ${\cal F}_i^{\rm Ham} (Q^j , P_j)$ is 
numerically equal to the Lagrangian one: ${\cal F}_i^{\rm Ham}(Q^j,P_j) = 
{\cal F}_i^{\rm Lag}(Q^j,\dot{Q}^j =  \pa 
H/\pa P_j)$\,.

\subsection{What do we know about the radiation reaction force?}
\label{subsec3.1}
The radiation reaction force ${\cal F}$ 
was computed explicitly, at lowest (Newtonian) fractional order, in 
harmonic Cartesian-like coordinates,
as part of the complete 2.5PN equations of motion, by Damour and Deruelle 
\cite{DD81a,D82,D83}. An equivalent result was also 
derived within the ADM canonical formalism by Sch\"afer \cite{S85,S86,SGauss}. 
At higher post-Newtonian orders one has only 
an incomplete knowledge of the equations of motion, and one has to rely 
on the (assumed) balance between energy and angular momentum losses in 
the system and at infinity \cite{IW,rad}. To get an idea of the 
generic structure of the radiation damping (in various coordinate 
systems, and at various PN approximations) let us consider the general 
radiation reaction force written (at 1PN fractional accuracy; and
setting $ G = 1$) by Iyer and 
Will \cite{IW}.
\bea
\label{3.8}
&& {\cal F}_i^{\rm Lag} = \mu\,\left [ \alpha(R,v)\,\dot{R}\,n^i 
+ \beta(R,v)\,v^i \right ]\,, \\
\label{3.9}
&& \alpha(R,v) = \frac{8}{5}\,\nu\,\frac{M}{R^2}\,
\frac{M}{R}\,(A_{5/2} + A'_{7/2} + \cdots )\,,\\
\label{3.10}
&& \beta(R,v) = -\frac{8}{5}\,\nu\,\frac{M}{R^2}\,
\frac{M}{R}\,(B_{5/2} + B'_{7/2} + \cdots )\,,
\eea
where $R$ is the relative radius and 
$v$ is the velocity.
Then, using post-Newtonian expressions for the energy and the angular 
momentum 
flux at 
infinity, and assuming energy and angular momentum 
balance, they obtained at lowest (Newtonian) fractional order
\bea
\label{3.11}
&& A_{5/2} = 3\,(1 + \overline{\beta})\,v^2 + \frac{1}{3}\,(23 + 
6\,\overline{\alpha} - 9\,\overline{\beta})\,\frac{M}{R} - 
5\,\overline{\beta}\,\dot{R}^2\,, \\
\label{3.12}
&& B_{5/2} = (2 + \overline{\alpha})\,v^2 + (2 - \overline{\alpha})\,
\frac{M}{R} - 3\,(1 + \overline{\alpha})\,\dot{R}^2\,.
\eea
See Ref.~\cite{IW} for the expressions of the 1PN-accurate radiation 
damping terms $A_{7/2}$ and $B_{7/2}$ in the equations of motion 
(equivalent, after some reshuffling, with the Lagrangian contributions 
$A'_{7/2}$, $B'_{7/2}$ in Eqs.~(\ref{3.9}), (\ref{3.10})).

The coefficients $\overline{\alpha}$ and $\overline{\beta}$ that appear
in Eqs.~(\ref{3.11}) and (\ref{3.12}) are two arbitrary 
gauge parameters that cannot be fixed by the energy balance method.
Iyer and Will \cite{IW} showed that this gauge freedom 
is equivalent to shifting the (conservative) coordinate 
system by small radiative corrections. Let us notice  that the gauge 
dependence is reduced when considering 
quasi-circular orbits. Indeed, in that case $\dot{R}^2 \simeq 0$,
$M/R \simeq v^2$ and Eqs.~(\ref{3.9}), ~(\ref{3.10}) become 
(considering only the 5/2PN terms which are sufficient for the point we 
wish to make)
\beq
\label{3.13}
\alpha_{\rm circ} \simeq \frac{8}{5}\,\nu\,\frac{M}{R^2}\,\left 
(\frac{M}{R}
\right )^2\,\left (\frac{32}{3} + 2\,\overline{\alpha} \right )\,,
\quad \quad 
\beta_{\rm circ} \simeq  -\frac{32}{5}\,\nu\,\frac{M}{R^2}\,\left 
(\frac{M}{R}
\right )^2\,.
\eeq
Hence, in the quasi-circular case the only gauge-dependence left 
is in the coefficient $\alpha (R , v)$ multiplying the {\it radial} 
component of the damping force $(\propto n^i)$. We can use this gauge 
arbitrariness to set the ratio
\beq
\label{3.14}
\left (\frac{\alpha}{\beta} \right )_{\rm circ.} \simeq  -\frac{1}{2}\,
\left (\frac{16}{3} + \overline{\alpha} \right )\,, 
\eeq
to any value we like. For example, by choosing $\overline{\alpha} = 
-16/3$ we can set ${\alpha}_{\rm circ.} =0$ or 
by choosing $\overline{\alpha} = 
-10/3$ we can set $({\alpha + \beta})_{\rm circ.} =0$.

Having understood the gauge dependence of the coefficient $\alpha$ in 
Eq.~(\ref{3.8}) let us come back to the general structure (\ref{3.8}) 
(considered at any PN accuracy, with some (unknown) coefficients $\alpha$ 
and $\beta$). The polar-coordinate version (for planar motion $\theta = 
\pi / 2$) of the Cartesian-like Lagrangian damping force (\ref{3.8}) 
reads (\ref{3.2}), (\ref{3.3}) with 
\bea
&& {\cal F}_R^{\rm Lag} = {\cal 
F}_i^{\rm Lag} \, \frac{\pa Q^i}{\pa R} = n^i \, {\cal F}_i^{\rm Lag}\,,\\
&& {\cal F}_{\varphi}^{\rm Lag} = {\cal F}_i^{\rm Lag} \, 
\frac{\pa Q^i}{\pa \varphi} = Q^x \, {\cal F}_y^{\rm Lag} - Q^y \, {\cal 
F}_x^{\rm Lag}.
\eea 
This yields
\beq
\label{3.15}
{\cal F}_\vphi^{\rm Lag} = \mu\,\beta\,R^2\,\dot{\vphi} \,, \quad 
\quad {\cal F}_R^{\rm Lag} = \mu\,(\alpha + \beta)\,\dot{R} \, .
\eeq

The important information for our present purpose is the difference 
between the $\varphi$-component of the damping force, which contains only 
$\beta$ and is, therefore, gauge-independent~\footnote{\label{f4}The discussion 
above concerns only the lowest-order term in $\beta$, but we shall see 
below that, to all orders, the crucial combination $\beta \, R^2$ can, 
for circular orbits, be expressed in terms of invariant quantities.}, and 
the $R$-component which contains the gauge-dependent combination $\alpha 
+ \beta$. Let us note, in particular, the expression of the ratio
\beq
\label{3.15b}
\frac{{\cal F}_R^{\rm Lag}}{{\cal F}_\vphi^{\rm Lag}} = 
\left (\frac{\alpha}{\beta} + 1 \right 
)\,\frac{\dot{R}}{R^2\,\dot{\vphi}}\,.
\eeq

In the following we shall be interested in quasi-circular motions with 
$\dot R \ll R \, \dot{\varphi}$. [We shall see that this condition 
remains satisfied even during part of the plunge phase.] As we see from 
Eq.~(\ref{3.15b}), for such motions the radial component of the damping 
force will contain one power of the small dimensionless quantity $\dot R  
/ (R \, \dot{\varphi})$. But we learned above, from the gauge dependence 
of the lowest-order damping force, that we can change the definition of 
the radial coordinate so as to set, for instance, the quantity $(\alpha / 
\beta) + 1$ to zero (for circular orbits). This means that the RHS of 
Eq.~(\ref{3.15b}) can be arranged, in the case of quasi-circular orbits, 
to contain three powers of the small parameter $\dot R/(R\,\dot{\varphi})$. 
({}From Eqs.~(\ref{3.11}), (\ref{3.12}) we see that for 
quasi-circular orbits $\alpha + \beta \propto {\dot R}^2$.) We have 
checked that the reasoning made above, using the lowest-order gauge 
dependence, can be formally extended to all higher PN orders.

The conclusion is that there should exist a special coordinate gauge 
where, for quasi-circular motions, an excellent approximation to the 
damping force is obtained by replacing the radial component simply by 
zero:
\beq
\label{3.16}
{\cal F}_R^{\rm Lag} = 0 = {\cal F}_R^{\rm Ham}\,.
\eeq
To test, a posteriori, the robustness of the approximation (\ref{3.16}), 
we shall also consider another special gauge: namely that where $(\alpha 
/ \beta)_{\rm circ} = 0$. [As we said above, this can be achieved at 
lowest order by a suitable choice of $\overline{\alpha}$, and this can be 
extended to higher PN orders by suitable choices of higher gauge 
parameters.] Finally, this means that there exists another coordinate 
gauge where, to an excellent approximation, the radial damping force is 
given as
\beq
\label{3.17}
{\cal F}_R^{\rm Lag} = {\cal F}_R^{\rm Ham} 
= \frac{\dot{R}}{R^2\,\dot{\vphi}}\,{\cal F}_{\vphi}^{\rm Ham}\,.
\eeq
The results in the two gauges are compared and discussed at the end of 
Sec.~\ref{sec6}.

What is important for the following is that in both gauges (\ref{3.16}) 
or (\ref{3.17}), the knowledge of the full damping force can be deduced 
from the sole knowledge of ${\cal F}_{\varphi}$.

\subsection{Non-perturbative estimate of the angular momentum reaction 
force along quasi-circular orbits}

The analysis of the previous subsection has shown that the crucial 
equation in which one should accurately incorporate radiation reaction 
effects is
\beq
\label{3.17b} 
\frac{d P_\vphi}{d t} = {\cal F}_\vphi^{\rm Ham}(R,\varphi,P_R,P_\vphi) 
\, .
\eeq
As $P_\vphi$ is just the total angular momentum of the binary system, 
Eq.~(\ref{3.17b}) expresses the rate of loss of angular momentum under 
gravitational radiation reaction. As usual we shall estimate the RHS 
${\cal F}_\vphi = {\cal F}_\vphi^{\rm Ham} = {\cal F}_\vphi^{\rm Lag}$ 
(remember that ${\cal F}^{\rm Ham}$ and ${\cal F}^{\rm Lag}$ differ only 
in the arguments in which they are expressed) by assuming that there is a 
balance between the mechanical angular momentum lost by the system, and 
the flux of angular momentum at infinity in the form of gravitational 
waves. In the case of interest here of quasi-circular orbits we expect 
that, to a good approximation, ${\cal F}_\vphi$ will not depend 
explicitly on $\varphi$ and will, therefore, be expressible in terms of 
the {\it orbit-averaged} flux of angular momentum. Moreover, in the case 
of quasi-circular orbits there is a simple relation between
angular-momentum-loss and energy-loss. 
Indeed, the rate of energy-loss along any orbit, in polar coordinates, 
is given by
\beq
\label{3.18}
\frac{d {\cal E}}{d t} = 
\frac{d H}{d t} = \dot{R}\,{\cal F}_{R} +
\dot{\vphi}\,{\cal F}_{\vphi}\,,
\eeq
and in particular along quasi-circular orbit we have (remembering 
Eq.~(\ref{3.15b}))
\beq
\label{3.19}
\left (\frac{d H}{d t} \right )_{\rm quasi-circ.} 
\simeq \dot{\vphi}\,{\cal F}_{\vphi}^{\rm circ.} + {\cal O}(\dot{R}^2)\,.
\eeq
Finally, if we know some good estimate of the (averaged) energy-loss 
along circular orbits, say
\beq
\label{3.20}
\left (\frac{d H}{d t} \right )_{\rm circ.} 
\simeq - \Phi_{\rm circ.}(\dot{\vphi})\,,
\eeq
we can obtain a good estimate of the needed $\varphi$-reactive force 
\beq
\label{3.21}
{\cal F}_{\vphi}^{\rm circ.} \simeq -\frac{\Phi_{\rm 
circ.}(\dot{\vphi})}{\dot{\vphi}}\,.
\eeq
The problem of giving a non-perturbative, re-summed estimate of the 
energy-loss-rate (or ``flux function'') along circular orbits, say 
$\Phi_{\rm circ}$, has been recently tackled by Damour, Iyer and 
Sathyaprakash \cite{DIS}. By combining several of the non-perturbative 
techniques recalled above (to work with an invariant function $F(v)$, to 
use some global information about $F(v)$ in the complex $v$-plane, to use 
Pad\'e approximants) Ref.~\cite{DIS} came up with the following 
expression for $\Phi_{\rm circ}$, considered as a function of the 
gauge-invariant observable
\beq
\label{3.21b}
v_{\omega} \equiv ({GM} \, \omega / c^3)^{1/3}\,; \quad \quad \omega \equiv 
\dot{\varphi} \, ,
\eeq
namely,
\beq
\label{3.22}
\Phi_{\rm circ.} = F_{\rm DIS}(v_{\omega}) = 
\frac{32}{5G}\,\nu^2\,v_\omega^{10}\,
\frac{\widehat{f}_{\rm DIS}(v_\omega;\nu)}{1 - v_\omega/v_{\rm pole}(\nu)}\,.
\eeq
Here, and in the following, we set $c=1$ to simplify formulas. The 
function $\widehat{f}_{\rm DIS} (v_{\omega} ; \nu)$ entering 
Eq.~(\ref{3.22}) is the ``factored flux function'' of \cite{DIS}, scaled 
to the Newtonian (quadrupole) flux (hence the caret on $f_{\rm DIS}$). It 
was shown in \cite{DIS} that the sequence of near-diagonal Pad\'e 
approximants of $\widehat{f}_{\rm DIS} (v)$ exhibits a very good 
convergence (at least in the $\nu = 0$ limit where high-order PN 
expansions are known \cite{TTS}) toward the exact result (numerically 
known when $\nu = 0$ \cite{P95}). On this basis, it was argued in 
\cite{DIS} that, in the comparable-mass case, $\nu \ne 0$, our ``best 
estimate'' of $\widehat{f}$ is obtained by Padeeing the currently most 
complete post-Newtonian results, namely the 2.5PN ones \cite{2.5PN}. This 
yields a result of the form
\beq
\label{3.23}
\widehat{f}_{\rm DIS}(v;\nu) = 
\frac{1}{1 + \frac{c_1\,v}{1 + \frac{c_2\,v}{1 + 
\frac{c_3\,v}{1 +\frac{c_4\,v}{1 + c_5\,v}}}}}\,,
\eeq
where the dimensionless coefficients $c_i$ depend only on $\nu$. The 
$c_k$'s are some explicit functions of the coefficients $f_k$ of the 
straightforward Taylor expansion of $\widehat{f} (v)$. In turn, the 
$f_k$'s, being defined by the identity (where $T$ means ``Taylor 
expansion'')
\beq
\label{3.27n}
T \, [\widehat{f} (v)] \equiv T \left[ \left( 1 - \frac{v}{v_{\rm pole}} 
\right) \, \widehat{F} (v) \right] = 1 + f_1 v + f_2 v^2 + \cdots
\eeq
are given by
\beq
\label{3.28n}
f_k = F_k - F_{k-1} / v_{\rm pole}
\eeq
in terms of the Taylor coefficients of the usual (Newton-normalized) flux 
function
\beq
\label{3.29n}
T \, [\widehat{F} (v)] \equiv T \left[ \frac{5 G}{32 \, \nu^2 \, 
v^{10}} \, F(v) \right] = 1 + F_2 v^2 + F_3 v^3 + \cdots \,.
\eeq
[Note that $F_1 = 0$, but that $f_1 = -1 / v_{\rm pole} \ne 0$.] More explicitly 
we have
\beq
F_2 = -\frac{1247}{336} - \frac{35}{12}\,\nu\,,\quad \quad 
F_3 = 4 \pi\,, 
\eeq
\beq
F_4 = -\frac{44711}{9072} + \frac{9271}{504}\,\nu + 
\frac{65}{18}\,\nu^2\,, \quad \quad F_5 = - 
\left (\frac{8191}{672} + \frac{535}{24}\,\nu \right )\,\pi\,,
\eeq
and 
\bea
&& c_1 = - f_1\,, \quad \quad c_2 = f_1 - \frac{f_2}{f_1}\,, 
\quad \quad c_3 = \frac{f_1\,f_3 - f_2^2}{f_1\,(f_1^2 - f_2)}\,,\\
&& c_4 = -\frac{f_1\,(f_2^3 + f_3^2 + f_1^2\,f_4 - f_2\,(2\,f_1\,f_3 + f_4))}
{(f_1^2 -f_2)\,(f_1\,f_3 - f_2^2)}\,,\\
&& c_5 = -\frac{(f_1^2 - f_2)\,(-f_3^3 + 2f_2\,f_3\,f_4 - f_1\,f_4^2 -
f_2^2\,f_5 + f_1\,f_3\,f_5)}{(f_1\,f_3 - f_2^2)\,
(f_2^3 + f_3^2 + f_1^2\,f_4 - f_2\,(2\,f_1\,f_3 + f_4))}\,.
\eea
As is clear from these expressions, they depend on the definition used 
for the quantity $v_{\rm pole} (\nu)$ which represents a $\nu$-dependent 
estimate of the location of the ``pole'' in $\Phi_{\rm circ}$, which 
coincides (see the discussion in \cite{DIS}) with the location of the 
``light-ring'' or last unstable circular orbit (
$R_{\rm light-ring}^{\rm Schw.} = 3 {GM}$ in the $\nu \rightarrow 0$ limit). 
Actually, as we shall use the Pad\'e representation only above and around 
the LSO ($R_{\rm LSO}^{\rm Schw.} = 6 {GM}$ when $\nu = 0$) the 
precise choice of $v_{\rm pole} (\nu)$ is probably not crucial (as long 
as it stays near its known $\nu = 0$ limit: $v_{\rm pole} (\nu = 0) = 1 / 
\sqrt 3$). In this work, we shall follow Ref.~\cite{DIS} and use the pole 
location they obtained from Padeing their ``new'' energy function $e(x)$, 
namely
\beq
\label{3.24}
v_{\rm pole}^{\rm DIS} = \frac{1}{\sqrt{3}}\,\sqrt{\frac{1 + 
\frac{1}{3}\,\nu}{1 - \frac{35}{36}\,\nu}}\,. 
\eeq 
Then, combining Eqs.~(\ref{3.21}), (\ref{3.21b}) and (\ref{3.22}) we 
define our best estimate of the $\varphi$-component of the radiation 
reactive force along quasi-circular orbits as:
\beq
\label{3.25}
{\cal F}_{\vphi}^{\rm circ.} \equiv - \frac{GM}{v_\omega^3}\,\Phi_{\rm 
DIS}(v_\omega)
= - \frac{32}{5}\,\mu\,\nu\,v_\omega^7\,
\frac{\widehat{f}_{\rm DIS}(v_\omega;\nu)}{1 - v_\omega/v_{\rm pole}^{\rm DIS} 
(\nu)}\,.
\eeq
To ease the notation we shall work in the following with reduced 
quantities, 
that is:
\bea
\label{3.26}
&& r \equiv \frac{R}{G\,M} \,, \quad \quad p_r \equiv \frac{P_R}{\mu} \,, 
\quad \quad p_\vphi \equiv \frac{P_\vphi}{\mu\,G\,M} = \frac{{\cal 
J}}{\mu\,G\,M} \equiv j \,, \\
\label{3.27}
&& \wt \equiv \frac{t}{G\,M} \,, 
\quad \quad \wH \equiv \frac{H_{\rm real}^{\rm improved}}{\mu} \,,
\quad \quad \wH_{\rm eff} \equiv \frac{H_{\rm eff}}{\mu} \,.
\eea
Finally, the dynamics, including radiation reaction, in re-scaled 
coordinates, is explicitly described by the 
following system of equations (in the ``canonical'' case where 
Eq.~(\ref{3.16}) holds)
\bea 
\label{3.28}
&&\frac{dr}{d \wt} = \frac{\pa \wH}{\pa p_r}
(r,p_r,p_\vphi)\,, \\
\label{3.29}
&& \frac{d \vphi}{d \wt} = \ww \equiv \frac{\pa \wH}{\pa p_\vphi}
(r,p_r,p_\vphi)\,, \\
\label{3.30}
&& \frac{d p_r}{d \wt} + \frac{\pa \wH}{\pa r}
(r,p_r,p_\vphi)=0\,, \\
&& \frac{d p_\vphi}{d \wt} = \wF_\vphi(\ww (r,p_r,p_{\varphi}))\,,
\label{3.31}
\eea
with
\beq
\label{3.32}
\wH = \frac{1}{\nu}\,\sqrt{1 + 2\nu\,\left [\sqrt{A(r)\,\left (1 + 
\frac{p_r^2}{B(r)} + 
\frac{p_\vphi^2}{r^2} \right )} -1 \right ]}\,,\\
\eeq
\beq
\label{3.33}
\wF_\vphi (v_{\omega} \equiv \ww^{1/3}) = \frac{{\cal F}_{\vphi}}{\mu} = 
- \frac{32}{5}\,\nu\,v_\omega^7\,
\frac{\widehat{f}_{\rm DIS}(v_\omega;\nu)}{1 - v_\omega/v_{\rm pole}^{\rm DIS} 
(\nu)}\,,
\eeq
and where in Eq.~(\ref{3.32}) we use the scaled versions of our current 
best estimate of the effective metric 
coefficients $A(r)$, $B(r)$, see \cite{BD99} and 
Eqs.~(\ref{2.13})--(\ref{2.15}) above, that is
\beq
\label{3.34}
A(r) \equiv 1 - \frac{2}{r} + \frac{2\nu}{r^3} \,, 
\quad \quad B(r) \equiv \frac{1}{A(r)}\,\left (1 - \frac{6\nu}{r^2} 
\right )\,. 
\eeq
Note that the argument $v_{\omega}$ entering $\widehat{\cal F}_{\vphi}$, 
Eq.~(\ref{3.33}), is simply defined as $v_{\omega} \equiv \ww^{1/3}$, 
where $\ww \equiv \omega\,(G M)$ is the function of $r$, $p_r$ and $p_{\vphi}$ defined by 
Eq.~(\ref{3.29}), i.e. $\ww (r,p_r,p_{\vphi}) \equiv \pa \widehat H 
(r,p_r,p_{\vphi}) / \pa p_{\vphi}$.

\section{Transition between inspiral and plunge}
\label{sec4}

\begin{figure}
\begin{center}
\begin{tabular}{cc}   
\hspace{-0.8cm} 
\epsfig{file=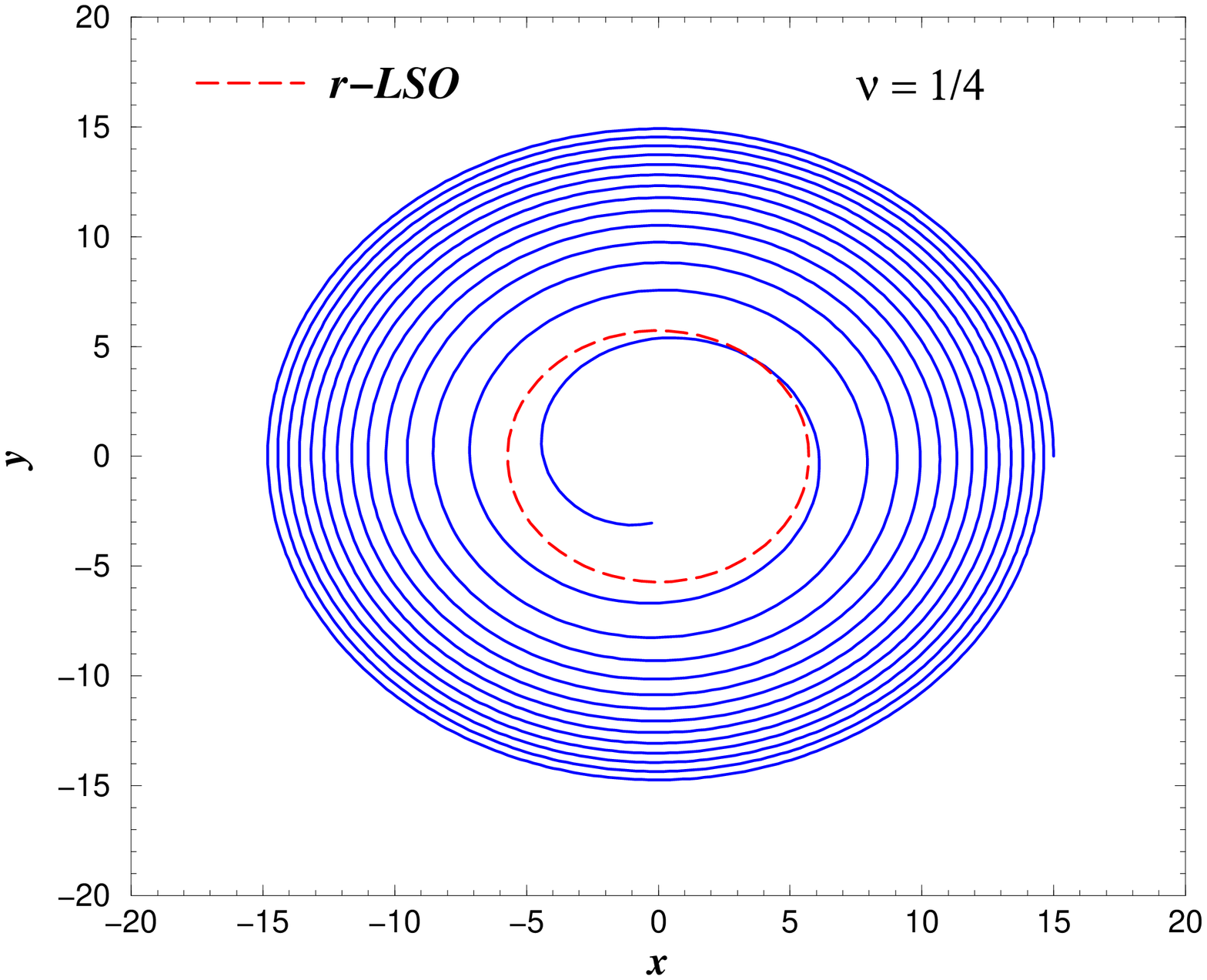,width=0.39\textwidth,height = 
0.39\textwidth,angle=-90} & 
\hspace{-0.0cm}
\epsfig{file=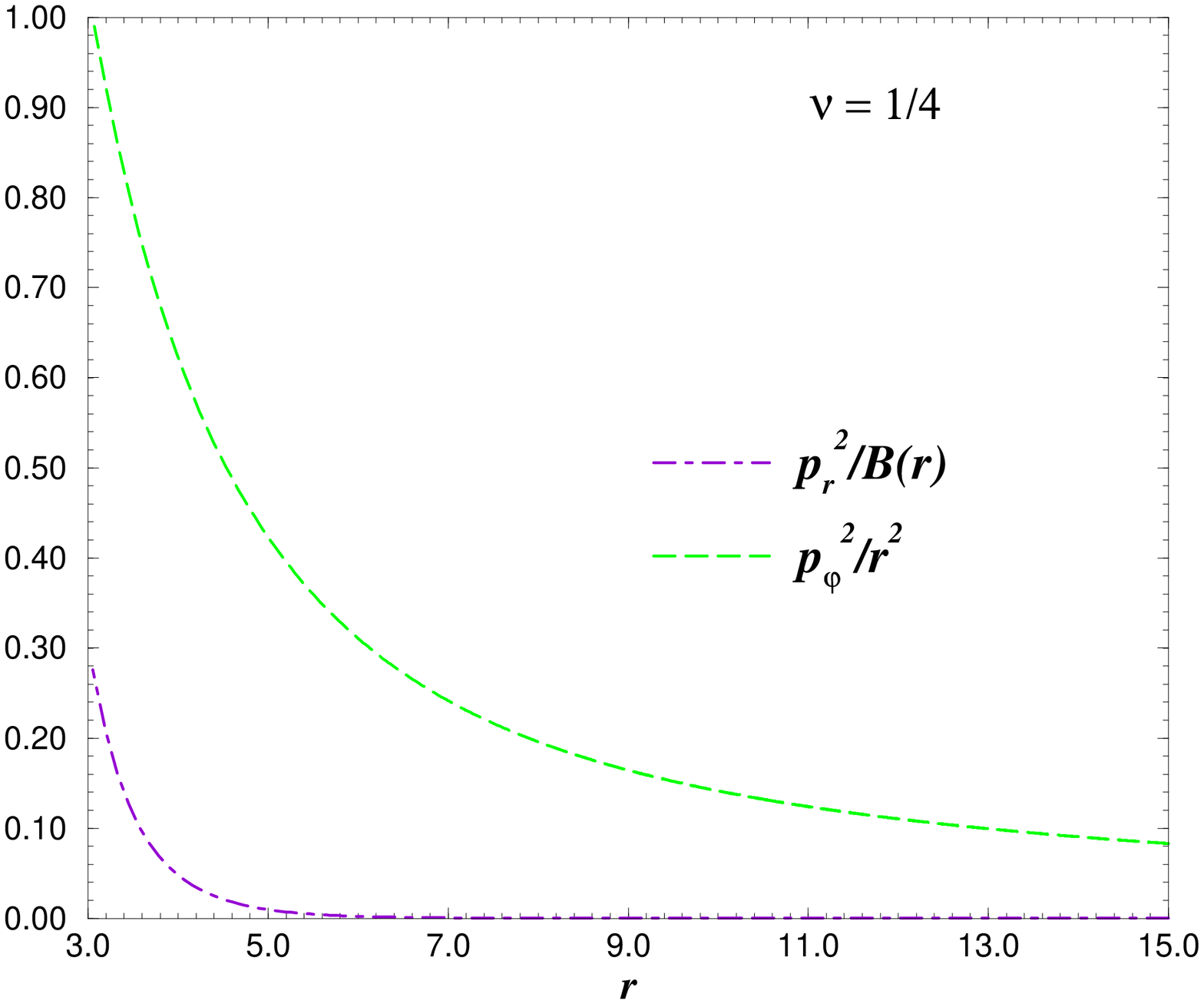,width=0.39\textwidth,height = 
0.39\textwidth,angle=-90}
\end{tabular}
\caption{\sl On the left panel we show the inspiraling circular (relative) orbit 
for $\nu = 1/4$. The location of the $r$-LSO, defined by the conservative 
part of the dynamics, is also indicated.
On the right panel we compare the two kinetic contributions
that enter the  Hamiltonian: the ``radial'' and 
the ``azimuthal'' one. The figure shows that the assumption 
we made of quasi-circularity, i.e. $p_r^2 / B(r) \ll p_{\vphi}^2 
/ r^2$, is well satisfied throughout the transition 
from the adiabatic phase to the plunge.}
\label{Fig1}
\end{center}
\end{figure}

The first-order evolution system (\ref{3.28})--(\ref{3.31}) defines our 
proposed best estimate for completing the usually considered 
``adiabatic'' inspiral evolution into a system which exhibits a smooth 
transition between inspiral and plunge. 
The rest of this paper will be 
devoted to extracting some of the important information contained in this 
new evolution system. Before coming to grips with such detailed 
information, it is useful to have a first visual impression of the 
physics contained in our system (\ref{3.28})--(\ref{3.31}). To do this we 
plot on the left panel of Fig.~\ref{Fig1} the result of a full numerical 
evolution of Eqs.~(\ref{3.28})--(\ref{3.31}) in the equal-mass case ($\nu 
= 1/4$). 
We started the evolution at $r = 15$, $\vphi = 0$ and used  
as initial values for $p_\vphi$ and $p_r$ the ones provided by the 
adiabatic approximation (see Eqs.~(\ref{4.4}) and (\ref{4.10}) 
below).
The dashed circle in this 
plot indicates the radial coordinate location of the 
LSO defined by the conservative part of the dynamics, 
i.e. by the Hamiltonian $\widehat{H} (r,p_r,p_{\vphi})$. More precisely 
this ``$r$-LSO'' is simply defined (for any $\nu$) by $r = r_{\rm LSO} 
(\nu)$, where $r_{\rm LSO} (\nu)$ is the solution of Eq.~(\ref{2.16}). In 
particular, $r_{\rm LSO} \left( \frac{1}{4} \right) = 5.718$, as recalled 
in Eq.~(\ref{2.17}). Note that, in presence of radiation reaction 
effects, there is an arbitrariness in what one would like to mean by 
saying: ``the system is crossing the LSO''. Indeed, we could define the 
``LSO-crossing'' in several inequivalent ways, notably: (i) $r$-LSO: the 
time when $r = r_{\rm LSO} (\nu)$; (ii) $j$-LSO: the time when $p_{\vphi} 
\equiv j = j_{\rm LSO} (\nu)$; (iii) $\omega$-LSO: the time when $d\vphi 
/ d\wt \equiv \ww = \ww_{\rm LSO} (\nu)$. [The ``LSO'' functions of 
$\nu$ being defined by solving Eq.~(\ref{2.16}); see Eq.~(\ref{2.17}).] 
This arbitrariness is not a problem. Our new evolution system 
(\ref{3.28})--(\ref{3.31}) describes a smooth transition ``through'' the 
formally defined ``old'' LSO, and does not care about old definitions. In 
other words, when $\nu$ is finite, and especially when $\nu \simeq 1/4$ 
(which, one should remember, is expected to be an {\it accumulation point} 
of observed values of $\nu$; see footnote~\ref{f2} above) the smooth transition 
process blurs the notion of LSO. It is only for $\nu \ll 1$ (see below) 
that one recovers a sharp transition near the $H$-defined LSO. On the 
right panel of Fig.~\ref{Fig1} we compare the two kinetic contributions 
to the Hamiltonian (\ref{3.32}): the ``azimuthal'' contribution 
$p_{\vphi}^2 / r^2$, and the ``radial'' contribution $p_r^2 / B(r)$. One 
sees on this Figure that our basic assumption of quasi-circularity 
(which, at the level of $\widehat H$, means $p_r^2 / B(r) \ll p_{\vphi}^2 
/ r^2$) is well satisfied throughout the transition. In fact, even down to 
$r \simeq 3.79$ one has $p_r^2 / B(r) \, < \,0.1 \, p_{\vphi}^2 / r^2$. The 
radial kinetic energy would become equal to the azimuthal one only below 
$r=3$. We shall, anyway, not use, in the following, our system below the 
(usual) ``light-ring'' $r \simeq 3$ (where $p_r^2 / B(r) \simeq 0.30 \, 
p_{\vphi}^2 / r^2$).

\begin{figure}
\begin{center}
\begin{tabular}{cc}   
\hspace{-0.8cm} 
\epsfig{file=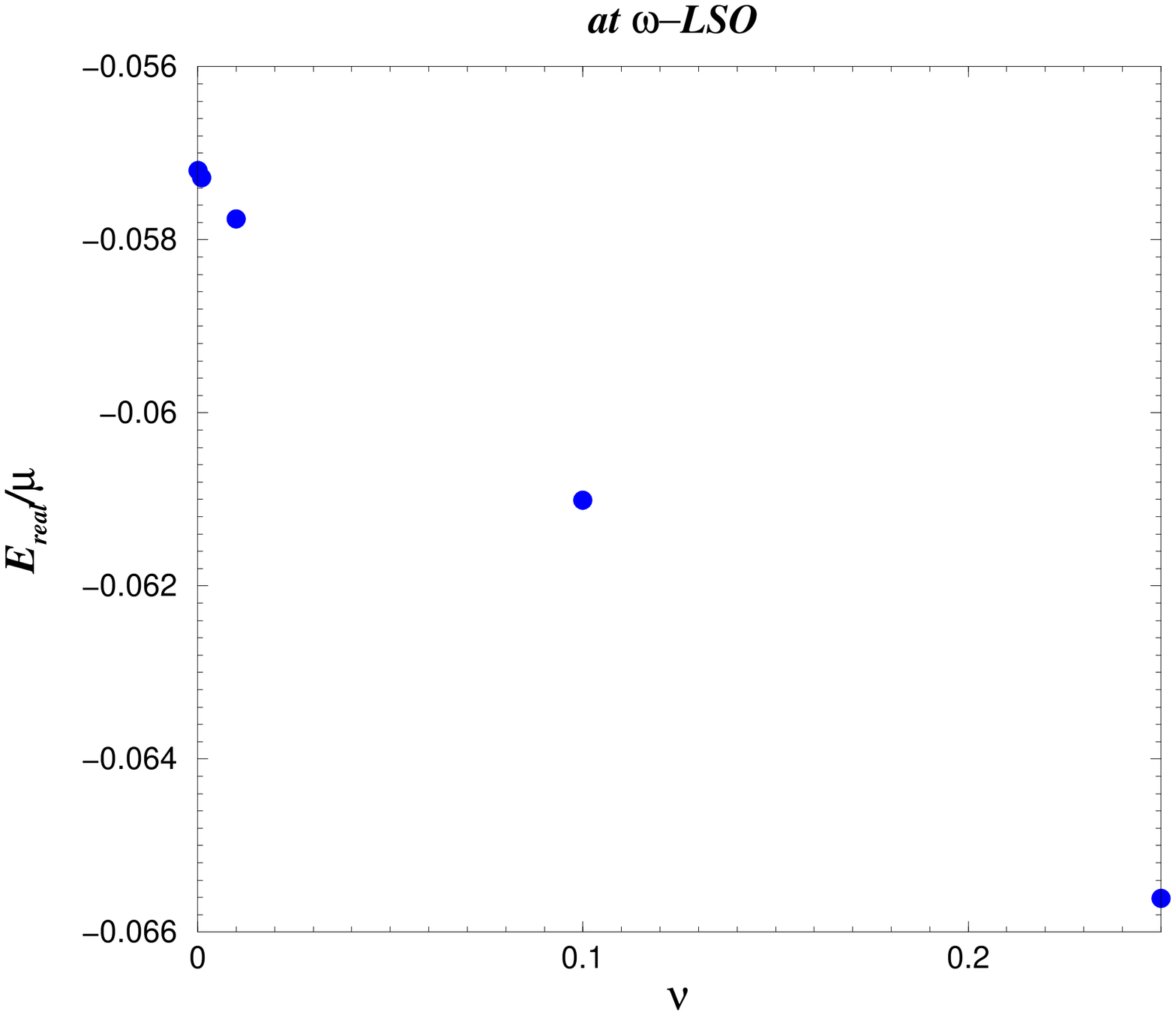,width=0.5\textwidth,height = 
0.5\textwidth,angle=-90} & 
\hspace{-0.0cm}
\epsfig{file=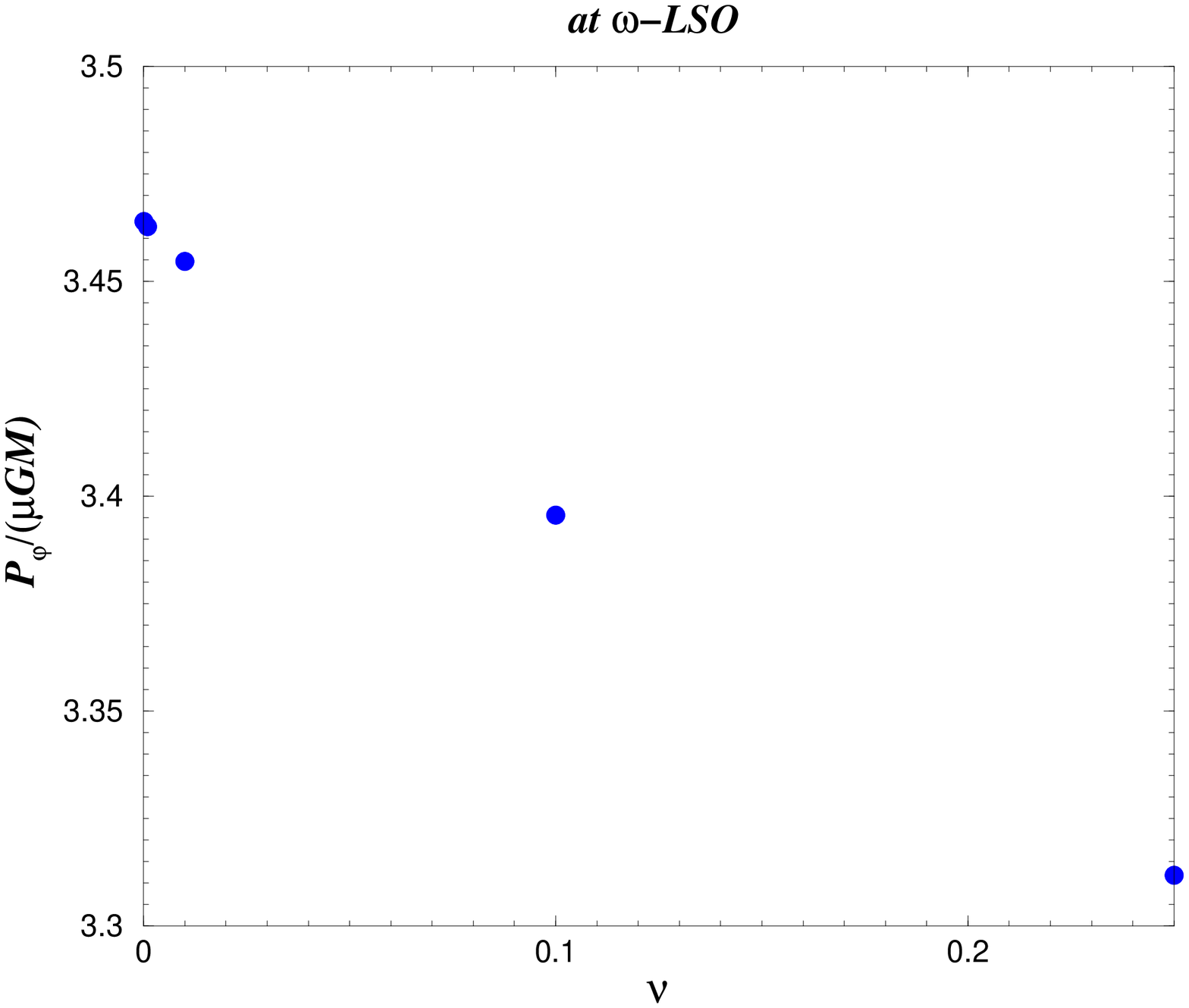,width=0.5\textwidth,height = 
0.5\textwidth,angle=-90}
\end{tabular}
\caption{\sl Variation with $\nu$ of the $\omega$-LSO values of the real 
reduced non-relativistic energy $ E_{\rm real}^{\rm NR}/\mu  = 
({\cal E}_{\rm real} - M)/\mu $ (on the left), and of the real angular momentum 
$j = P_\varphi/(\mu G M)$ (on the right), 
computed integrating the full dynamics, 
i.e. with radiation reaction effects included.}
\label{Fig2}
\end{center}
\end{figure}

\begin{figure}
\begin{center}
\begin{tabular}{cc}   
\hspace{-0.8cm} 
\epsfig{file=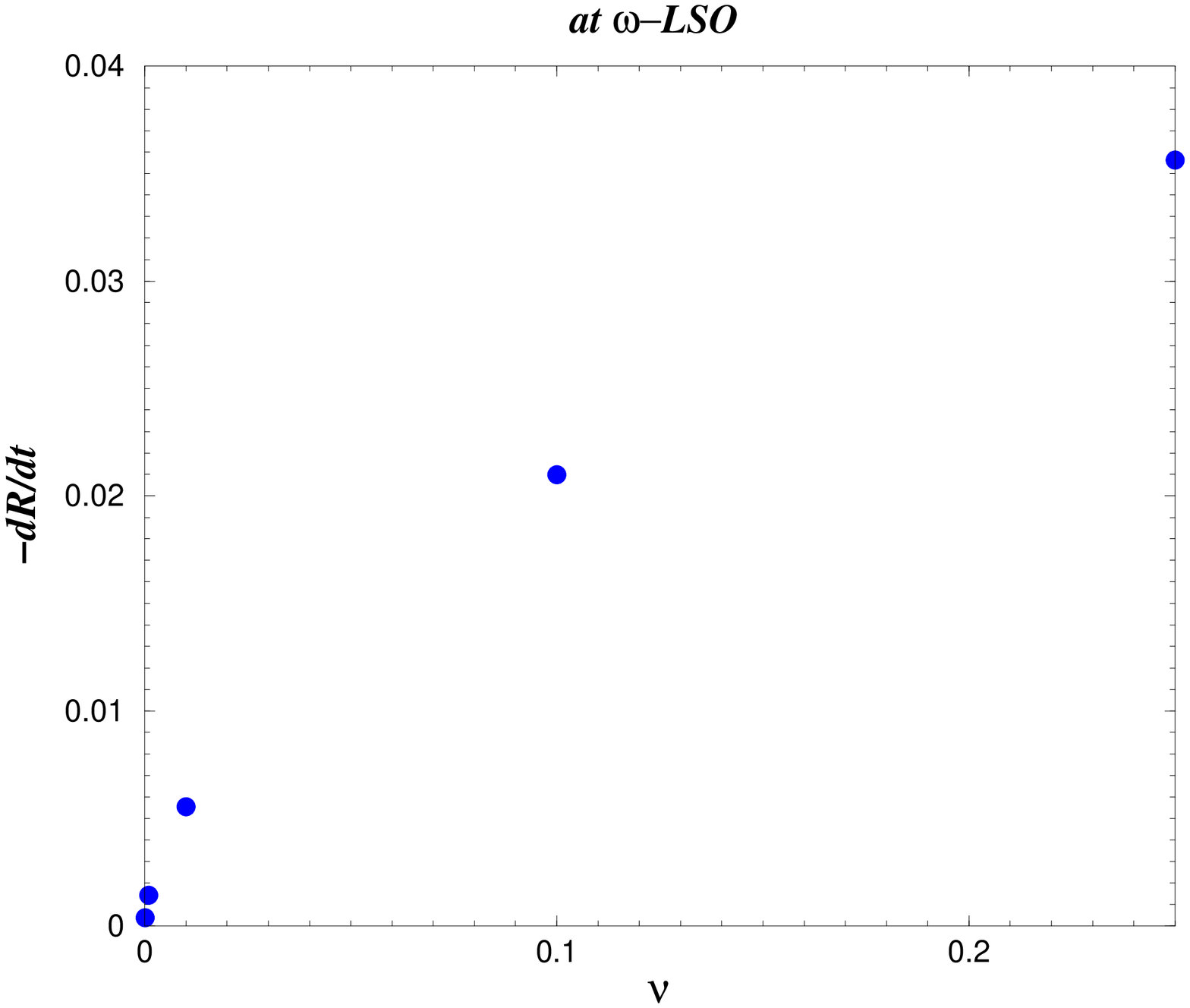,width=0.5\textwidth,height = 
0.5\textwidth,angle=-90} & 
\hspace{-0.0cm}
\epsfig{file=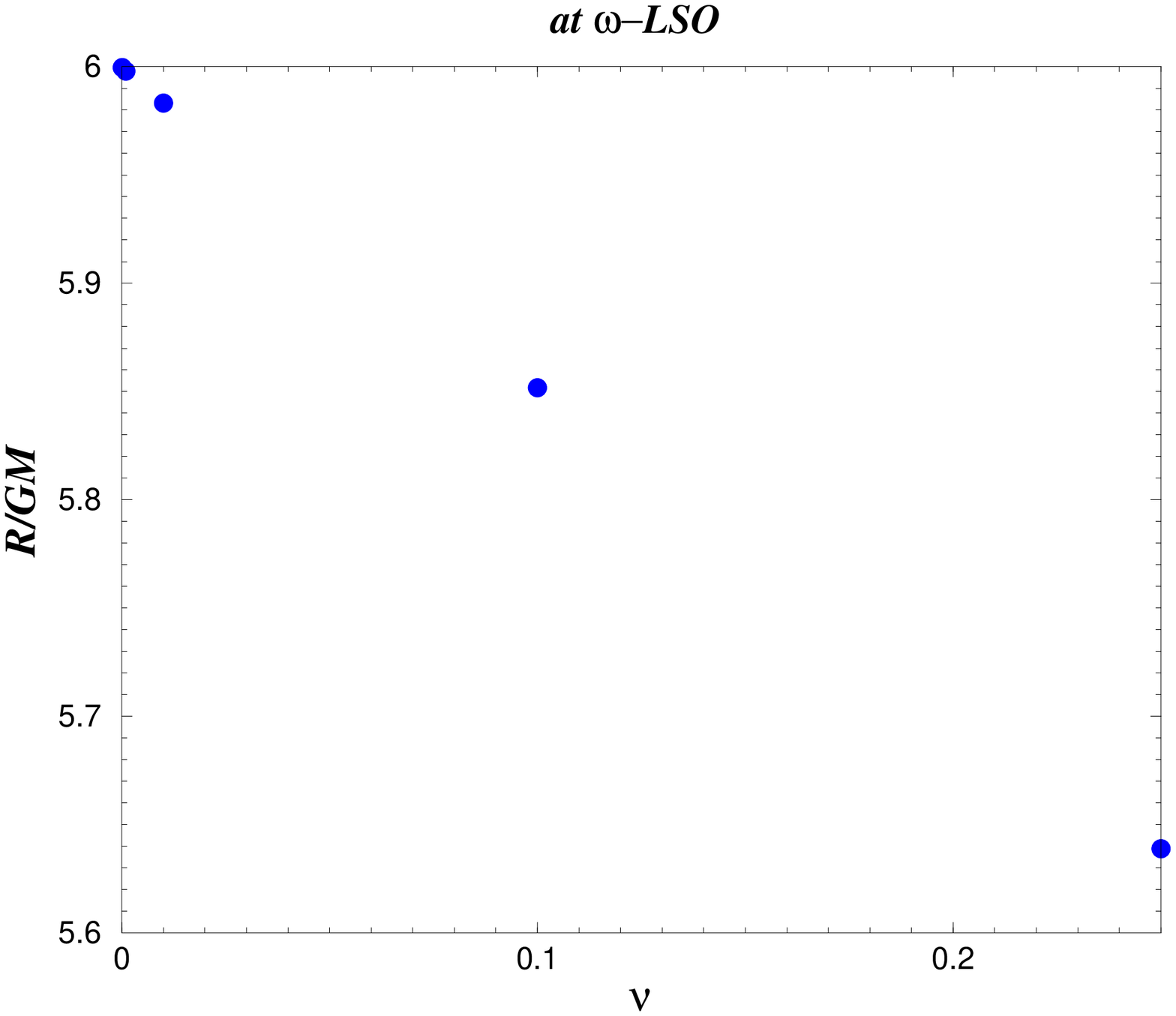,width=0.5\textwidth,height = 
0.5\textwidth,angle=-90}
\end{tabular}
\caption{\sl $\omega$-LSO values of the 
radial velocity (on the left) and of the radial position 
(on the right) versus $\nu$, derived integrating the full 
dynamical evolution.}
\label{Fig3}
\end{center}
\end{figure}

We exhibit more quantitative results on the transition between the inspiral and 
the plunge in Figs.~\ref{Fig2} and \ref{Fig3}. These figures plot the values of 
several 
physical quantities (energy, angular momentum, radial velocity and radial 
coordinate) 
computed at the $\omega$-LSO (i.e. when $\omega = \omega_{\rm LSO} (\nu)$) after 
integration of the system (\ref{3.28})--(\ref{3.31}). The energy which is 
plotted is 
the reduced non-relativistic real energy, i.e. $({\cal E}_{\rm real} - M) / 
\mu$. [In the test-mass limit, this reduced energy equals $\sqrt{8/9} - 1
= - 0.057191$.]

Having obtained, through Figs.~\ref{Fig2} and \ref{Fig3}, a first impression 
of the physics of the inspiral $\rightarrow$ plunge transition, we shall now 
study in more detail this transition, notably by comparing it with 
various analytical approximations. The first approximation we shall 
consider is the current standard one used for dealing with the inspiral 
phase: the adiabatic approximation.

\subsection{Comparison with the adiabatic approximation}
\label{subsec4.1}

Let us compare the exact numerical evolution  with the usual 
adiabatic approximation to inspiral motion. This approximation is  
defined by saying that the (effective) body follows an adiabatic 
sequence of exact circular orbits whose energy is slowly drained out by 
gravitational radiation. It is obtained from  
Eqs.~(\ref{3.28}), (\ref{3.31}), by neglecting $p_r^2$, i.e. by setting 
$p_r=0$. Noticing that $\pa \wH/\pa p_r = 2 p_r\,\pa \wH/\pa p_r^2 
\propto p_r$ we get that $dr/d \wt$ vanishes linearly with $p_r$.  The 
first equation (\ref{3.28}) is then formally satisfied with $p_r = 0 = 
\dot r$.
Imposing now $p_r=0$ in Eqs.~(\ref{3.29}) and (\ref{3.30}) we obtain two 
further equations:
\bea
\label{4.3.1}
&& \frac{\pa \wH_0}{\pa r}(r,p_\vphi) = 0\,, \\
\label{4.3.2}
&& \ww = \frac{\pa \wH_0}{\pa p_\vphi}(r,p_\vphi)\,, 
\eea
where we define 
\beq
\wH_0(r,p_\vphi) \equiv \wH(r,p_r = 0,p_\vphi)  = 
\frac{1}{\nu}\,\sqrt{1 + 2\nu\,\left [\sqrt{A(r)\,\left (1 + 
\frac{p_\vphi^2}{r^2} \right )} -1 \right ]}\,.
\eeq
Eq.~(\ref{4.3.1})
provides a link between $r$ and $p_\vphi \equiv j$ 
in the adiabatic limit. 
{}From the structure (\ref{3.32}) of $\widehat H$, it is easily seen that 
Eq.~(\ref{4.3.1}) is equivalent to looking for the minimum, say (for 
convenience) in the variable $u \equiv 1/r$, of the ``radial potential''
\beq
\label{4.3b}
W_j (u) = A (u) \, [1 + j^2 \, u^2 ] \, .
\eeq
Solving $\pa_u \, W_j (u)=0$ gives a parametric representation of $j^2$ in 
terms of $u$:
\beq
\label{4.3c}
j_{\rm adiab}^2 (u) = - \frac{A' (u)}{(u^2 \, A(u))'} \, ,
\eeq
where the prime denote $d/du$. In the case where the function $A$ is 
given by Eq.~(\ref{3.34}), i.e. $A(u) = 1 - 2u + 2\nu \, u^3$, 
Eq.~(\ref{4.3c}) yields, in term of the orignal (reduced) radial variable 
$r = 1/u$
\beq
\label{4.4}
j_{\rm adiab.}^2(r) = \frac{r^2\,(r^2 - 3\,\nu)}
{r^3 -3r^2 + 5\nu}\,. 
\eeq
Note that there exist real circular orbits (though possibly unstable 
ones) as long as $j_{\rm adiab}^2 (r) > 0$, i.e. as long as 
$ r^3 - 3r^2 + 5\nu > 0$.
In fact the positive, real solution in $r$ of 
\beq
\label{4.5}
[r^3 - 3r^2 + 5\nu]_{\rm light-ring} =0
\eeq
defines the light-ring or last unstable circular orbit (with $j^2 
(r_{\rm light-ring}) = + \infty$). We 
find $r_{\rm light-ring} \simeq 2.84563$ in the case $\nu = 1/4$. 
Eq.~(\ref{4.3.2}) then gives the parametric 
representation of $\ww = \omega\,(G M)$ throughout the adiabatic phase for circular 
orbits:
\beq
\label{4.6}
\ww_{\rm adiab.}(r) = \frac{1}{r^{3/2}}\,\frac{\sqrt{1 - 3\nu/r^2}}{
\sqrt{1 + 2\nu(\sqrt{z(r)} -1)}}\,, 
\eeq
where $z(r)$ denotes the following quantity
\beq
\label{4.7}
z(r) \equiv \wH_{\rm eff}^2(r,p_r=0,p_\vphi = j_{\rm adiab.}) = 
\frac{r^3\,A^2(r)}{r^3 - 3\,r^2 + 5\nu}\,.
\eeq 
Note that the effective one-body description seems to become 
somewhat unsatisfactory at the light-ring (at least for exactly 
circular orbits).
 Indeed, we see from Eqs.~(\ref{4.6}) and (\ref{4.7}) that the blow 
up of $z(r)$, i.e. of the effective energy, at the light-ring, Eq.~(\ref{4.5}), 
implies that the real orbital frequency of circular 
orbits, $\widehat{\omega}_{\rm circ.} (r)$, Eq.~(\ref{4.6}), tends to zero at 
the light-ring. 
This is probably an unphysical behaviour [from the test-mass limit, one expects 
the orbital frequency to have a non-zero limit at the light-ring; see, e.g., 
Ref.~\cite{DIS} where Pad\'e approximants are used to compute a finite value of 
$\widehat{\omega}_{\rm light-ring} (\nu)$].
The other factors in Eq.~(\ref{4.6}) imply, as expected, a regular {\it 
increase} of 
$\widehat{\omega} (r)$ as $r$ decreases below the LSO. Pending the construction 
of an improved version of the effective one-body approach which would be better 
behaved, we have decided, when dealing with the evolution of the system 
(\ref{3.28})--(\ref{3.31}), 
to stop the simulation at the light-ring.
[In our simulations of plunging orbits the effective energy stays bounded,
but the orbital frequency  $\widehat{\omega} (\wt)$  levels off very close
to the light-ring.] 

Finally Eq.~(\ref{3.31}) becomes in the adiabatic limit 
\beq
\label{4.8}
\frac{d j}{d \wt} = 
\wF_\vphi\left (\frac{\pa \wH_0}{\pa p_\vphi}(r,j) \right )\,.
\eeq
Then using $d j/d \wt = (dj/dr)\,(dr/d \wt)$ 
and $d \vphi = \ww\, d\wt $ we can solve the motion in the adiabatic 
limit by quadratures:
\beq
\label{4.8b}
d \wt_{\rm adiab.} = \left (\frac{d j_{\rm adiab.}}{d r} \right )
\,\frac{dr}{\wF_\vphi(\ww_{\rm adiab.}(r))}\,, 
\eeq
\beq
\label{4.9}
d \vphi_{\rm adiab.} = \left (\frac{d j_{\rm adiab.}}{d r} \right )
\,\frac{\ww_{\rm adiab.}(r)}{\wF_\vphi(\ww_{\rm adiab.}(r))}\,dr\,. 
\eeq
The radial velocity $v_r \equiv dr/ d\wt$, as a function of the parameter 
$r$,  in the adiabatic limit, is given by:
\beq
\label{4.10}
v_r^{\rm adiab.} = \frac{\wF_\vphi(\ww_{\rm adiab.}(r))}{{d j_{\rm 
adiab.}}/{d 
r}}\,.
\eeq
Note that $v_r^{\rm adiab.}$ formally tends to $-\infty$ when $r \rightarrow 
r_{\rm 
LSO}$ (indeed, $j_{\rm adiab.} (r)$ reaches, by definition, a minimum at $r = 
r_{\rm 
LSO}$). This shows that the adiabatic approximation is meaningful only during 
the 
inspiral phase (i.e. ``above'' the LSO). 
\begin{figure}
\begin{center}
\begin{tabular}{cc}   
\hspace{-0.8cm} 
\epsfig{file=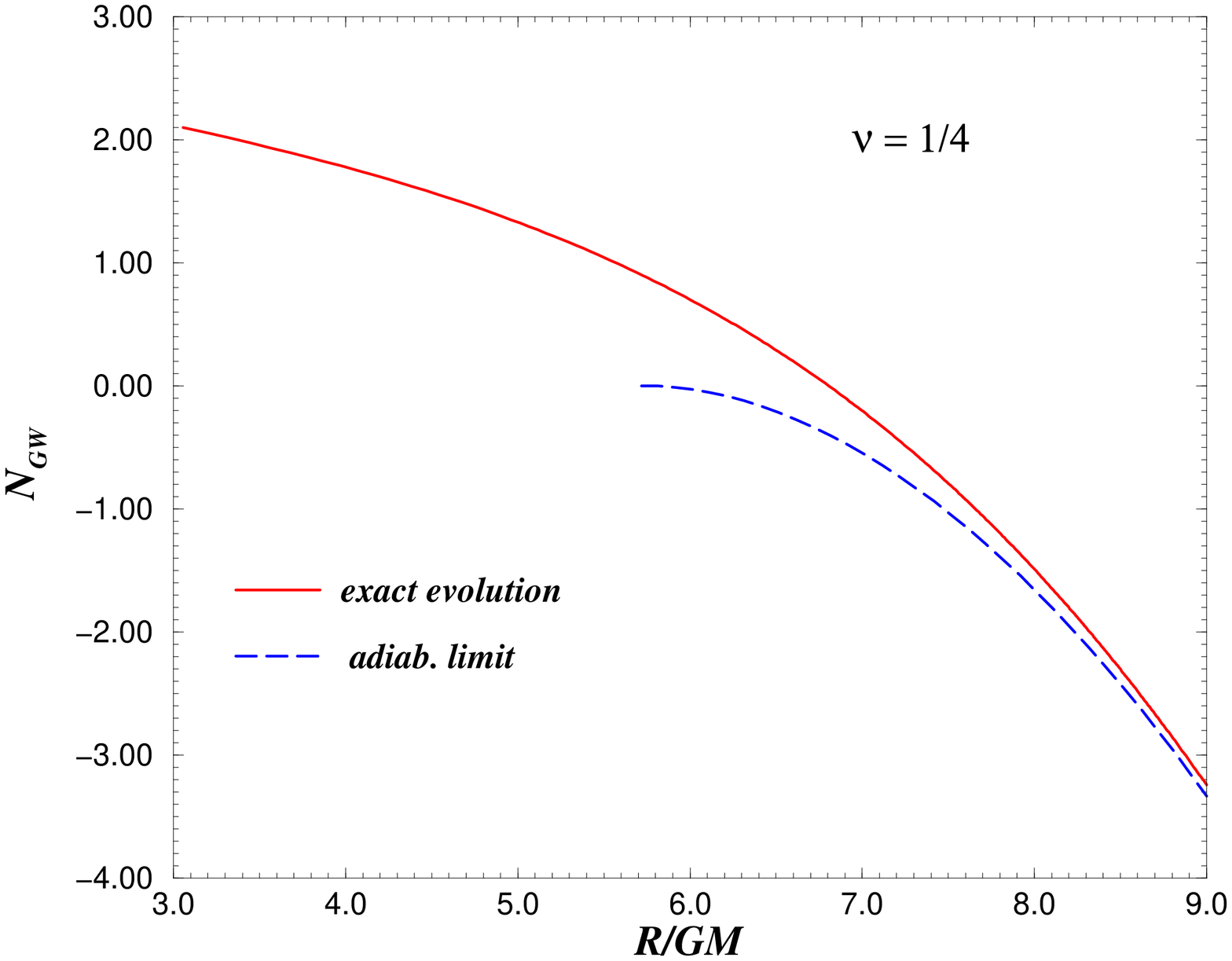,width=0.48\textwidth,height = 
0.48\textwidth,angle=-90} & 
\hspace{-0.0cm}
\epsfig{file=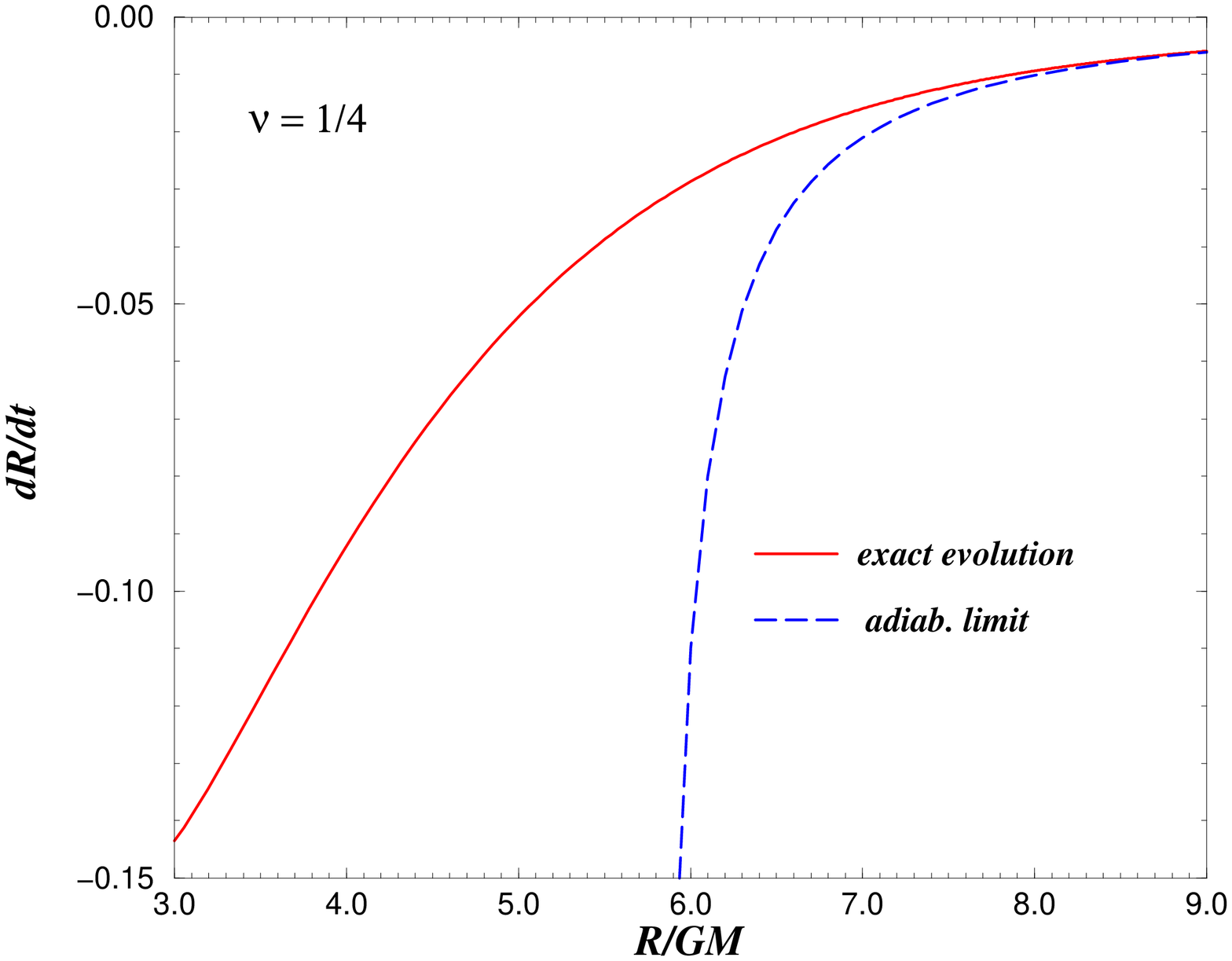,width=0.48\textwidth,height = 
0.48\textwidth,angle=-90}
\end{tabular}
\caption{\sl We compare the number of gravitational cycles  
(on the left) and the radial velocity (on the right), computed with the 
exact evolution and within the adiabatic approximation, versus $R/GM$.}
\label{Fig4}
\end{center}
\end{figure}
In Figs.~\ref{Fig4}, \ref{Fig5} we compare, for 
$\nu = 1/4$, the number of gravitational cycles, defined by ${\cal N}_{\rm GW} = 
\vphi_{\rm GW}/(2\,\pi) = \vphi/\pi$, the orbital angular frequency $\omega$
(or, equivalently,
 the gravitational wave frequency, $f_{\rm GW} = \omega_{\rm GW}/(2\pi) = 
\omega/\pi$), and the radial velocity, computed with the exact equations of 
motion and in the adiabatic limit, as well as the gravitational waveform. These 
figures show that, in the equal-mass case $\nu = 1/4$, the adiabatic 
approximation 
starts to significantly deviate from the exact evolution quite before one 
reaches the LSO. 
Fig.~\ref{Fig4} is normalized so that ${\cal N}_{\rm GW}^{\rm adiab}$ and
${\cal N}_{\rm GW}^{\rm exact}$ coincide for large values of $ R/G M$, and that
${\cal N}_{\rm GW}^{\rm adiab}$ 
be zero at the $r$-LSO. For instance, 
 we find that the number of GW cycles given by the 
adiabatic approximation differs from the exact number already by 0.1 when $r 
\simeq 8.8$,
and that ${\cal N}_{\rm GW}^{\rm exact} (r_{\rm LSO}) = 0.9013$ . 
The left panel of Fig.~\ref{Fig5} contrasts $\omega/\omega_{\rm LSO}$
($ = f_{\rm GW}/f_{\rm GW\,LSO}^{\rm Schw.}$
where $f_{\rm GW\,LSO}^{\rm Schw.} = 6^{-3/2}/G M \pi$ is the fiducial 
Schwarzschild LSO GW frequency), computed with the exact evolution and 
within the adiabatic approximation, as a function of  time. 
Note that, for the horizontal axis we use $\widehat{\omega}_{\rm LSO} (0) (\wt - 
{\wt}_{\rm LSO})$, 
where $\widehat{\omega}_{\rm LSO} (0) = \pi \widehat{f}_{\rm GW \,LSO}^{\rm Schw.} = 
6^{-3/2}$ (provided by the $\nu \rightarrow 0$ limit of Eq.~(\ref{6.12})) 
and $\wt_{\rm LSO}$ is defined as the time at which the adiabatic solution 
reaches 
the $r$-LSO position. Finally, 
on the right panel of Fig.~\ref{Fig5} we compare the last few GW cycles of the 
exact  
and the adiabatic {\it restricted} waveform, i.e. $h(t) \equiv v^2 
\, \cos \, \phi_{\rm GW} (t)$,  with $ v = (d \varphi/d \wt \,)^{1/3}$ and 
$\phi_{\rm GW} = 2 \varphi$, in 
the crucial interesting region around the LSO. 
By {\it adiabatic} restricted waveform we mean the restricted waveform 
in which $\varphi (\wt) = \varphi_{\rm adiab.} (\wt)$ is 
derived by integrating the two equations (\ref{4.8b}) and (\ref{4.9}) (which 
give a  
parametric representation of $\wt_{\rm adiab.} (r)$ and $\widehat{\varphi}_{\rm 
adiab.} (r)$ in terms of the auxiliary parameter $r$).  

Note in Fig.~\ref{Fig5} that the dephasing between the two waveforms becomes 
visible somewhat before the LSO 
(we shall dwell more on this subject in Section \ref{sec6}).
Note also that the time when the adiabatic evolution reaches the LSO 
(``adiabatic LSO'')
corresponds to a time when the exact evolution reaches a frequency
$\omega \simeq 0.80\, \omega_{\rm LSO} (0)$, i.e. a time significantly {\it before}
the $\omega$-LSO. This is why there are more cycles after the adiabatic LSO in 
Fig.~\ref{Fig5} (more than two cycles), than there will be after the (exact)
 $\omega$-LSO (we shall see below that $N_{\rm GW}^{\rm after LSO} 
= 2 N_{\rm orbit}^{\rm after LSO} = 1.2048$ for $\nu = 1/4$).
\begin{figure}
\begin{center}
\begin{tabular}{cc}   
\hspace{-0.8cm} 
\epsfig{file=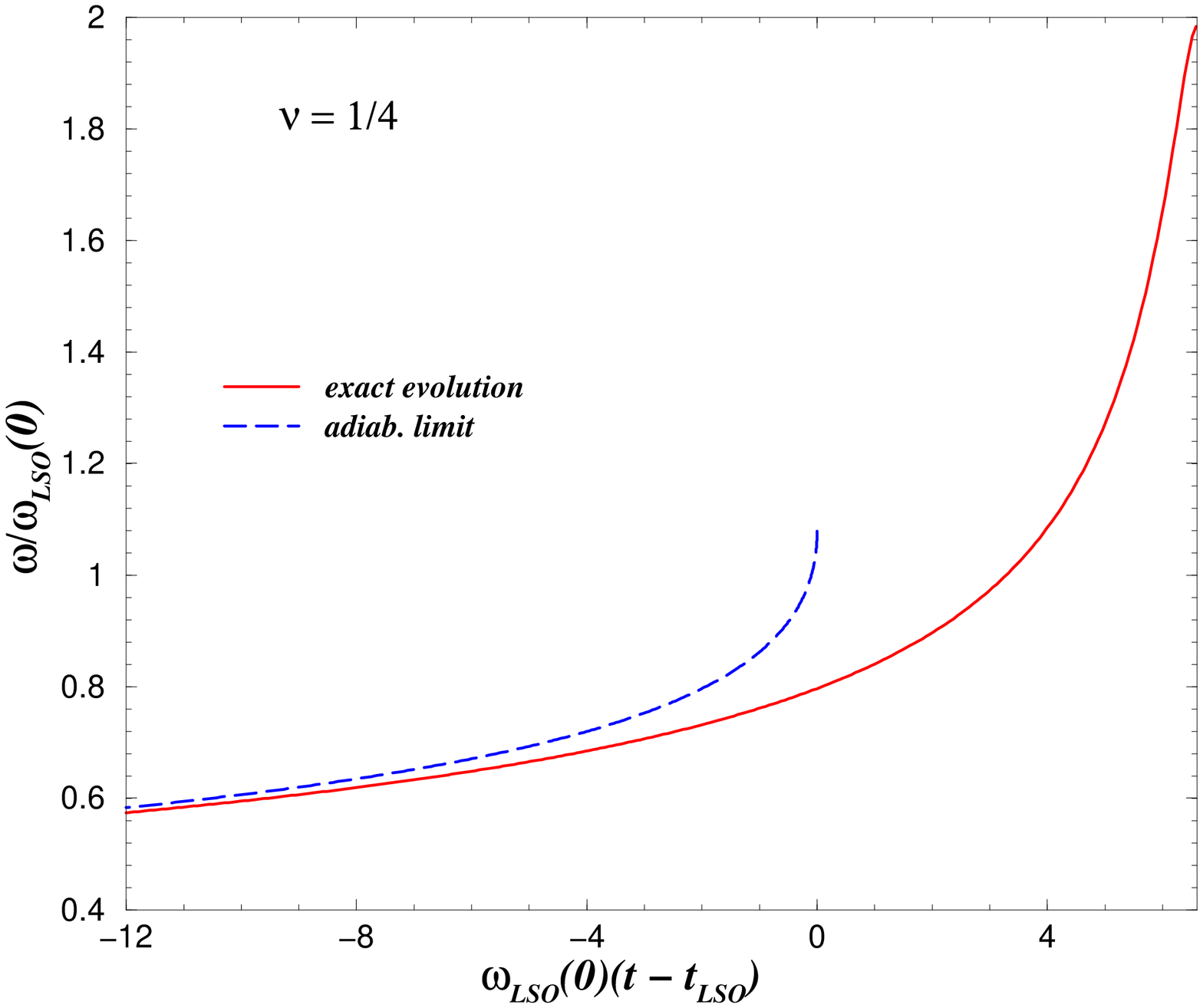,width=0.48\textwidth,height = 
0.48\textwidth,angle=-90} & 
\hspace{-0.0cm}
\epsfig{file=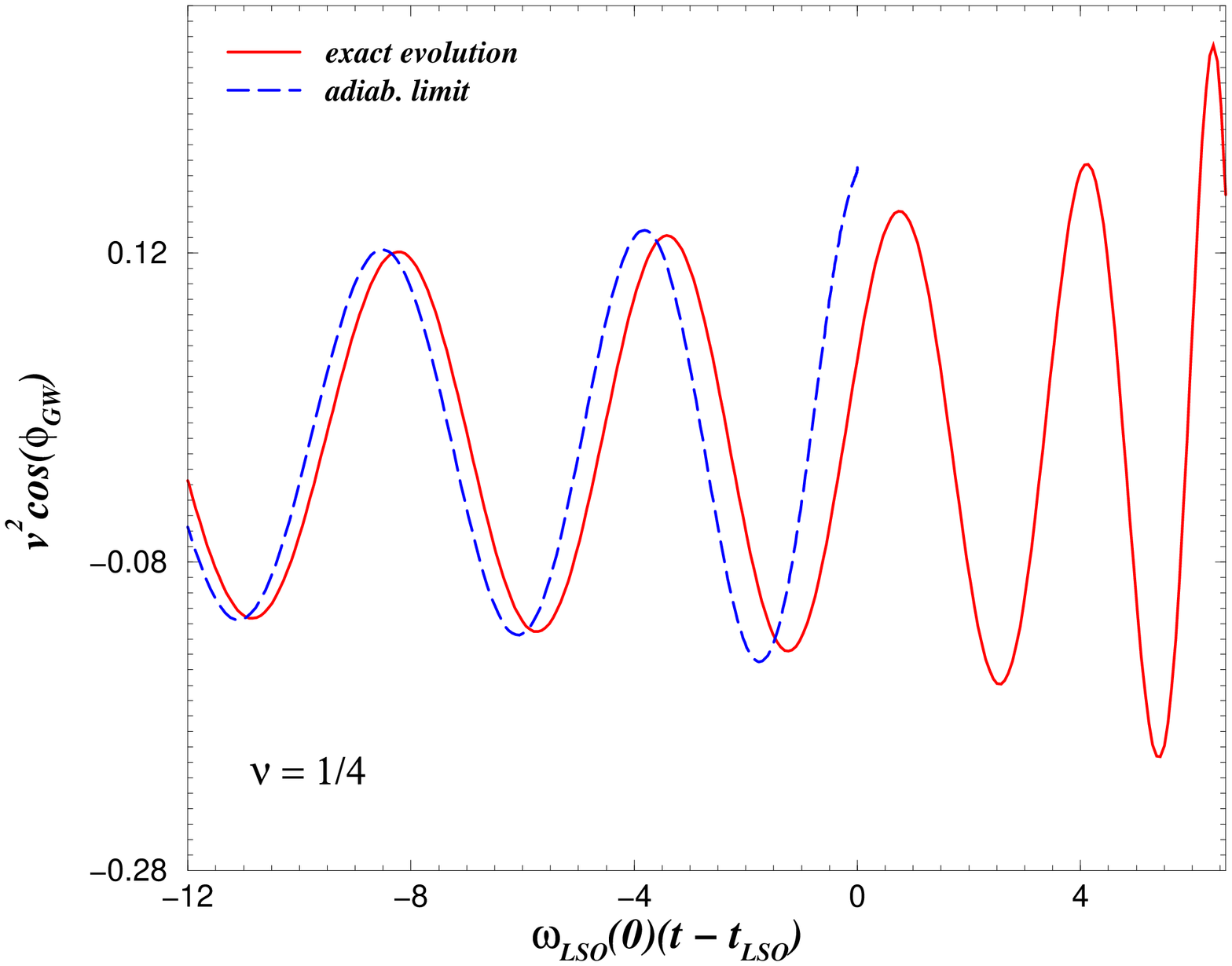,width=0.48\textwidth,height = 
0.48\textwidth,angle=-90}
\end{tabular}
\caption{\sl We contrast the orbital frequency (on the left), 
divided by the Schwarzschild value $\widehat{\omega}_{\rm LSO}(0) = 
6^{-3/2}$, and the  restricted waveform (on the right),   
evaluated with the exact dynamical system and within the adiabatic 
approximation. Note that in both plots the quantities are given 
as a function of the rescaled time variable $\widehat{\omega}_{\rm LSO}(0)(\wt - 
\wt_{\rm LSO})$, where  $\wt_{\rm LSO}$ is defined as the time at which 
the adiabatic solution reaches the $r$-LSO position.}
\label{Fig5}
\end{center}
\end{figure}

\subsection{The $\dot{r}$-linearized approximation}
\label{subsec4.2}
The previous subsection has shown the severe shortcomings of the adiabatic 
approximation. Let us now consider a second analytical approximation which is 
more 
accurate than the adiabatic one, and which, in particular, allows one to see 
analytically what happens during the transition between the inspiral and the 
plunge. 
This approximation is based on a simple linearization with respect to the radial 
velocity $dr/d\wt$, which is small during the inspiral, as well as the 
beginning of the plunge.

As $\wH$ depends quadratically on $p_r$ and $p_r \ll 1$ we pose
\beq
\label{4.11}
C_r(r,j) \equiv \left [ \frac{1}{p_r}\,\frac{\pa \wH}{\pa p_r}
(r,p_r,j) \right ]_{p_r \rightarrow 0} = 
\frac{1}{\nu\,\wH_0(r,j)}\,\frac{1}{\wH^0_{\rm 
eff}(r,j)}\,\frac{A^2(r)}{1-6\nu/r^2}\,, 
\eeq
(note that $C_r$ is a positive quantity), where 
\beq
\wH^0_{\rm eff}(r,j) = \wH_{\rm eff}(r,p_r = 0,j) = \sqrt{A(r)\,\left (1 + 
\frac{j^2}{r^2}\right )}\,. 
\eeq
Then, modulo $p_r^2$ fractional effects that we neglect, we can write
\beq
\label{4.12}
\frac{dr}{d \wt} \simeq C_r(r,j)\,p_r\,.
\eeq
Differentiating twice the above equation with respect to time, we obtain
\beq
\label{4.13}
\frac{d^2 p_r}{d \wt^2} \simeq \frac{1}{C_r(r,j)}\,\frac{d^3r}{d \wt^3} \, ,
\eeq
when neglecting some nonlinear terms $\propto (dr/d \wt)^2$ and $(dr/d \wt) \, 
(dj/d \wt)$. On the 
other hand, taking the derivative with respect to time of 
Eq.~(\ref{3.30}) and neglecting fractional corrections of ${\cal O}(p_r^2)$,
we end up with
\beq
\label{4.14}
\frac{d^2 p_r}{d \wt^2} = - \frac{d}{d \wt}\, \frac{\pa \wH}{\pa r}
(r,p_r,p_\vphi)\simeq - \frac{\pa^2 \wH_0}{\pa r^2}\,\frac{d r}{d \wt} - 
\frac{\pa^2 \wH_0}{\pa r \pa j}\,\wF_\vphi\,.
\eeq
To get an autonomous system we further approximate $j$ by solving for $j$ in the 
lowest-order approximation to Eq.~(\ref{3.30}), obtained by neglecting both 
$p_r$ and 
$dp_r / d \wt$. In other words, $j(r)$ is obtained, as in adiabatic 
approximation, by solving Eq.~(\ref{4.3.1}). Finally, $j \simeq j_{\rm 
adiab.} 
(r)$, as given by Eq.~(\ref{4.4}). 
We define
\bea
\label{4.15}
\omega_r^2 (r) &\equiv&  C_r(r,j_{\rm adiab.}(r))\,\frac{\pa^2 \wH_0}
{\pa r^2}(r,j_{\rm adiab.} (r)) \,, \nonumber \\
&=& \frac{1}{\nu^2\,\wH_0^2(r,j_{\rm adiab.})}\,\frac{r^5 - 6\,r^4 + 3\nu\,r^3 + 
20\nu\,r^2 - 
30\,\nu^2}{r^6\,(r^2-6\nu)}\,, 
\eea
\bea
\label{4.16}
B_r(r) &\equiv& C_r(r,j_{\rm adiab.}(r))\,\frac{\pa^2 \wH_0}
{\pa r \pa j}(r,j_{\rm adiab.} (r))\,\wF_\vphi(\ww_{\rm adiab.} (r))\,, 
\nonumber \\
&=&-\frac{2\,j_{\rm adiab.}(r)}{\nu^2\,\wH_0^2(r,j_{\rm adiab.})}\,
\frac{(r^3 - 3r^2 + 5\nu)^2}{r^7\,(r^2-6\nu)}
\,\wF_\vphi(\ww_{\rm adiab.} (r)) \,,
\eea
(where the replacements $j \rightarrow j_{\rm adiab.} (r)$ are done after the 
partial differentiations). It is easily seen that the quantity $\partial^2 
\widehat{H}_0 
/ 
\partial r \, \partial j$ is negative, so that ($\widehat{\cal F}_{\varphi}$ 
being 
also negative) the quantity $B_r$ given by Eq.~(\ref{4.16}) is positive.

Combining Eqs.~(\ref{4.13}) and ~(\ref{4.14}), we finally derive 
the following third order differential equation in $r$:
\beq
\label{4.17}
\frac{d^3r}{d \wt^3} + \omega_r^2(r)\,\frac{d r}{d \wt}
\simeq - B_r(r)\,.
\eeq 
We shall often refer to Eq.~(\ref{4.17}) as the ``linear $\dot r$-equation'' 
because 
it was obtained by working linearly in the radial velocity $\dot r = dr / 
d \wt$. [Note, however, that this is a third-order {\it nonlinear} differential 
equation 
in $r$.] It is easily seen that the quantity $\omega_r^2 (r)$ defines the square 
of 
the frequency of the radial oscillations. As seen in Eq.~(\ref{4.15}) it is 
proportional to the curvature of the effective radial potential $H_0 (r,j)$ 
determining the radial motion. Above the LSO, i.e. when $r > r_{\rm LSO} (\nu)$, 
the 
radial potential has a {\it minimum} (defining the stable circular orbit with 
angular 
momentum $j$) and, therefore, $\omega_r^2 (r)$ is positive. When $r = r_{\rm 
LSO} 
(\nu)$, the radial potential has an inflection point (see Eq.~(\ref{2.16})), 
and, 
therefore, $\omega_r^2 (r)$ vanishes. When $r < r_{\rm LSO} (\nu)$, the radial 
potential is concave, and $\omega_r^2 (r)$ becomes negative. [See, e.g., 
Fig.~\ref{Fig1} of \cite{BD99} for a plot of the shape of the radial potential.]

Within the same approximation used above (i.e., essentially, neglecting terms 
which 
are {\it fractionally} of order $p_r^2$), we can finally write the angular 
frequency 
along our quasi-circular orbits as
\beq
\label{4.18}
\frac{d \vphi}{d \wt} \simeq \frac{\pa {\wH}_0}{\pa j} \, (r, j_{\rm 
adiab.} (r))\,.
\eeq
Note that $\varphi$ is obtained from this equation by a quadrature,
 once the radial motion 
$r(\wt)$ is known from the integration of Eq.~(\ref{4.17}).
\begin{figure}
\begin{center}
\begin{tabular}{cc}   
\hspace{-0.8cm} 
\epsfig{file=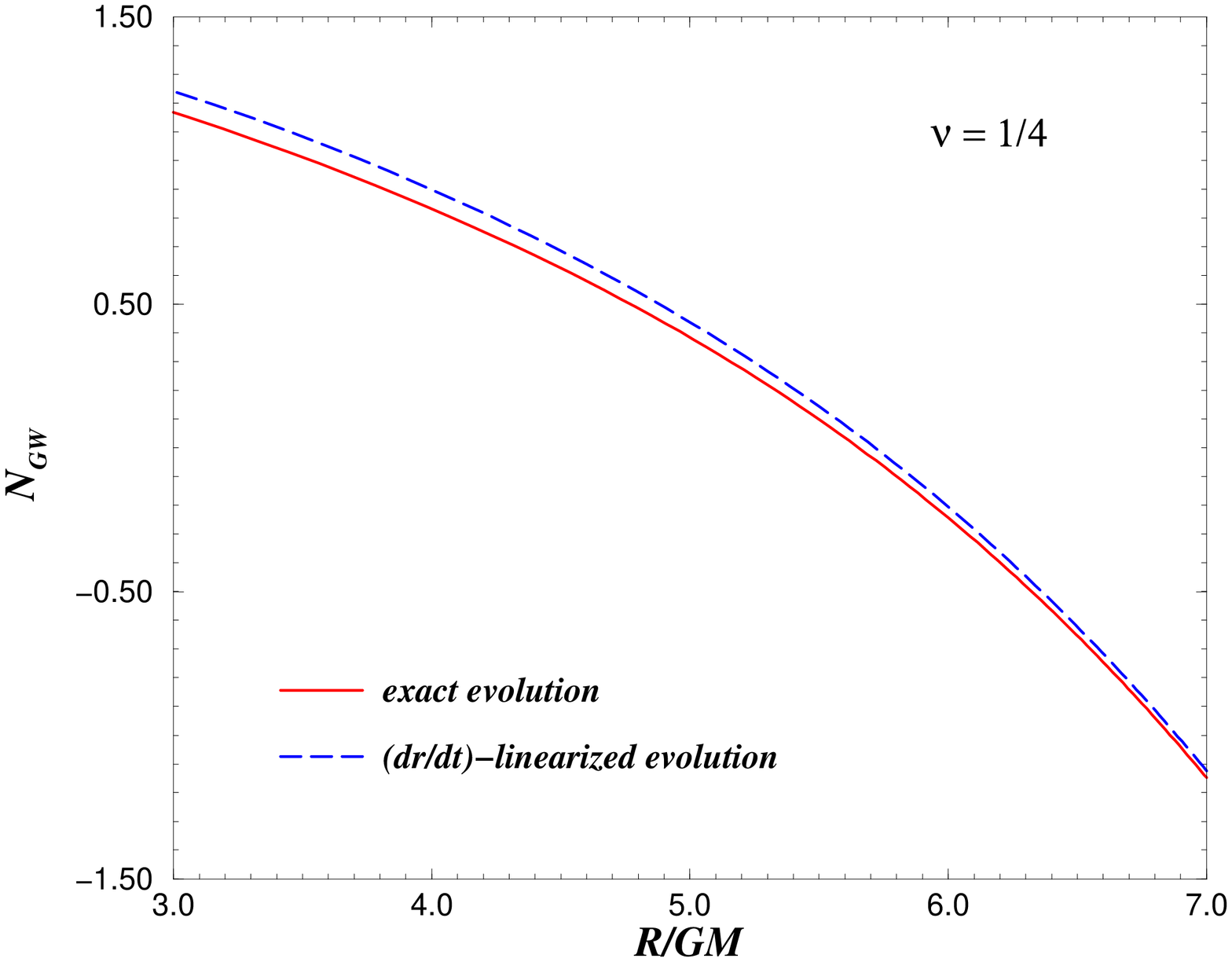,width=0.5\textwidth,height = 
0.5\textwidth,angle=-90} & 
\hspace{-0.0cm}
\epsfig{file=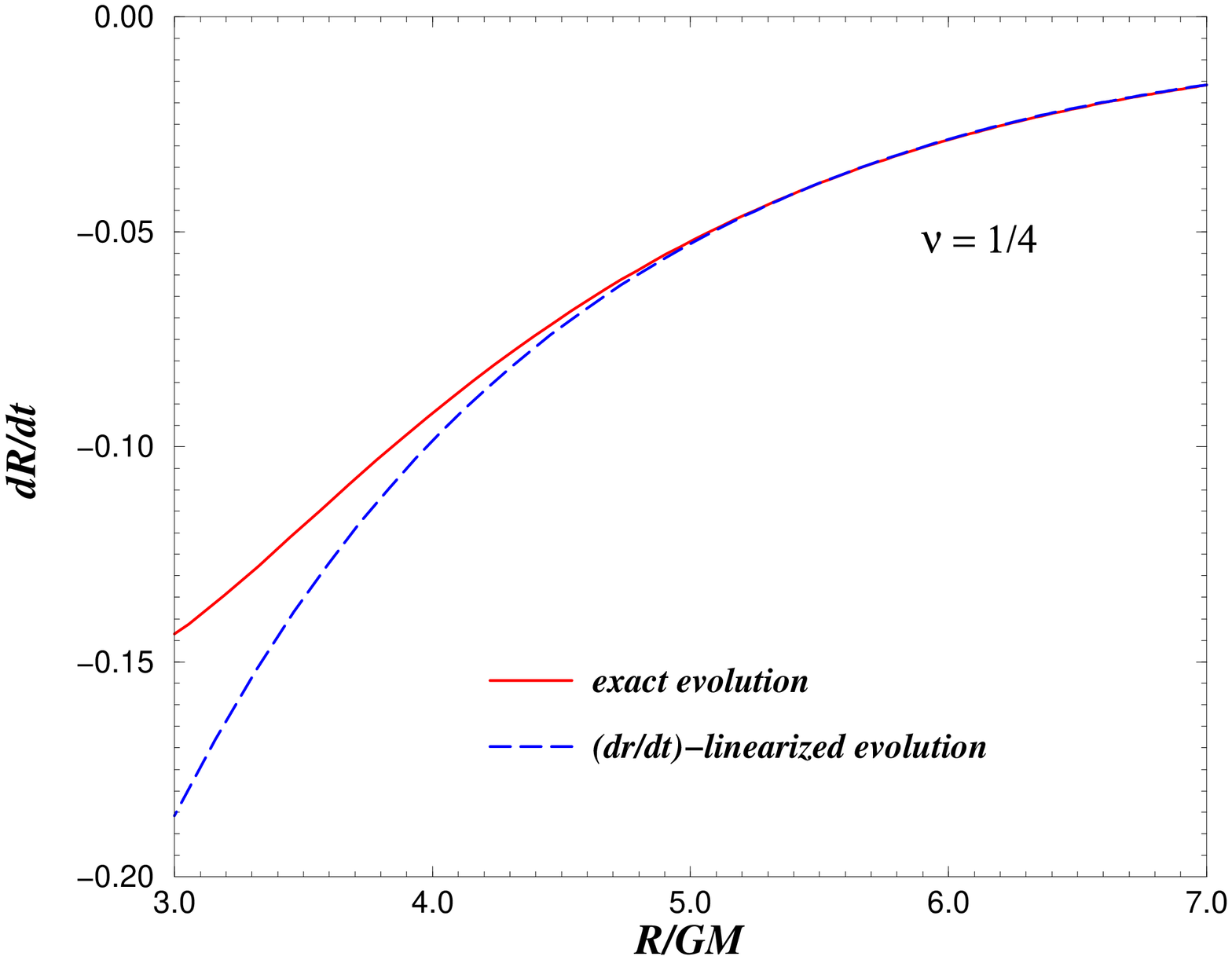,width=0.5\textwidth,height = 
0.5\textwidth,angle=-90}
\end{tabular}
\caption{\sl Contrast of the number of gravitational cycles  
(on the left) and the radial velocity (on the right), computed with the 
exact evolution and the linear-$\dot{r}$ equation, versus $R/GM$.}
\label{Fig6}
\end{center}
\end{figure}

The conceptually interesting feature of the above ``$\dot r$-linearized'' 
approximation is the structure of Eq.~(\ref{4.17}). The previously considered 
``adiabatic'' approximation corresponds to neglecting $d^3 r / d {\wt}^3$ 
in 
Eq.~(\ref{4.17}). We now see that this is a good approximation only when the 
characteristic frequency of variation of the radial motion, defined, say, by 
$\omega_{\rm caract.}^2 \equiv (d^3 r / d{\wt}^3) / (dr / d\wt)$ is 
much 
smaller than the frequency of radial oscillations $\omega_r^2$ (determined by 
the 
restoring radial force ensuring the existence of stable circular orbits). As 
$\omega_r^2$ tends to zero, before changing sign, at the LSO, it is clear that 
the 
adiabatic approximation must break down somewhat above the LSO. When it breaks 
down 
the ``inertia term'' $d^3 r / d{\wt}^3$ in Eq.~(\ref{4.17}) becomes 
comparable 
to both the ``restoring force'' term $\omega_r^2 \, dr / d\wt$ and the 
``driving force'' $-B_r$ coming from gravitational radiation damping.

In Figs.~\ref{Fig6} we compare the number of gravitational cycles and 
the radial velocity evaluated with the exact evolution 
and the $\dot r$-linearized equations. 
We start the evolution at $r = 15$ and fix the initial values of $dr/d \wt$ and  
$d^2r/d \wt^2$ 
in the ``adiabatic approximation'' defined by neglecting in Eq.~(\ref{4.17}) 
the ``inertia term'' $d^3 r / d {\wt}^3$ (and then by differentiating again the 
resulting approximate equation). 
Moreover, we normalize ${\cal N}_{\rm GW}^{\rm linear}$ 
to be zero at the $r$-LSO. We derive from the exact evolution 
${\cal N}_{\rm GW}^{\rm exact}(r_{\rm LSO}) = -0.04223$.
The main conclusion drawn from Figs.~\ref{Fig6} is that the $\dot r$-linearized 
approximation is quite good both during the inspiral phase and, more 
importantly, during the transition to the plunge taking place near the LSO. This 
is 
interesting to 
know because it shows that the crucial physical effect that is lacking in the 
usually 
considered adiabatic approximation is the simple ``inertia term'' $d^3 r / d 
{\wt}^3$ in Eq.~(\ref{4.17}). Note, however, that in order to add this inertia 
term 
it is 
necessary to have in hand the Hamiltonian describing at least the slightly 
non-circular orbits (the normalization of Eq.~(\ref{4.17}) crucially depends on 
the 
knowledge of $\omega_r^2$ which depends both on $\partial \widehat H / \partial 
p_r^2$ and on $\partial ^2 \widehat H / \partial r^2$). This being said, we do 
not, 
however, recommend to use in practice the $\dot r$-linearized approximation. 
Indeed, 
we think that the ``exact'' system (\ref{3.28})--(\ref{3.31}) is a more accurate 
description of the evolution of the system because it keeps all the nonlinear 
effects 
in $p_r^2$. Numerically speaking, it is essentially as easy to integrate the 
``exact'' 
system than its $\dot r$-linearized approximation, so that there would be anyway 
no 
practical advantage in downgrading the accuracy of the system 
(\ref{3.28})--(\ref{3.31}). However, we shall see next that the $\dot 
r$-linearized 
system can be further used to lead to a simple analytical approach to the 
transition 
to the plunge in the case where $\nu \ll 1$.

\subsection{The universal $\rho$-equation}
\label{subsec4.3}
Until now we have been considering the general case where the symmetric mass 
ratio 
$\nu \equiv m_1 \, m_2 / (m_1 + m_2)^2$ can be of order of its maximum value 
$\nu_{\rm 
max} = 1/4$. As is clear from the results above when $4\nu$ is of order unity 
the {\it 
non-adiabatic} aspects of radiation damping effects become important in an 
extended 
region of order $\Delta (R / G M) \sim 1$, above the standard LSO. On the other 
hand, we 
expect that when $4\nu \ll 1$ the transition between the adiabatic inspiral and 
the 
plunge will be sharply localized around the standard LSO, defined by 
Eq.~(\ref{2.16}). 
Indeed, when $\nu$ is a small parameter, the damping force $\widehat{\cal 
F}_{\varphi}$, Eq.~(\ref{3.33}), being proportional to $\nu$, can be treated as 
a 
perturbatively small quantity in the evolution of the system. Consequently, the 
``driving force'' term, $-B_r$, in the $\dot r$-linearized equation (\ref{4.17}) 
contains the small parameter $\nu$. It is then clear that all the time 
derivatives of 
$r$ (being driven by $B_r$) will tend to zero with $\nu$. If the coefficient 
$\omega_r^2$ in Eq.~(\ref{4.17}) never vanishes it is easy to see how one would 
satisfy Eq.~(\ref{4.17}) by solving for $dr / d\wt$, while considering 
$d^3 r / 
d {\wt}^3$ as a fractionally small term (to be evaluated by further 
differentiating $dr / d\wt \simeq -B_r / \omega_r^2$). In that case, one 
sees 
that $dr / d\wt$ would be ${\cal O} (\nu)$ (and $d^3 r / d {\wt}^3 
= 
{\cal O} (\nu^3)$) as $\nu \rightarrow 0$. However, the fact that $\omega_r^2 
(r)$ 
vanishes when $r = r_{\rm LSO} (\nu)$ shows that the way $dr / d\wt$ 
tends to 
zero with $\nu$, near the LSO, is more subtle. Having understood from this 
reasoning 
that, when $\nu \rightarrow 0$, the interesting transition effects take place 
very 
near the LSO, we now turn to a precise analysis of this transition.

A first method for dealing (when $\nu \rightarrow 0$) with this transition would 
be 
(as just sketched) to continue working with the third-order equation 
(\ref{4.17}), 
considered in the immediate neighbourhood of $r = r_{\rm LSO} (\nu)$. However, 
it is 
better (in order not to increase the differential order) to go back to the exact 
system (\ref{3.28})--(\ref{3.31}) and to approximate it directly when $\nu 
\rightarrow 
0$ and $r \rightarrow r_{\rm LSO} (\nu)$.

Let us see the consequences of the evolution (\ref{3.28})--(\ref{3.31}) when $r$ 
is 
very near $r_{\rm LSO} (\nu)$. To do this it is convenient to introduce some 
notation.
Using, as we did in Sec.~\ref{subsec4.2}, the fact that 
$\wH$ depends quadratically on $p_r$ and that $p_r \ll 1$, we define:
\beq
\label{4.19}
C_r^{\rm LSO} (\nu) \equiv \left [ \frac{1}{p_r}\,\frac{\pa \wH}{\pa p_r}
(r,p_r,j) \right ]_{p_r \rightarrow 0}^{\rm LSO}\,. 
\eeq
Note that $C_r^{\rm LSO}$ is a number, which depends on 
$\nu$~\footnote{\label{f5} As we consider 
$\nu \ll 1$, we could further take the limit $\nu \rightarrow 0$ in all the 
quantities 
which have a finite limit as $\nu = 0$. However, in order not to unnecessarily 
loose 
accuracy we shall not do so. For instance we shall always consider that $r_{\rm 
LSO} 
(\nu)$ is computed for $\nu \ne 0$, though we shall see later that the direct 
$\nu$-dependence in $r_{\rm LSO} (\nu)$ (which is ${\cal O} (\nu)$) is 
parametrically 
small compared to the width ${\cal O} (\nu^{2 / 5})$ of the radial axis where 
the 
transition takes place.}. In terms of the previous definition (\ref{4.11}), one 
has simply $C_r^{\rm LSO} (\nu) = C_r (r_{\rm LSO} (\nu) , j_{\rm LSO} (\nu))$. 
Explicitly, it reads
\beq
C_r^{\rm LSO} (\nu) = \left [
\frac{A^2(r)}{\nu\,\wH_0(r,j)\,\wH^0_{\rm eff}(r,j)\,(1-6\nu/r^2)} \right ]_{\rm 
LSO}\, .
\eeq
In the $\nu = 0$ limit this simplifies to
\beq
C_r^{\rm LSO} (0) = \frac{\sqrt{2}}{3}  \, .
\eeq
The point in having introduced the notation (\ref{4.19}) is that 
Eq.~(\ref{3.28}) 
reads simply, when one is very near the LSO:
\beq
\label{4.20}
p_r \simeq \frac{1}{C_r^{\rm LSO}}\,\frac{dr}{d \wt}\, .
\eeq
This allows us to recast Eq.~(\ref{3.30}) in the form (after neglecting 
fractional $p_r^2$ terms on the RHS)
\beq
\label{4.21}
\frac{1}{C_r^{\rm LSO}}\,\frac{d^2r}{d \wt^2} \simeq - \frac{\pa 
\wH_0}{\pa r}(r,j) \,.
\eeq
Here, as above, $\wH_0 (r,j) \equiv \wH (r, p_r = 0, p_{\varphi} \equiv j)$.
Then we expand the RHS of the above equation around the LSO, 
i.e. we write
\beq
\label{4.22}
r = r_{\rm LSO}(\nu) + \delta r \,, \quad \quad  j = j_{\rm LSO}(\nu) + 
\delta j \,.
\eeq
Keeping the first nontrivial terms in the expansion in powers of $\delta r$ and 
$\delta j$ (and neglecting subleading terms, such as those of order ${\cal 
O}(\delta r 
\delta j)$, ${\cal O}((\delta j)^2)$ and ${\cal O}((\delta r)^3)$) one obtains
\beq
\label{4.23}
\frac{\pa \wH_0}{\pa r} \simeq \frac{1}{2} \,\left (\frac{\pa^3 
\wH_0}{\pa r^3} \right )_{\rm LSO}\,(\delta r)^2 + \left (\frac{\pa^2 \wH_0}{\pa 
r \pa 
j} \right )_{\rm LSO}\,(\delta j) \,.
\eeq
Moreover, near the LSO we can write Eq.~(\ref{3.31}) as:
\beq
\label{4.24}
\frac{d (\delta j)}{d \wt} = \frac{d j}{d\wt} \simeq \wF_\vphi(\ww_{\rm 
LSO}) 
\,, \quad \quad \mbox{\small with} \quad  \ww_{\rm LSO} = \left (\frac{\pa 
\wH_0}{\pa j}\right )_{\rm LSO}\,.
\eeq
This yields
\beq
\label{4.25}
\delta j \simeq   \wF_\vphi(\ww_{\rm LSO})\,(\wt - \wt_{\rm LSO})\,,
\eeq
where $\wt_{\rm LSO}$ is the time at which $j (\wt) = j_{\rm LSO} 
(\nu)$.
Let us also define
\beq
\label{4.26}
A_r^{\rm LSO} \equiv C_r^{\rm LSO}\,\left (\frac{\pa^3 \wH_0}
{\pa r^3}\right )_{\rm LSO} \,, \quad \quad 
\label{4.26a}
B_r^{\rm LSO} \equiv C_r^{\rm LSO}\,\left (\,\frac{\pa^2 \wH_0}
{\pa r \pa j} \right )_{\rm LSO}\,\wF_\vphi(\ww_{\rm LSO}) \,.
\eeq
The quantity $B_r^{\rm LSO}$ is the LSO value of the quantity $B_r (r)$ 
introduced in 
Eq.~(\ref{4.16}) above. The explicit values of these quantities are
\bea
\label{4.27n}
&& A_r^{\rm LSO} (\nu) = \left [
\frac{(r^3 - 2\,r^2 + 2\nu)\,(-210\nu\,j^2 - 60\nu\,r^2 + 
60j^2\,r^2-12j^2\,r^3 + 6r^4)}{r^7\,(r^2-6\nu)\,(j^2 + 
r^2)\,\nu^2\,\wH^2_0(r,j)}
 \right ]_{\rm LSO}\,, \nonumber \\
&& B_r^{\rm LSO} (\nu) = \left [ -
\frac{2j\,(r^3 - 2r^2 + 2\nu)\,(r^3 - 3r^2 + 5\nu)}{r^5\,(r^2-6\nu)\,(r^2 + j^2)
\,\nu^2\,\wH^2_0(r,j)}
\right ]_{\rm LSO}\,\wF_\vphi(\ww_{\rm LSO})\, .
\eea
In the $\nu = 0$ limit they simplify to
\beq
\label{4.28n}
A_r^{\rm LSO} (0) = \frac{1}{1296}\,, \quad  
\qquad {B_r^{\rm LSO} (\nu)} \stackrel{\nu \rightarrow 0}{=} 
-\frac{1}{72\sqrt{3}}\,\nu\,
\left [\frac{\wF_\vphi(\ww_{\rm LSO}(\nu);\nu)}{\nu} \right ]_{\nu \rightarrow 
0} = 
1.052 \cdot 10^{-4}\,\nu\,.
\eeq
Finally, inserting Eq.~(\ref{4.25}) into Eq.~(\ref{4.23}), and replacing 
everything in 
Eq.~(\ref{4.21}) yields the simple equation
\beq
\label{4.27}
\frac{d^2 \delta r}{d \wt^2} + \frac{1}{2}\,A_r^{\rm LSO}\,(\delta r)^2 = 
- B_r^{\rm LSO}\,(\wt - \wt_{\rm LSO})\,.
\eeq
This equation can be recast in a universal form by re-scaling 
the variables $\delta r$ and $\delta \wt = (\wt -\wt_{\rm LSO})$. Indeed, 
posing 
\beq 
\label{4.27a}
\delta r = k_r \, \rho\,, \quad  \qquad \wt - \wt_{\rm LSO} = 
\delta 
\wt = k_t \, \tau \, , 
\eeq
with
\beq 
\label{4.28}
k_r \equiv (B_r^{\rm LSO})^{2/5}\,(A_r^{\rm LSO})^{-3/5} \,,
\quad \quad k_t \equiv (A_r^{\rm LSO}\,B_r^{\rm LSO})^{-1/5} 
\,,
\eeq
it is straightforward to derive the following ``universal $\rho$-equation''
\beq
\label{4.29}
\frac{d^2 \rho}{d \tau^2} + \frac{1}{2}\,\rho^2 = -\tau \,.
\eeq
The explicit values of the scaling coefficients $k_r$ and $k_t$ are easily 
derived 
from our previous results. Let us only quote explicitly their $\nu = 0$ limit:
\beq
\label{4.29n}
k_r (0) =  1.890\,\nu^{2/5}\,, \quad  \qquad k_t (0) = 26.19\,\nu^{-1/5} \, .
\eeq
Note the interesting fractional scalings $k_r \propto \nu^{2/5}$, $k_t \propto 
\nu^{-1/5}$.

Let us also note the autonomous (time-independent) equation obtained by taking 
the time 
derivative of Eq.~(\ref{4.29}):
\beq
\label{4.29b}
\frac{d^3 \rho}{d \tau^3} + \rho \, \frac{d\rho}{d\tau} = -1 \, .
\eeq
Eq.~(\ref{4.29b}) could have been directly derived by considering the $\dot 
r$-linearized Eq.~(\ref{4.17}) close to $r = r_{\rm LSO}$. There is, however, 
more 
information in Eq.~(\ref{4.29}) because its derivation showed that $\tau = 0$ 
marks 
the moment where $j(t) = j_{\rm LSO} (\nu)$.
\begin{figure}
\begin{center}
\begin{tabular}{cc}   
\hspace{-0.8cm} 
\epsfig{file=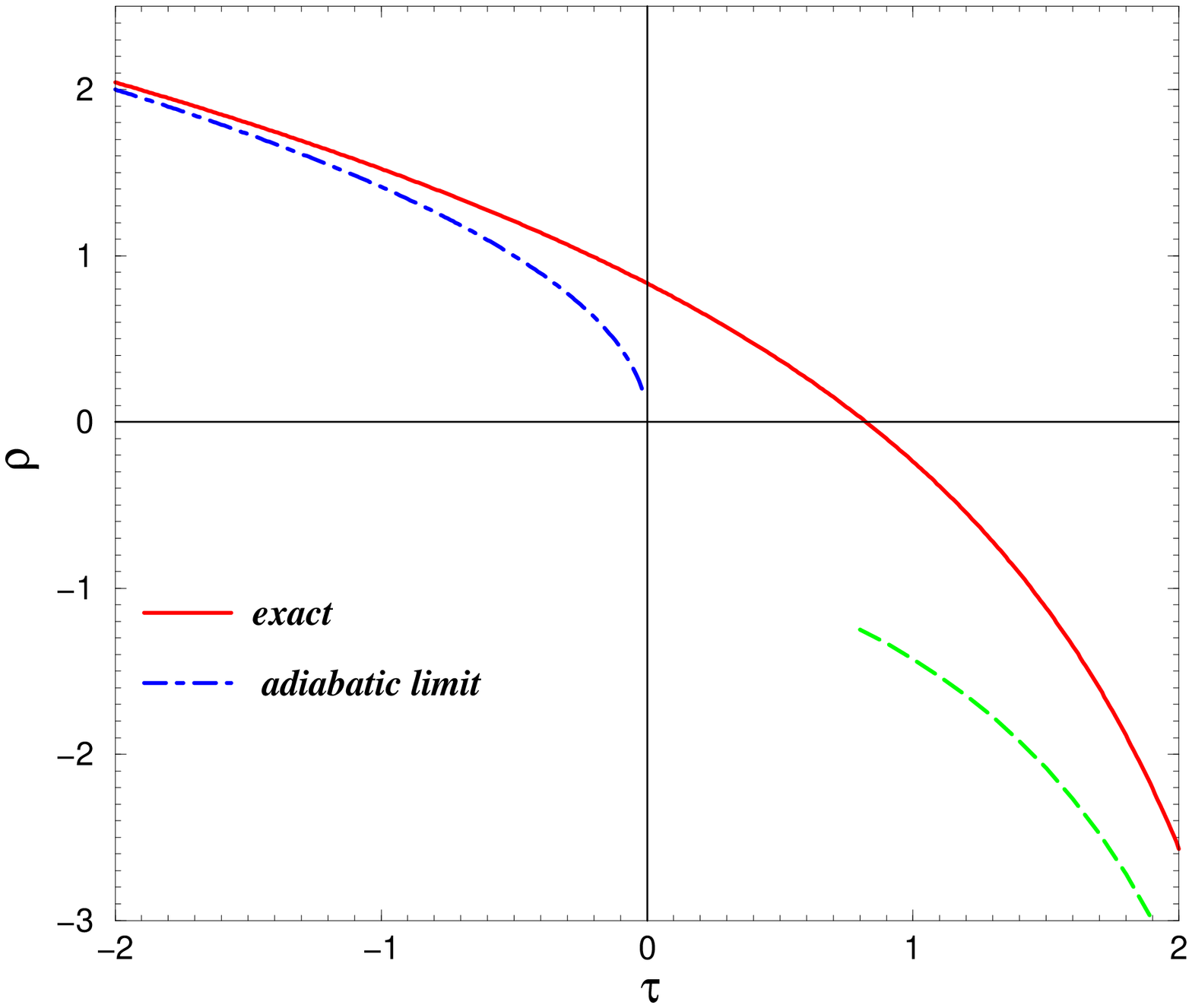,width=0.5\textwidth,height = 
0.5\textwidth,angle=-90} & 
\hspace{-0.0cm}
\epsfig{file=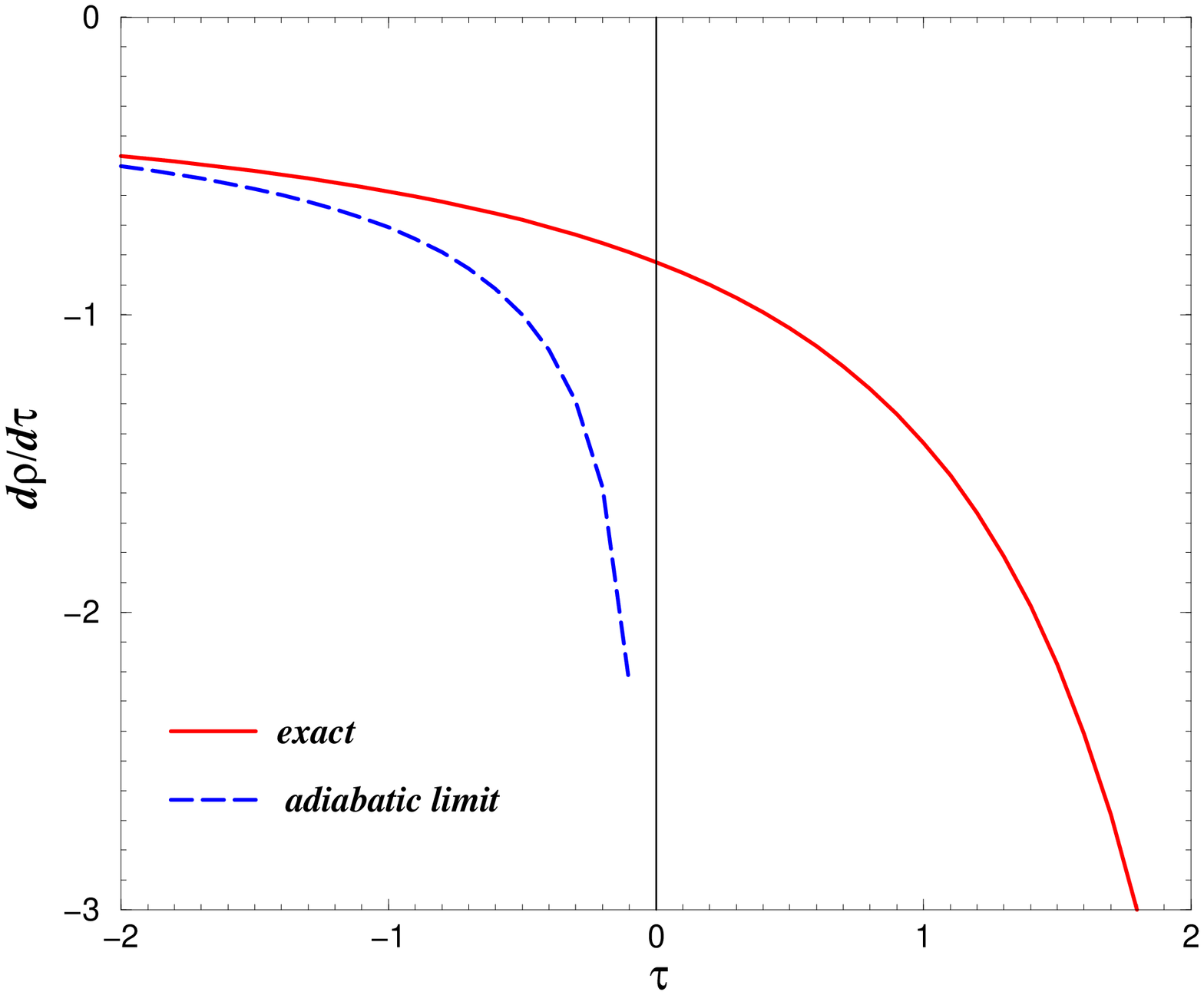,width=0.5\textwidth,height = 
0.5\textwidth,angle=-90}
\end{tabular}
\caption{\sl The universal $\rho$ and $\dot{\rho}$ curves and their adiabatic 
approximations. The long-dashed curve at the bottom 
of the left panel represents 
the approximate asymptotic solution (4.53).}
\label{Fig7}
\end{center}
\end{figure}

The adiabatic approximation is recovered by neglecting in 
Eq.~(\ref{4.29}) the first term on the RHS. This gives
\beq
\label{4.30}
\rho_{\rm adiab.} = \sqrt{-2 \tau} \,, \quad \quad \left( \frac{d \rho}{d \tau} 
\right)_{\rm 
adiab.} = - \frac{1}{\sqrt{-2\tau}} = - \frac{1}{\rho_{\rm adiab.}} \, .
\eeq
The universal $\rho$ and $\dot{\rho}$ curves and their adiabatic approximations 
are shown in Fig.~\ref{Fig7}. We have integrated Eq.~(\ref{4.29}) 
fixing the initial values (for large, negative $\tau$) 
of $\rho$ and $d \rho/ d\tau$ in the 
adiabatic limit provided by Eq.~(\ref{4.30}).
We see from Fig.~\ref{Fig7} that the adiabatic approximation begins to be 
unacceptably bad when $\tau \simeq -1$. 
{}From the integration of Eq.~(\ref{4.29}) we get the important numerical 
values: 
\bea
\label{new1}
&& \tau = 0: \quad  \rho = 0.8339 \,, \quad \quad \frac{d\rho}{d\tau} = 
-0.8233\,, \\
&& \rho = 0: \quad \tau = 0.8226 \,, \quad \quad \frac{d\rho}{d\tau} = -1.267\,. 
\label{new2}
\eea
We recall that $\tau=0$ marks the moment where $j(t) = j_{\rm LSO}(\nu)$, 
while $\rho=0$ corresponds to $r(t) = r_{\rm LSO}$.
The values given by Eqs.~(\ref{new1}) and (\ref{new2}) 
can then be used to compute corresponding values 
of the physical quantities $r$, $dr / d \wt$ and $j$ 
by using the following parametric representations derived from our treatment 
above:
\bea
\label{4.31}
&& r(\tau) = r_{\rm LSO}(\nu) + k_r \,\rho (\tau)\,, \quad \quad \quad \quad 
\quad
\wt (\tau) = \wt_{\rm LSO} + k_t \, \tau\,, \\
\label{4.32}
&& j (\tau) = j_{\rm LSO}(\nu) + {\wF}_{\vphi}(\ww_{\rm LSO}) \, k_t \,\tau\,,
\quad \quad 
\left( \frac{dr}{d\wt}\right)(\tau) = \frac{k_r}{k_t} \, \frac{d \rho}{d 
\tau}\,.
\label{4.33}
\eea
Correspondingly to these approximate results for $r$, $\wt$, $j$ and $dr/ d 
\wt$, one can also write an approximate result for the angular frequency, 
namely
\bea
\label{4.33n}
\left( \frac{d\varphi}{d\wt}\right)(\tau) = \widehat{\omega} (\tau) && = 
\frac{\partial \widehat{H}_0}{\partial j} \, (r(\tau) , j(\tau)) 
\nonumber \\
&&\simeq \widehat{\omega}_{\rm LSO} (\nu) + \left( \frac{\partial^2 
\widehat{H}_0}{\partial 
r \, \partial j} \right)_{\rm LSO} \, k_r \, \rho (\tau) 
+ \left( \frac{\partial^2 \widehat{H}_0}{\partial j^2} \right)_{\rm 
LSO} \, \widehat{\cal F}_{\varphi}^{\rm LSO} \, k_t \, \tau \, . 
\eea
In the approximation where we replace $\nu$ by zero in all quantities which have 
a 
finite limit when $\nu \rightarrow 0$, the above parametric results give the 
following 
explicit numerical links [except for $\wt_{\rm LSO}$ which is an 
arbitrary 
integration constant]
\bea
\label{4.34n}
&& r(\tau) = 6 + 1.890\,\nu^{2/5} \, \rho (\tau) + {\cal O} (\nu) \,, \quad 
\quad \quad 
\wt (\tau) =  \wt_{\rm LSO} + 26.19\, \nu^{-1/5} \, \tau \,, \\ 
&& j(\tau) = \sqrt{12} -0.3436\, \nu^{4/5} \, \tau + {\cal O} (\nu) \,, \quad 
\quad
\left( \frac{dr}{d \wt} \right)(\tau) \simeq 0.07216\,\nu^{3/5} \, 
\frac{d\rho}{d\tau} \, ,\\
&& \widehat{\omega} (\tau) =  \frac{1}{6 \sqrt 6} -0.03214\, \nu^{2/5} \, 
\rho (\tau) - 0.005062\, \nu^{4/5} \, \tau + {\cal O} (\nu) \, . 
\eea
Note that these explicit results are less accurate than our previous implicit 
expressions Eqs.~(\ref{4.31})--(\ref{4.33}) (because of the ${\cal O} (\nu)$ 
error terms entailed by 
$r_{\rm 
LSO} (\nu) = r_{\rm LSO} (0) + {\cal O} (\nu)$, etc.). For consistency with the 
rest 
of the paper, we have used here (as in Eq.~(\ref{4.28n}))
 the $\nu \rightarrow 0$ limit of the value of 
$\nu^{-1} \, {\cal F}_{\varphi}^{\rm LSO}$ defined by the 2.5PN Pad\'e estimate 
(\ref{3.33}), namely $\nu^{-1} \, {\cal F}_{\varphi}^{\rm LSO} \simeq -0.01312$. 
Note 
that a 
more accurate value of this quantity is, according to Poisson's numerical 
results \cite{P95} $\nu^{-1} \, \widehat{\cal F}_{\varphi}^{\rm LSO} \simeq 
-0.01376$,
which is $ \simeq 5\%$ larger (in modulus). 
Note the various scalings with $\nu$ implied (when considering a point in the 
transition region parametrized by some fixed numerical values of $\rho$ and 
$\tau$) by 
Eqs.~(\ref{4.31})--(\ref{4.33}): notably $\delta r = {\cal O} (\nu^{2/5})$, 
$\delta j 
= {\cal O} (\nu^{4/5})$ and $p_r \sim \dot r = {\cal O} (\nu^{3/5})$. We shall 
discuss 
below in more details some of these scalings.

Fig.~\ref{Fig7} vividly illustrates the fact (mentioned above) that the 
definition of 
``LSO-crossing'' becomes ambiguous in presence of radiation damping. Indeed, for 
instance, the time where $r = r_{\rm LSO} (\nu)$ (``$r$-LSO''), i.e. the time 
where 
$\rho = 0$, differs from the time where $j = j_{\rm LSO} (\nu)$ (``$j$-LSO''), 
i.e. 
the time where $\tau = 0$ (see also Eq.~(\ref{new1})).
\begin{figure}
\begin{center}
\epsfig{file=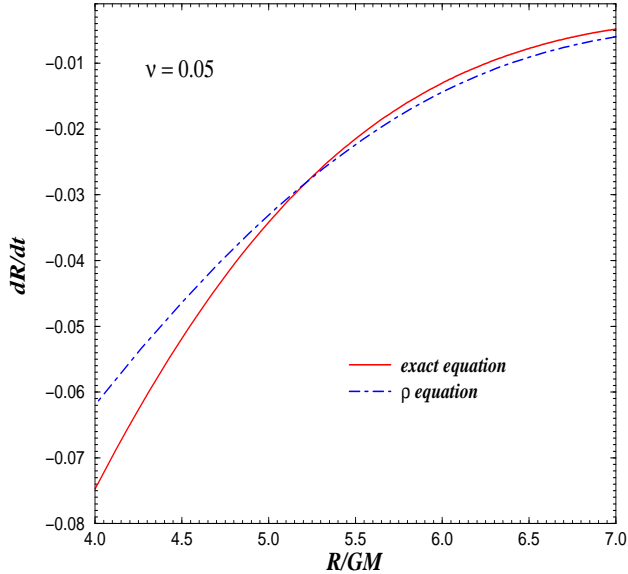,width=0.5\textwidth,height = 0.5\textwidth,angle=-90}
\caption{\sl Plot of the radial velocity computed both with the exact 
evolution and with the $\rho$-equation in the case $\nu = 0.05$. 
The fractional error in $dR / d t$ at the $r$-LSO is $\simeq 10\%$.}
\label{Fig8}
\end{center}
\end{figure}
An important issue is the domain of validity of the universal $\rho$-equation, 
i.e. 
the range of values of $\nu$ for which one can use 
Eqs.~(\ref{4.31})--(\ref{4.33}) to 
approximate the transition between inspiral and plunge. We have investigated 
this 
question numerically by comparing the radial velocity computed with the 
``exact'' evolution (\ref{3.28})--(\ref{3.31}), and with the $\rho$-equation 
(\ref{4.29}). 
Let us define the practical limit of the domain of validity of the 
$\rho$-approximation by 
requiring that the fractional error in $dr / d \wt$ at the (say) $r$-LSO 
be 10\%. We find that this limit is reached when $\nu$ gets as large as
\beq
\nu_{\rm max} \simeq 0.05 \, .
\eeq
Therefore, the explicit expressions above can be used to estimate numerically 
the physical quantities in the transition region only for $\nu \leq \nu_{\rm 
max}$. 
Note 
that the accuracy of the $\rho$-results above is, by construction, limited to 
some 
small neighbourhood of the LSO. They should not be used (even if $\nu <  
\nu_{\rm 
max}$) to estimate, for instance, the radial velocity at a radius which is 
significantly different from $r_{\rm LSO}$ (say at $r=5$ or $r=7$). To 
illustrate this 
we compare in Figs.~\ref{Fig8} and \ref{Fig9} the radial velocity computed with 
the exact 
evolution, with that deduced from the $\rho$-equation. We examine three cases: 
$\nu = \nu_{\rm max} = 0.05$, $\nu = 10^{-2}$ and  $\nu = 10^{-4}$. 

Note that, though the accuracy of the approximation defined by the
 $\rho$-equation increases
as $\nu \rightarrow 0$, its domain of validity actually shrinks as $\nu$ gets 
small. 
Indeed, if we keep $\rho$ finite we see that $\delta r \simeq 1.890\,\nu^{2/5} 
\, 
\rho$, tends to zero with $\nu$.

\begin{figure}
\begin{center}
\begin{tabular}{cc}   
\hspace{-0.8cm} 
\epsfig{file=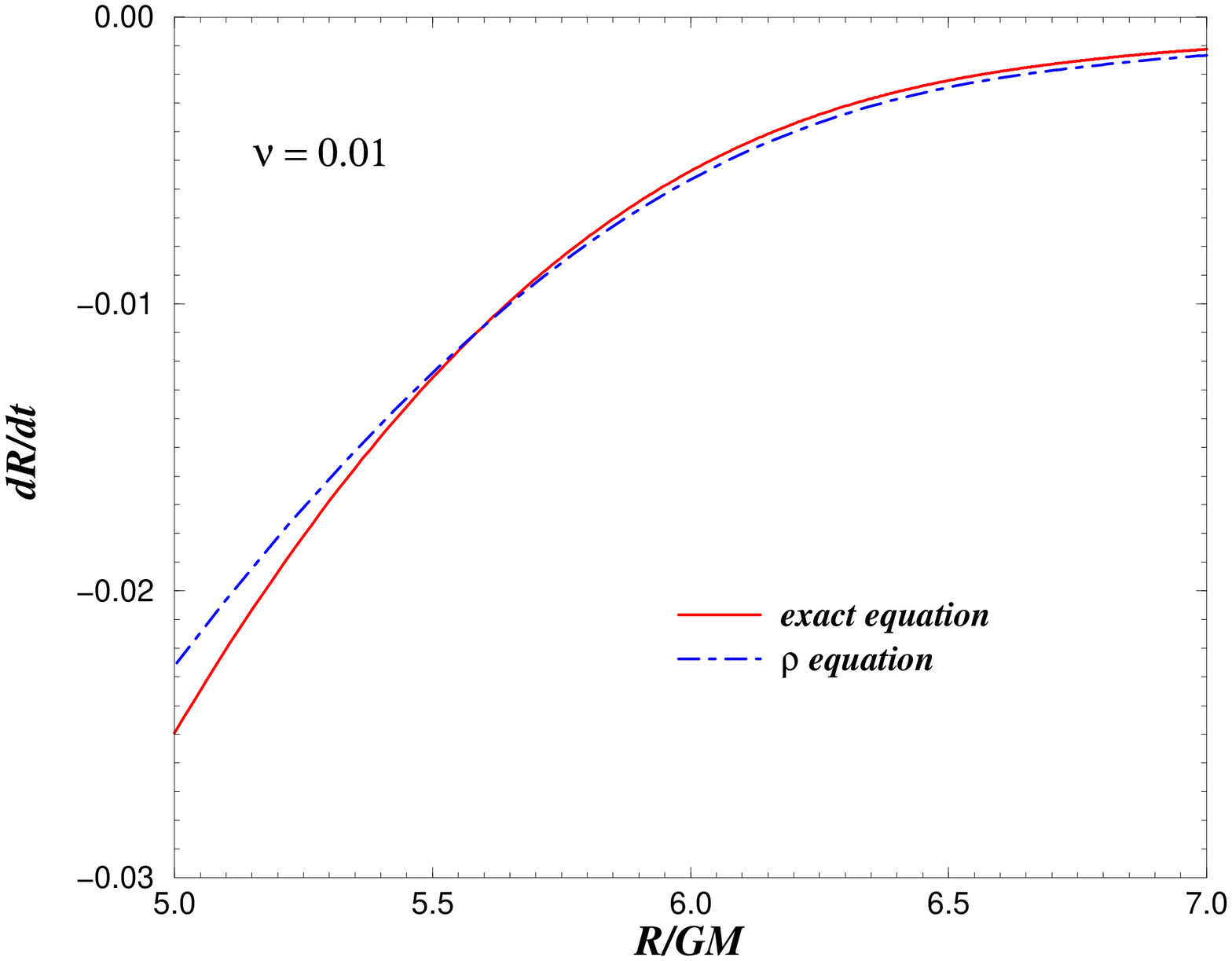,width=0.5\textwidth,height = 
0.5\textwidth,angle=-90} & 
\hspace{-0.0cm}
\epsfig{file=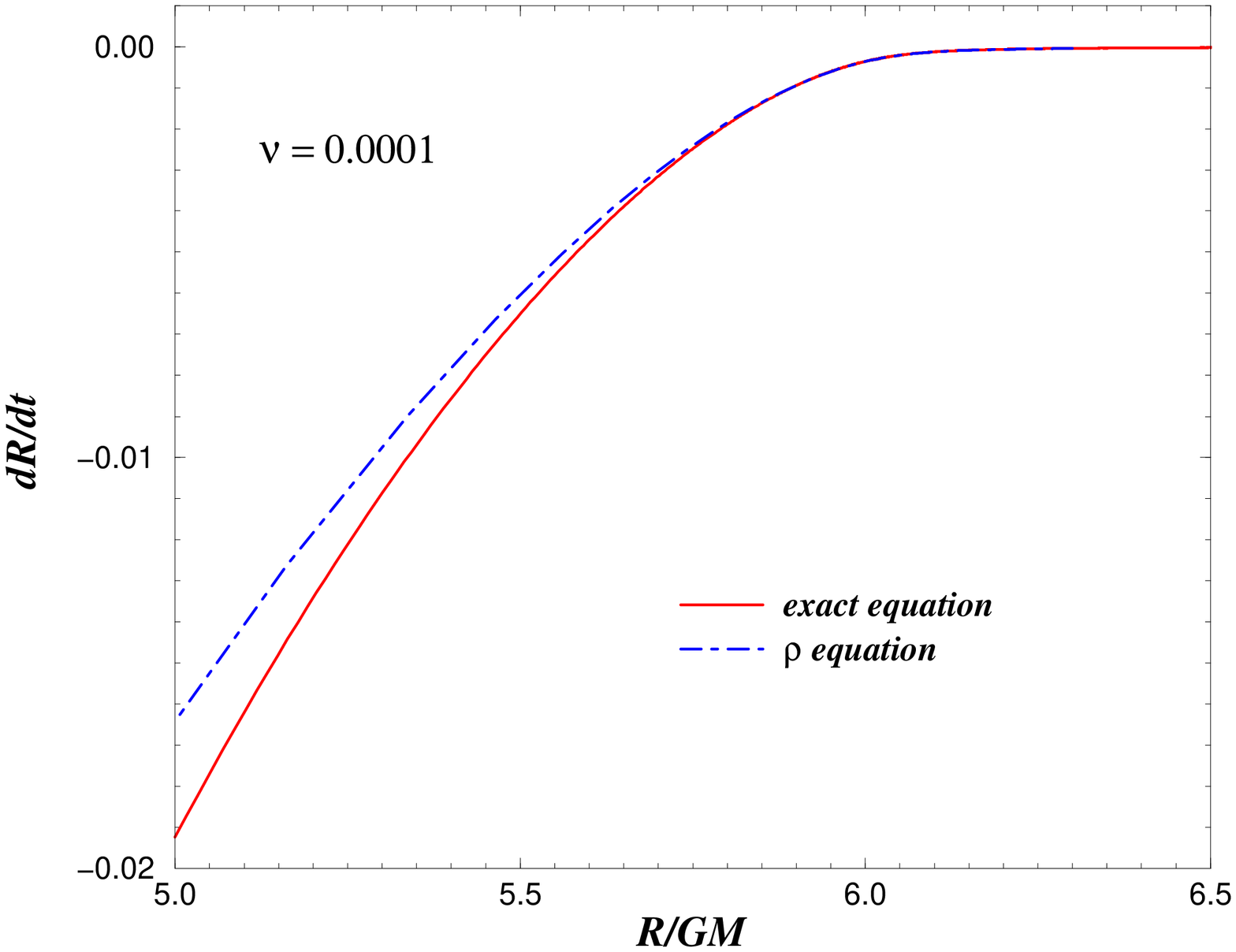,width=0.5\textwidth,height = 
0.5\textwidth,angle=-90}
\end{tabular}
\caption{\sl We compare the radial velocity evaluated with the exact 
dynamical system and with the $\rho$-equation in the cases $\nu = 0.01$ and $\nu 
= 0.0001$.}
\label{Fig9}
\end{center}
\end{figure}
Before discussing the scaling predictions made by the $\rho$-approximation, let 
us 
comment on the various possible definitions of ``LSO crossing''. We recall that 
we 
define: (i) the ``$r$-LSO'' (by the requirement $r(t) = r_{\rm LSO} (\nu)$), 
(ii) the 
``$j$-LSO'' ($j(t) = j_{\rm LSO} (\nu)$), and (iii) the ``$\omega$-LSO'' ($\ww 
(t) = 
\ww_{\rm LSO} (\nu)$). [In addition, one can also define an ``energy-LSO'', and 
a 
``naive'' LSO such that $R = 6{\rm GM}$.] We see from our results above that the 
$r$-LSO corresponds (in the $\rho$-approximation) to $\rho = 0$, while the 
$j$-LSO 
corresponds to $\tau = 0$, and the $\omega$-LSO to $\rho + 0.1575\, \nu^{2/5} \, 
\tau = 
0$. 
{}From these results and the results displayed in Fig.~\ref{Fig7} and 
Eqs.~(\ref{new1})
and (\ref{new2}), 
we have 
the following ordering between these LSO's: $\omega\hbox{-LSO} < r\hbox{-LSO} < 
j\hbox{-LSO}$, where the order symbols refer to the location on the radial axis. 
We 
see also that when $\nu^{2/5} \ll 1$ the $\omega$-LSO nearly coincides with the 
$r$-LSO. When discussing scaling relations it would be essentially equivalent to 
use 
any definition of LSO-crossing. For definiteness, and for consistency with the 
rest of 
this paper where we shall use it, we shall consider the $\omega$-LSO (because it 
is 
more invariantly defined than the $r$-LSO). To sufficient approximation for 
determining the leading scaling with $\nu$, we shall consider that the 
$\omega$-LSO 
corresponds to $\rho \simeq 0$.

\begin{figure}
\begin{center}
\begin{tabular}{cc}   
\hspace{-0.8cm} 
\epsfig{file=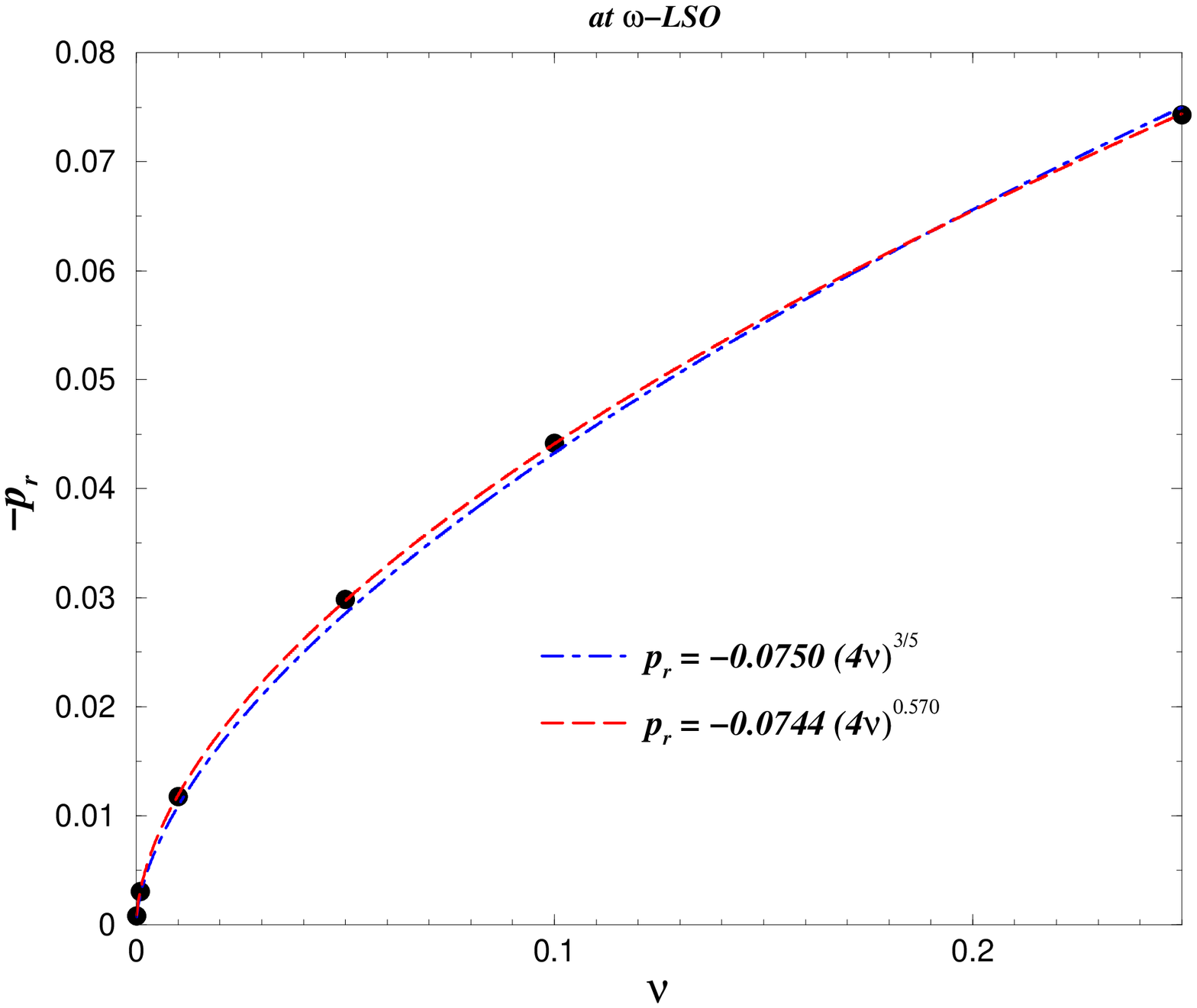,width=0.48\textwidth,height = 
0.48\textwidth,angle=-90} & 
\hspace{-0.0cm}
\epsfig{file=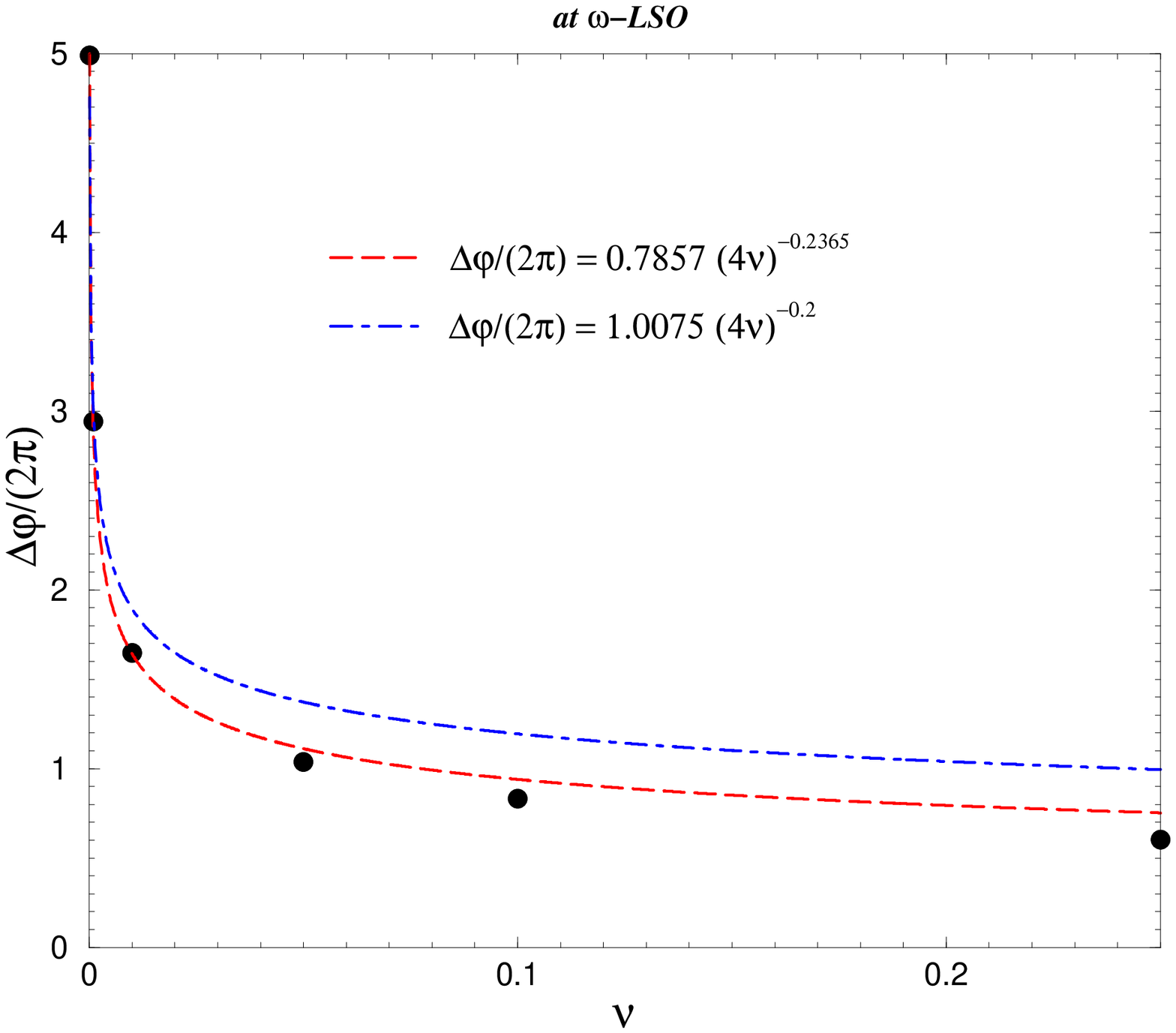,width=0.48\textwidth,height = 
0.48\textwidth,angle=-90}
\end{tabular}
\caption{\sl Scaling laws for the radial momentum and the post-LSO  
number of orbits provided by the $\rho$-approximation. 
On the left panel we show the exact numerical results 
for the radial momentum obtained by integrating the full evolution system 
down to the $\omega$-LSO. Two fits of the type 
$p_r =-a\,(4\nu)^{3/5}$ and $p_r = -a\, (4\nu)^b$ are also indicated.
On the right panel the number of orbits remaining ``after LSO-crossing''
is compared with the numerical results 
computed from the exact evolution. 
We have indicated both the best fit to a formula of the type $\Delta 
\varphi / 2\pi = a \, (4\nu)^{-1/5}$, and of the type $\Delta \varphi / 2\pi 
= a \, (4\nu)^b$. Note that, even if the figure covers the range of values 
of $\nu$ up to $1/4$, both fits have been evaluated including 
values only up to $\nu_{\rm max} = 0.05$.}
\label{Fig10}
\end{center}
\end{figure}

One of the most useful scaling law to consider is that concerning the radial 
momentum at the $\omega$-LSO. Combining Eqs.~(\ref{4.20}) and (\ref{4.33}) we 
get:
\beq
\label{4.39}
p_r = \frac{1}{C_r^{\rm LSO}} \, \frac{dr}{d \wt} = \frac{1}{C_r^{\rm 
LSO}}\,(A_r^{\rm LSO})^{-2/5}\,(B_r^{\rm 
LSO})^{3/5}\,\frac{d \rho}{d \tau}\,.
\eeq
{}From Eq.~(\ref{new2}) the value of $d\rho / d\tau$ at the $\omega$-LSO (i.e. 
$\rho 
\simeq 0$) 
is $d\rho / d\tau \simeq -0.8233$. Using also the numerical values (taken when 
$\nu 
\rightarrow 0$) of the coefficients entering Eq.~(\ref{4.39}), we get the 
predicted scaling
\beq
\label{4.40}
(p_r)_{\omega{\rm-LSO}} \simeq -0.0844 \,(4\nu)^{3/5} \, .
\eeq
In the left panel of Fig.~\ref{Fig10} we compare 
the analytical scaling prediction, $(p_r)_{\omega{\rm-LSO}} \propto 
\,(4\nu)^{3/5}$,  
with the numerical results obtained by integrating the full evolution system 
(\ref{3.28})--(\ref{3.31}) down to the $\omega$-LSO. 
We have also computed the best fits to the data using either a formula with one 
free parameter,  
of the type $p_r = -a \, (4\nu)^{3/5}$ or with two free parameters, $p_r = -a \, 
(4\nu)^b$.
Note that the predicted scaling is a   
surprisingly good fit to the exact results, even for values of $\nu$ much larger 
than the 
domain of validity of the $\rho$-equation. In fact, it is numerically quite 
accurate even 
for $\nu = 1/4$. [In the one-parameter fit, note that the best-fit coefficient
$a = 0.0750$ is $11 \%$ smaller than the calculated one, Eq.~(\ref{4.40}).
This is because the best-fit one takes into account the values of $p_r$ for 
larger
values of $\nu$ than the test-mass-limit result (\ref{4.40}).]

Another useful scaling law concerns the number of orbits remaining ``after 
LSO-crossing''. Let us define the number of orbits after LSO-crossing as $\Delta 
\varphi / 2\pi$ where $\Delta \varphi$ is the difference in orbital phase 
between the ``light-ring'' $r = r_{\rm light-ring} \, (\nu)$ (obtained 
from Eq.~(\ref{4.5})) and the $\omega$-LSO, $\omega = \omega_{\rm LSO} (\nu)$.
This quantity cannot be really estimated within the $\rho$-approximation, 
because this 
approximation assumes that $\delta r \ll 1$. However, we can formally say that, 
within 
the $\rho$-approximation, we wish to consider the asymptotic limit where $\rho$ 
tends 
to $-\infty$ proportionally to $\nu^{-2/5}$ (so that $\delta r$ is finite). The 
question is therefore: what is the asymptotic behaviour of the solution $\rho = 
\rho 
(\tau)$ of Eq.~(\ref{4.29}) when $\rho \rightarrow - \infty$? It seems that in 
this 
limit the ``source term'' $-\tau$ on the RHS of Eq.~(\ref{4.29}) is 
relatively 
negligible. Indeed, let us neglect it and solve the approximate equation 
$\frac{d^2 
\rho}{d \tau^2} + \frac{1}{2} \, \rho^2 = 0$. This equation describes the motion 
of a 
particle ($\ddot{\rho} = -\partial V (\rho) / \partial \rho$) with potential 
energy 
$V(\rho) = \rho^3 / 6$. This potential energy (which represents the effective 
radial 
potential near the inflection point corresponding to the LSO) is unboundedly 
negative 
when $\rho \rightarrow -\infty$. Writing the conservation of ``energy'', 
$\frac{1}{2} 
\, \dot{\rho}^2 + V(\rho) =$ const, one finds that, as $\rho \rightarrow 
-\infty$, the 
kinetic energy grows without bound and approximately satisfy $\frac{1}{2} \, 
\dot{\rho}^2 \approx - V(\rho)$ whose solution is
\beq
\label{4.37n}
\rho = -12 \, (\tau_{\infty} - \tau)^{-2}
\eeq
for some constant $\tau_{\infty}$. We conclude that, as $\rho \rightarrow 
-\infty$, the 
variable $\tau$ tends to a finite limit $\tau_{\infty}$. 
[We find $\tau_{\infty} \simeq 3.9$. The corresponding 
curve is shown in the left panel of Fig.~\ref{Fig7}.] Therefore, from 
Eq.~(\ref{4.37n}), 
the 
total time elapsed after the LSO, $\wt_{\infty} - \wt_{\rm 
LSO}$, 
scales like $\nu^{-1/5}$. Correspondingly, within the $\rho$-approximation, the 
leading 
approximation to the orbital phase (obtained by integrating the zeroth order 
term in Eq.~(\ref{4.37n})) reads
\beq
\label{4.37}
\frac{\Delta \vphi}{2 \pi} = \int_{\wt_{\rm LSO}}^{\wt_{\infty}} \frac{\ww}{2 
\pi}\,d \wt 
\simeq \frac{\ww_{\rm LSO}}{2 \pi}\,(\wt_{\infty} - \wt_{\rm LSO}) = 
 \frac{\ww_{\rm LSO}}{2 \pi}\,(A_r^{\rm LSO}\,B_r^{\rm 
LSO})^{-1/5}\,\tau_{\infty} \,.
\eeq
As $\ww_{\rm LSO}$ admits a finite limit as $\nu \rightarrow 0$, we expect from 
Eq.~(\ref{4.37}) the scaling law
\beq
\label{4.38}
\frac{\Delta \vphi}{2 \pi} \propto (4\,\nu)^{-1/5}\,.
\eeq
This prediction is compared in Fig.~\ref{Fig10} with the numerical results 
obtained by 
integrating the full system (\ref{3.28})--(\ref{3.31}). As expected from the 
necessity to inconsistently consider parametrically large values of $\rho 
\propto 
\nu^{-2/5}$, this prediction is less accurate than that obtained for the radial 
momentum at 
the $\omega$-LSO. We have indicated both the best fit to a formula of the type 
$\Delta 
\varphi / 2\pi = a \, (4\nu)^{-1/5}$, and the best fit to $\Delta \varphi / 2\pi 
= a \, (4\nu)^b$. Note that, both fits have been evaluated including 
values of $\nu$ only up to $\nu_{\rm max} = 0.05$. Indeed, as discussed above, 
beyond this value the fractional error in the radial velocity at the $r_{\rm 
LSO}$ 
is $\sim 10\%$.

Some comments are in order concerning these results. First, we note that 
although 
$N_{\rm LSO}^{\rm after} = \Delta \varphi / 2\pi$ tends to infinity when $\nu 
\rightarrow 0$, it does so very slowly so that the total number of orbits after 
the LSO 
is always quite small compared to the number of orbits ``just before and around 
the 
LSO''. Let us define the latter number as $N_{\rm LSO}^{\rm around} \equiv 
f_{\rm 
orbit.}^2 / \dot{f}_{\rm orbit.} = \frac{1}{2} \, f_{\rm GW}^2 / \dot{f}_{\rm 
GW}$ 
where $f_{\rm orbit.} = \frac{1}{2} \, f_{\rm GW} = \omega / 2\pi$ denotes the 
orbital 
frequency, and $\dot{f}_{\rm orbit.}$ the time derivative of the orbital 
frequency 
caused by GW damping. In the adiabatic approximation, combined with a Newtonian 
approximation for both the orbital energy and the GW flux, this number reads 
(see, 
e.g., \cite{DIS2})
\beq
N_{\rm LSO}^{\rm around} \simeq \frac{2.924}{4\nu} \, .
\eeq
The ratio $N_{\rm LSO}^{\rm after} / N_{\rm LSO}^{\rm around} \simeq 0.3446\, 
(4\nu)^{4/5}$ (derived using the result of the fit, i.e.  
$N_{\rm LSO}^{\rm after} = 1.0075\,(4\nu)^{-1/5}$) 
is therefore {\it parametrically} small as $\nu \rightarrow 0$. This suggests 
that, 
when 
$4\nu \ll 1$, the existence of even a formally parametrically large $(\propto 
\nu^{-1/5})$ absolute number of cycles left after the LSO will have only a 
fractionally 
negligible effect on the extraction of a GW signal from the noise by means of 
relativistic filters built on the adiabatic approximation, and terminated at the 
LSO 
\cite{DIS}, \cite{DIS2}. On the other hand, when $4\nu \sim 1$ the ratio $N_{\rm 
LSO}^{\rm after} / N_{\rm LSO}^{\rm around}$ is not very small. In particular, 
when 
$\nu = 1/4$ the number of orbits after the $\omega$-LSO is equal to $N_{\rm 
LSO}^{\rm after} (\nu = 1/4) = 0.6024$ (computed from the 
exact evolution), while $N_{\rm LSO}^{\rm around} (\nu = 1/4) = 2.924$. The 
ratio between the two is $N_{\rm LSO}^{\rm after} / N_{\rm LSO}^{\rm around} = 
0.2060$. 
As 
recently emphasized in Ref.~\cite{DIS2}, the fact that $N_{\rm LSO}^{\rm 
around}$ is 
not large means that the filtering of such a signal out of the noise is a 
delicate 
matter which sensitively depends on the modeling of the phase evolution near 
the LSO, 
and on the modeling of what happens to the signal after LSO crossing. In 
Ref.~\cite{DIS2} it was assumed that the signal is abruptly terminated at the 
LSO. In a 
later section we shall use the tools introduced here to go beyond such an 
approximation 
and study the part of the waveform which is emitted after LSO crossing.

\section{Initial data for numerical relativity}
\label{sec6}

One of the main aims of this paper is to use the improved approach to the 
transition 
from the inspiral to the plunge introduced above to compute initial dynamical 
data 
(i.e. initial positions and momenta) for binary black holes that have just 
started 
their plunge motion. Ideally, we wish to give dynamical data for two black holes 
$(\mbox{\boldmath$q$}_1 , \mbox{\boldmath$q$}_2 , \mbox{\boldmath$p$}_1 , 
\mbox{\boldmath$p$}_2)$ such that the coordinate distance $\vert 
\mbox{\boldmath$q$}_1 
- \mbox{\boldmath$q$}_2 \vert$ is: (i) large enough that one can trust the 
re-summed 
non-perturbative technique allowing one to compute these data; (ii) large enough 
to 
allow one to hope to complete the present work by constructing the initial 
gravitational data $(g_{ij} (x) , K_{ij} (x))$ determined (in principle) by 
$(\mbox{\boldmath$q$}_a , \mbox{\boldmath$p$}_a)$; and, finally (iii) small 
enough to 
leave only less than one orbit (at least when $\nu \sim 1/4$) to evolve by means 
of 
a full 
3D numerical relativity code. We think that the point (i) is satisfied if we use 
the 
Pad\'e-type \cite{DIS} plus effective-one-body \cite{BD99} methods we have 
combined 
above, {\it and} if we stop the evolution of quasi-circular orbits anywhere 
around the 
LSO. We shall leave the point (ii), i.e. the important task of completing the 
present 
work by constructing gravitational data to future work. However, in preparation 
for 
this task we shall show how one can compute the dynamical data 
$(\mbox{\boldmath$q$}_a 
, \mbox{\boldmath$p$}_a)$ in the convenient ADM coordinates. Indeed, the 
coordinate 
conditions introduced by Arnowitt, Deser and Misner \cite{ADM60}
have the double advantage: (a) to be linked to the $3+1$ formulation 
which 
is used in numerical relativity, and (b) to be linked to explicit, high-order 
post-Newtonian calculations \cite{JS98}.
Concerning 
the point (iii), the work above shows that if we stop the inspiral $+$ plunge 
evolution 
at the (invariantly defined) $\omega$-LSO (i.e. when $d\varphi / dt = 
\omega_{\rm LSO} 
(\nu)$) there indeed remains (when $4\nu \sim 1$) less than one orbit to go 
before 
reaching the light-ring (see next section for a discussion of the importance of 
the 
light-ring). Note that there is nothing sacred about giving data precisely at 
the 
$\omega$-LSO. Because of the points (i) and (ii) above we wish to stay ``as high 
as 
possible''. Because of point (iii) we must, however, be just after LSO crossing. 
As was 
already discussed, there are several possible definitions of ``LSO crossing''. 
The 
$\omega$-LSO is the innermost LSO (see below) and is therefore a convenient 
choice 
(however, there would be nothing wrong in giving data at a slightly different 
place; in 
fact we recommend to do it to check the robustness of the numerical spacetimes 
evolved 
from our data).

As just recalled we wish to (numerically) compute complete dynamical data at the 
$\omega$-LSO, and in ADM coordinates. The evolution system 
(\ref{3.28})--(\ref{3.31}) 
given above allows one to compute dynamical data $(r,\varphi,p_r,p_{\varphi})$ 
for the 
relative motion described in (reduced) {\it effective} coordinates (i.e. the 
coordinates used in the effective-one-body description). In Ref.~\cite{BD99} we 
have 
shown how to map the ADM positions and momenta $(q^{\rm ADM} , p^{\rm ADM})$ 
onto the 
effective positions and momenta $(q,p)$ by means of a generating function $G 
(q^{\rm 
ADM} , p)$. Let us first recall, in order to avoid any confusion, the trivial 
transformations linking Cartesian-like to polar-like coordinates, as well as 
those 
linking the original to the scaled coordinates. We recall that we work in the 
center of 
mass frame and that we consider planar motion in the equatorial plane $\theta = 
\pi / 
2$:
\beq
Q^i = q_1^i - q_2^i\,, \quad \qquad P_i = p_{1i} = - p_{2i} \,, 
\eeq
\beq
P_R = n^i \, P_i\,, \quad \qquad P_{\varphi} = Q^x \, P_y - Q^y \, P_x \,, 
\eeq
\beq
q^i = \frac{Q^i}{{GM}}\,, \quad \qquad \quad p_i = \frac{P_i}{\mu} \,, 
\eeq
\beq
p_r = \frac{P_R}{\mu} = n^i \, p_i\,, \quad \quad \quad p_{\varphi} = 
\frac{P_{\varphi}}{\mu \, {GM}} = q^x \, p_y - q^y \, p_x \, .
\eeq
Here $n^i = Q^i / R = q^i / r$ is the radial unit vector ($R = \vert 
\mbox{\boldmath$Q$} \vert$, $r = \vert \mbox{\boldmath$q$} \vert$). We have also 
$Q^x = 
R \cos \varphi$, $Q^y = R \sin \varphi$, $q^x = r \cos \varphi$, $q^y = r \sin 
\varphi$. The relations above hold both in effective coordinates (denoted by 
$(q^i 
, p_i)$ 
without extra labels) and in ADM coordinates $(q_{\rm ADM}^i , p_i^{\rm ADM})$. 
The 
link between $(q^i , p_i)$ and $(q_{\rm ADM}^i , p_i^{\rm ADM})$ is defined by a 
generating function $G (q_{\rm ADM}^i , p_i)$ and reads
\bea
\label{6.1}
&& q^i = q_{\rm ADM}^i + \frac{\partial \, G(q^{\rm ADM},p)}{\partial \, 
p_i} \,, \\
\label{6.2}
&& p^{\rm ADM}_i = p_i + 
\frac{\partial \, G(q^{\rm ADM},p)}{\partial \, q_{\rm ADM}^i} \, . 
\eea
The generating function $G$ has been derived up to 2PN order
in \cite{BD99} (see Ref.~\cite{DJS1} for the determination of $G$ at the 3PN 
level)
\beq
\label{6.3}
G (q^{\rm ADM},p) = \frac{1}{c^2} \, 
G_{1 {\rm PN}} (q^{\rm ADM},p) + \frac{1}{c^4} \, G_{2 {\rm PN}} (q^{\rm 
ADM},p) \, . 
\eeq
The partial derivatives needed in Eqs.~(\ref{6.1}), (\ref{6.2}) read
\bea
\label{6.4}
\frac{\partial G_{\rm 1PN}(q,p)}{\partial q^i} &=& 
p_i\,\left [-\frac{\nu}{2}\,\mbox{\boldmath$p$}^2 + 
\left ( 1 + \frac{\nu}{2} \right )\,\frac{1}{q}\right ] - 
q_i\,(\mbox{\boldmath$q$}\cdot \mbox{\boldmath$p$})\,\left ( 1 + \frac{\nu}{2} 
\right )\,
\frac{1}{q^3}\,,\\
\label{6.5}
\frac{\partial G_{\rm 1PN}(q,p)}{\partial p_i} &=&
q^i\,\left [-\frac{\nu}{2}\,\mbox{\boldmath$p$}^2 + 
\left (1 + \frac{\nu}{2} \right )\,\frac{1}{q} \right ] - 
p^i\,(\mbox{\boldmath$q$}\cdot \mbox{\boldmath$p$})\,\nu\,, 
\eea
\bea
\label{6.6}
\frac{\partial G_{\rm 2PN}(q,p)}{\partial q^i} &=& 
p_i\,\left [\frac{1}{8}\,\nu\,(1 + 3\,\nu)\,
\mbox{\boldmath$p$}^4 + \frac{\nu}{8}\,(2 - 5\,\nu)\,
\frac{\mbox{\boldmath$p$}^2}{q} + 
\frac{3}{8}\,\nu\,(8 + 3\,\nu)\,\frac{
(\mbox{\boldmath$q$}\cdot\mbox{\boldmath$p$})^2}{q^3} \right . \nonumber \\
&&\left . + \frac{1}{4}\,(1 - 7\,\nu + \nu^2)\,\frac{1}{q^2} \right ]  
+ q_i\,(\mbox{\boldmath$q$}\cdot\mbox{\boldmath$p$})\,
\left [- \frac{3}{8}\,\nu\,(8 + 3\,\nu)\,\frac{
(\mbox{\boldmath$q$}\cdot\mbox{\boldmath$p$})^2}{q^5} \right . \nonumber \\
&& \left . - \frac{\nu}{8}\,(2 - 5\,\nu)\,\frac{
\mbox{\boldmath$p$}^2}{q^3} 
- \frac{1}{2}\,(1-7\,\nu+\nu^2)\,\frac{1}{q^4} \right ]\,, \\
\label{6.7}
\frac{\partial G_{\rm 2PN}(q,p)}{\partial p_i} &=& 
q^i\,\left [\frac{1}{8}\,\nu\,(1 + 3\,\nu)\,\mbox{\boldmath$p$}^4 + 
\frac{\nu}{8}\,(2 - 5\,\nu)\,\frac{\mbox{\boldmath$p$}^2}{q} 
+ \frac{3}{8}\,\nu\,(8+3\,\nu)\,
\frac{(\mbox{\boldmath$q$}\cdot\mbox{\boldmath$p$})^2}{q^3} \right .
\nonumber \\
&& \left . + \frac{1}{4}\,(1 - 7\,\nu + \nu^2)\,\frac{1}{q^2}
\right ] + p^i\,(\mbox{\boldmath$q$}\cdot\mbox{\boldmath$p$})\,\left [
\frac{\nu}{2}\,(1 + 3\,\nu)\,\mbox{\boldmath$p$}^2 
+\frac{\nu}{4}\,(2 - 5\,\nu)\,\frac{1}{q} \right ]\,. 
\eea
Given $q^i$ and $p_i$, we use first Eq.~(\ref{6.1}), and the values of the 
partial 
derivatives (\ref{6.4})--(\ref{6.7}), to solve numerically for $q_{\rm ADM}^i$. 
Then 
we use Eq.~(\ref{6.2}) to compute $p_i^{\rm ADM}$: 

The initial data we start with are the results of the 
numerical integration of the system (\ref{3.28})--(\ref{3.31}), i.e. the values 
of $r$, 
$\varphi$, $p_r$ and $p_{\varphi}$ at some time in the evolution (which we 
choose to be 
the time when $\omega (t) = \omega_{\rm LSO} (\nu)$). Actually, the value of 
$\varphi$ 
is without significance and we renormalize it to the convenient value 
$\varphi_{\rm 
new} = 0$ so that we work with Cartesian-like data of the simple form (remember 
that we 
work in the $x-y$ plane, $q^z = 0 = p_z$, and that we simplify the writing by 
denoting 
$q_i \equiv q^i$ when working in Cartesian-like coordinates)
\beq
\label{6.8}
q_x = r\,, \quad \quad q_y = 0\,, \quad \quad p_x = p_r\,,\quad \quad 
p_y = \frac{p_\vphi}{r}\,.
\eeq
When solving, as indicated above, Eqs.~(\ref{6.1}), (\ref{6.2})  to derive 
$q_x^{\rm 
ADM}$, $q_y^{\rm ADM}$ and $p_x^{\rm ADM}$, $p_y^{\rm ADM}$, we get these 
quantities in 
a not optimally oriented coordinate system (i.e. though we started with $q_y = 
0$, we 
end up with $q_y^{\rm ADM} \ne 0$ because there is a rotation between the two 
coordinate systems). As the global orientation is of no physical significance, 
it is 
convenient to turn the ADM coordinate system by an angle $\alpha$ so that 
$\varphi_{\rm 
new}^{\rm ADM} = \varphi_{\rm old}^{\rm ADM} - \alpha = 0$. In other words, 
after 
this 
rotation one has, as in Eq.~(\ref{6.8}) above,
\beq
\label{5.14n}
q_x^{\rm ADM \, new} = r^{\rm ADM} \, , \quad q_y^{\rm ADM \, new} = 0 \, , 
\quad 
p_x^{\rm ADM \, new} = p_r^{\rm ADM} \, , \quad p_y^{\rm ADM \, new} = 
\frac{p_{\varphi}^{\rm ADM}}{r^{\rm ADM}} \, .
\eeq
The angle of rotation $\alpha$ is determined by
\beq
\label{5.15n}
\tan \alpha = \frac{q_y^{\rm ADM \, old}}{q_x^{\rm ADM \, old}} \, ,
\eeq
while the more invariant quantities $r^{\rm ADM}$ and $p_r^{\rm ADM}$ are given 
by
\beq
\label{6.9}
r^{\rm ADM} \equiv \sqrt{(q_{x \, {\rm old}}^{\rm ADM})^2 + (q_{y \, {\rm 
old}}^{\rm 
ADM})^2} \,, \quad \quad \\
p_r^{\rm ADM} \equiv \frac{1}{r^{\rm ADM}}\,(q_{x \, {\rm old}}^{\rm ADM}\,p_{x 
\, {\rm 
old}}^{\rm ADM} + q_{y \, {\rm old}}^{\rm ADM}\,p_{y \, {\rm old}}^{\rm ADM})\,.
\eeq
Note that (because of the rotational invariance of $G$) all the angular momenta 
coincide:
\beq
p_{\varphi} = p_{\varphi}^{\rm ADM} = q_x \, p_y - q_y \, p_x = q_{x \, {\rm 
old}}^{\rm 
ADM} \, p_{y \, {\rm old}}^{\rm ADM} - q_{y \, {\rm old}}^{\rm ADM} \, p_{x \, 
{\rm 
old}}^{\rm ADM} = q_{x \, {\rm new}}^{\rm ADM} \, p_{y \, {\rm new}}^{\rm ADM} - 
q_{y 
\, {\rm new}}^{\rm ADM} \, p_{x \, {\rm new}}^{\rm ADM} \, .
\eeq
This relation is a useful check on the numerical precision of the solution of 
Eqs.~(\ref{6.1}), (\ref{6.2}). 

In Tab.~\ref{Tab1} we give initial data in ADM coordinates
at the $\omega$-LSO for five values of the parameter $\nu$. We give the more 
invariant 
quantities corresponding to the ``new'' ADM coordinate system Eq.~(\ref{5.14n}). 
The 
quantity 
$p_t^{\rm ADM}$ denotes the ``transverse'' momentum, i.e. simply $p_t^{\rm ADM} 
\equiv 
p_{\varphi}^{\rm ADM} / r^{\rm ADM} \equiv p_{y \, {\rm new}}^{\rm ADM}$. For 
completeness, we give also the value of the angle $\alpha$, Eq.~(\ref{5.15n}).  
\begin{table}
\begin{center}
\begin{tabular}{cccccc}
$\nu $  & $r^{\rm ADM}$ & $p_r^{\rm ADM}$ & $p_t^{\rm ADM}$& $p_\varphi^{\rm 
ADM}$ &
$\alpha$ \\ 
\hline
$0.25 $ & $4.717$  &  $-0.07570$ & $0.7021$ & $3.312$ & $-0.006256$  \\
$ 0.1 $ & $4.853$  &  $-0.04425$ & $0.6997$ & $3.396$ & $-0.001524$   \\
$ 0.01$ & $4.938$  &  $-0.01163$ & $0.6996$ & $3.455$ & $-4.088\cdot10^{-5}$\\
$ 0.001$ & $4.948$ &  $-0.002992$& $0.6998$ & $3.463$ & $-1.054\cdot10^{-6}$\\
$ 0.0001$ &$4.949$ &  $-0.0007592$& $0.6999$& $3.464$ & $-2.675\cdot 10^{-8}$
\end{tabular}
\caption{\sl Initial data in ADM coordinates at $\omega$-LSO for five 
representative values 
of $\nu$.}
\label{Tab1}
\end{center}
\end{table}

So far all the results we have discussed considered the evolution system 
(\ref{3.28})--(\ref{3.31}) as the ``exact'' description of the transition 
through the 
LSO. However, as discussed in Sec.~III this system is more like a convenient 
fiducial 
system within a class of systems obtained by shifting (by ${\cal O} (v^5 / c^5)$ 
terms) 
the coordinate system. To test the robustness of our predictions for physical 
quantities at the LSO we shall now compare the results of the fiducial system 
(\ref{3.28})--(\ref{3.31}) with the results obtained by the more general system 
(\ref{3.4})--(\ref{3.7}), with a radial force ${\cal F}_R$ given (in terms of 
${\cal 
F}_{\varphi}$) by Eq.~(\ref{3.17}). For simplicity, we consider only the (most 
crucial) 
equal-mass case, $\nu = 1/4$. We find that our fiducial system (with ${\cal F}_R 
= 0$) 
yields the following numerical values at the $\omega$-LSO (when starting with an 
orbital phase $\varphi = 0$ at $r = 15$)
\bea
\label{6.14}
&& r = 5.639\,,\quad \quad p_r = -0.07432\,,\quad \quad \dot{r} = 
-0.03563\,,\\
\label{6.15}
&& \vphi = 82.72 \,,\quad \quad j = 3.312 \,,\quad \quad \frac{{\cal 
E}_{\rm real}^{\rm NR}}{M} = -0.01640\,.
\eea
On the other hand, the system including the non-zero radial force (\ref{3.17}) 
yields 
at the $\omega$-LSO (still starting with an orbital phase $\varphi = 0$ at 
$r=15$)
\bea
\label{6.17}
&& r = 5.638\,,\quad \quad p_r = -0.07388\,,\quad \quad \dot{r} = 
-0.03542\,,\\
\label{6.18}
&& \vphi = 82.77 \,,\quad \quad j = 3.311 \,,\quad \quad \frac{{\cal 
E}_{\rm real}^{\rm NR}}{M} = -0.01643\,.
\eea
As we see the differences in the numerical results are quite small. For 
instance, the 
fractional change in the (crucial) radial momentum is less than $6 \times 
10^{-3}$. We note also that the dephasing at the LSO is only 0.05 radians. This 
analysis 
indicates 
that the results based on our fiducial system are quite robust, mainly because 
our 
basic assumption of ``quasi-circularity'' ($\dot R \ll R \, \dot\varphi$) is 
well 
satisfied during the transition to the plunge.

\section{Gravitational wave-forms from inspiral to ring-down}
\label{sec5}

In this section, we provide, for data analysis purposes, an estimate of the 
complete 
waveform emitted by the coalescence of two black holes (with negligible spins). 
This 
estimate will be less accurate than our results above because we shall extend 
the 
integration of our basic system (\ref{3.28})--(\ref{3.31}) beyond its range of 
validity. We think, however, that even a rough estimate of the complete waveform 
(exhibiting the way the inspiral waveform smoothly transforms itself in a 
``plunge 
waveform'' and then into a ``merger plus ring-down'' waveform) is a very 
valuable 
information for designing and testing effectual gravitational wave templates. 
[See, in 
particular, the recent work \cite{DIS2} which emphasizes the importance of the 
details 
of the transition to the plunge for the construction of faithful GW templates 
for massive binaries.]

Our (rough) assumptions in this section will be the following: (i) we use the 
basic evolution system (\ref{3.28})--(\ref{3.31}) to describe the dynamics of 
the 
binary system from deep into the inspiral phase (say $r \simeq 15$) down to  
the ``light-ring'' $r = r_{\rm light-ring} (\nu) \simeq 3$; (ii) we estimate the 
waveform emitted during the inspiral and the plunge by means of the usual 
``restricted waveform'' approximation
\beq
\label{5.1}
\wt \leq \wt_{\rm end}: \quad h_{\rm inspiral} (\wt) = {\cal 
C} \, v_{\omega}^2 (\wt) \cos (\phi_{\rm GW} (\wt))\,, \quad v_{\omega} 
\equiv \left( \frac{d \varphi}{d \wt} \right)^{\frac{1}{3}}\,, \quad \phi_{\rm 
GW} \equiv 2\varphi \,,
\eeq
and (iii) we estimate the waveform emitted during the coalescence and ring-down 
by matching, at a time $\wt = \wt_{\rm end}$ where the light-ring is crossed, 
the 
inspiral $+$ plunge waveform (\ref{5.1}) to the least-damped quasi-normal mode 
of a Kerr black hole with mass and spin equal to the total energy and angular 
momentum of the plunging binary (at $\wt = \wt_{\rm end}$):
\beq
\label{5.6}
\wt \geq \wt_{\rm end}: \quad h_{\rm merger}(\wt) = {\cal A}\,e^{-(\wt 
-\wt_{\rm end})/\tau}\, \cos(\omega_{\rm qnm}\,(\wt -\wt_{\rm end}) + {\cal 
B})\,.
\eeq
For convenience, we shall normalize the waveform by taking ${\cal C} = 1$ in 
Eq.~(\ref{5.1}). The amplitude ${\cal A}$ and the phase ${\cal B}$ of the merger 
waveform (\ref{5.6}) are then determined by requiring the continuity of 
$h(\wt)$ 
and $dh / d\wt$ at the matching point $\wt = \wt_{\rm 
end}$. 

Before giving technical details let us comment on our assumptions (i)--(iii). 
First, we 
recall that Fig.~\ref{Fig1} had shown that the quasi-circularity condition 
$p_r^2 / 
B(r) \ll p_{\varphi}^2 / r^2$ (which is the basic condition determining the 
validity of 
our evolution system) was satisfied with good accuracy during the inspiral and 
the 
beginning of the plunge, and was still satisfied, though with less accuracy 
($p_r^2 / B 
\lesssim 0.3 \, p_{\varphi}^2 / r^2$ in the worst case $\nu = 1/4$) down to the 
light-ring $r \simeq 3$. In other words, our work is showing that the so called 
``plunge'' 
following the inspiral phase is better thought of as being still a 
quasi-circular 
inspiral motion, even down to the light-ring. We therefore expect that the usual 
restricted waveform (\ref{5.1}) (valid for circular motion)
will be an acceptable description of the GW 
emission 
during the plunge. Note that we consider that the description of the amplitude 
of the 
gravitational wave in terms of $v_{\omega}^2 \equiv \dot{\varphi}^{2/3}$, being 
simpler 
and more invariant, has a better chance of being correct than a description in 
terms of 
some other Newtonian-like approximation to the ``squared velocity'' such as $(r 
\, 
\dot{\varphi})^2$ or $1/r$. Some evidence for this faith is given by the fact 
that the 
GW flux is surprisingly well approximated (within 10\% down to the LSO) by the 
usual 
``quadrupole formula'' {\it if} the velocity used to define the quadrupole 
formula is 
the invariant $v_{\omega} = \dot{\varphi}^{1/3}$ (see, e.g., Fig.~3 of 
\cite{DIS}).

Concerning the choice of the light-ring for shifting the description between a 
(quasi-circular) binary motion and a deformed Kerr black hole, our motivation is 
twofold. First, in the test-mass limit, $\nu \ll 1$, it has been realized long 
ago, in 
the first work \cite{Davis} which found the existence of a merger signal of 
the type 
(\ref{5.6}) following a plunge event, that the basic physical reason underlying 
the 
presence of a ``universal'' merger signal was that when a test particle falls 
below $R 
\simeq 3 {GM}$, the GW it generates is strongly filtered by the potential 
barrier, 
centered around $R \simeq 3 {GM}$, describing the radial propagation of 
gravitational waves. It was then realized \cite{Press} that the peaking of the 
potential barrier around $R \simeq 3 {GM}$ is itself linked to the presence 
of an 
unstable ``light storage ring'' (i.e. an unstable circular orbit for massless 
particles) precisely at $R = R_{\rm light-right} = 3 {GM}$. A second 
argument 
(applying now in the equal-mass case, $\nu = 1/4$) indicating that 
$r_{\rm light-right} (1/4) \simeq 2.84563$ is an acceptable 
divide 
between the 
two-body and the perturbed-black-hole descriptions comes from the works on the, 
so 
called, ``close limit approximation'' \cite{PP94}. Indeed, recent 
work (see the review \cite{Pullin99}) suggests a matching between the two-body 
and the 
perturbed-black-hole descriptions when the distance modulus $\mu_0 \simeq 2$. 
Using the 
formulas of Ref.\cite{AP97} one finds that $\mu_0 \simeq 2$ 
corresponds to a 
coordinate distance in isotropic coordinates of $r^{\rm iso} \simeq \sqrt 2 {x
GM}$. 
This corresponds to a Schwarzschild-like radial distance $R \simeq r \, (1 + 
{GM} / 
2r)^2 \simeq 2.59 {GM}$ which is not very far from $R_{\rm light-ring} (1/4) 
\simeq 2.84 {GM}$.

In keeping with our prescription of setting the divide between a 
binary-black-hole 
description and a perturbed-single-black-hole one, at the time $\wt_{\rm 
end}$, when $r \simeq r_{\rm light-ring} (\nu)$, it is natural to assume that 
the 
final 
hole formed by the merger is a Kerr hole with mass $M_{\rm BH}$ and angular 
momentum ${\cal J}_{\rm BH}$ given by:
\beq
\label{5.4}
\frac{M_{\rm BH}}{\mu} \equiv \wH_{\rm end} = \frac{1}{\nu}\,
\sqrt{1 + 2\nu\,(\wH^{\rm end}_{\rm eff} -1)}\,,
\quad \quad j_{\rm end} \equiv  \frac{{\cal J}_{\rm BH}}{\mu\,G\,M}\,,
\eeq
while the dimensionless rotation parameter $\widehat{a}$ is:
\beq
\label{5.5}
\widehat{a}_{\rm BH} \equiv \frac{{\cal J}_{\rm BH}}{G\,M^2_{\rm BH}}=
\frac{\nu\,j_{\rm end}}{1 + 2\nu\,(\wH^{\rm end}_{\rm eff} -1)}\,.
\eeq
As the system reaches the stationary Kerr state, 
the non-linear dynamics of the merger becomes more and more describable 
in terms of oscillations of the black hole quasi-normal modes~\cite{QNR}.
During this phase, often called the ring-down phase, 
the gravitational signal will be a superposition of exponentially damped
sinusoids. The gravitational waveform 
will be dominated by the $l=2, m=2$ quasi-normal mode, which is the 
most slowly damped mode.

As a rough approximation we assume that the full merger $+$ ring-down signal 
(starting 
when the light-ring is reached) can be represented in terms of this least damped 
quasi-normal mode. If $\omega_{qnm}$ denotes the circular frequency of this 
mode, and 
$\tau$ its damping time, this leads to the simple description (\ref{5.6}). The 
quantities $(\omega_{qnm} , \tau)$ are functions of $(M_{\rm BH} , 
\widehat{a}_{\rm 
BH})$ which have been investigated numerically \cite{QNR}, \cite{L85}. Using 
analytic 
fits the following expressions for the frequency and the decay time of 
the quasi-normal modes were obtained \cite{E89}
\bea
\label{5.7}
&& M_{\rm BH}\,\omega_{\rm qnm}= \left [ 1 -0.63\,(1 - 
\widehat{a})^{3/10} 
\right ]\,f_f(\widehat{a}) \,,\\
\label{5.8}
&&\tau\, \omega_{\rm qnm} = 4 \left [ 1 - \widehat{a} \right 
]^{-9/20}\,f_Q(\widehat{a})\,,
\eea
where $f_f(\widehat{a})$ and $f_Q(\widehat{a})$ are correction factors 
provided by Tab.~2 of \cite{E89}. Note that $f_f = 0.9587$ and 
$f_Q = 1.0501$ for $\widehat{a} = 10^{-4}$.  

We have numerically studied only the equal-mass case $\nu = 1/4$. We have chosen 
the matching point $\wt_{\rm end}$ such that $r \, (\wt_{\rm end}) = 
r_{\rm light-ring} (1/4) = 2.84563$. 
With this value of $\wt_{\rm end}$ we obtain the following values for the 
characteristics of the formed black hole:
\bea
\label{5.9}
&& \widehat{a}_{\rm BH} = 0.7952\,, \quad \quad 
E_{\rm BH} = 0.9761\,M\,, \\
\label{5.10}
&& M\,\omega_{\rm qnm} = 0.5976\,, \quad \quad M/\tau = 0.07795\,.
\eea
\begin{figure}
\begin{center}
\epsfig{file=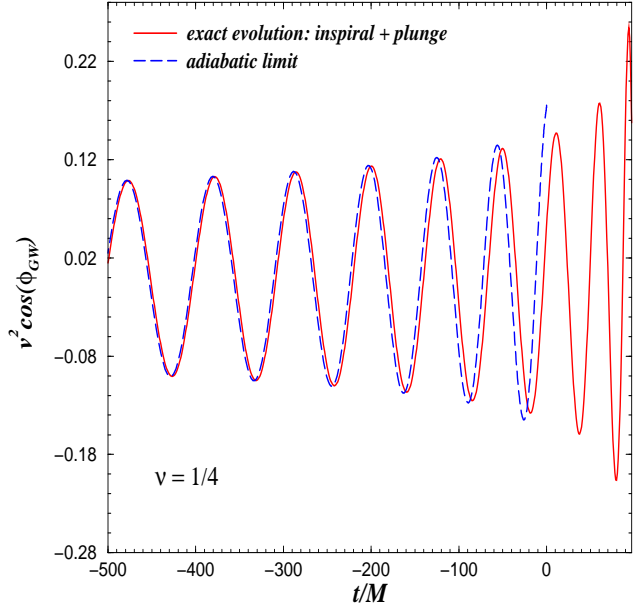,width=0.5\textwidth,height = 0.5\textwidth,angle=-90} 
\caption{\sl We compare the inspiral $+$ plunge waveform, 
terminated at the light-ring, to the adiabatic waveform, terminated at the 
adiabatic LSO.}
\label{Fig11}
\end{center}
\end{figure}
Note the numerical value of the quasi-normal mode frequency 
\beq
\label{5.12}
f_{\rm qnm} \simeq \frac{\omega_{\rm qnm}}{2 \pi} = 1885 \,\left 
(\frac{10\,M_\odot}{M_{\rm BH}} \right) \, {\rm Hz} \,.
\eeq
Our results for the waveform are shown in Figs.~\ref{Fig11} and \ref{Fig12}. In 
Fig.~\ref{Fig11} we compare the inspiral $+$ plunge waveform (\ref{5.1}) 
(terminated at 
the light-ring) to the usually considered adiabatic waveform (terminated at the 
``adiabatic LSO''). As already discussed in Section \ref{sec4},  
by ``adiabatic waveform'' we mean a restricted waveform (\ref{5.1}) (with ${\cal 
C} = 
1$) in which $\varphi (\wt) = \varphi_{\rm adiab.} (\wt)$ is 
defined by 
integrating the two equations (\ref{4.8b}) and (\ref{4.9}).  This Figure shows 
that there is a 
significant dephasing of the adiabatic waveform with respect to the (more) exact 
one 
already before the LSO. Moreover, the real inspiral signal continues to increase 
and oscillate for $ \simeq 2.35$ cycles after the adiabatic LSO. 

In Fig.~\ref{Fig12} we plot our estimate of the complete waveform: inspiral and 
plunge 
(solid line) followed by merger and ring-down (dashed line). We also indicated 
the 
locations of several possible definitions of LSO crossing (see Section 
\ref{sec4} 
above). In 
addition to the definitions mentioned above we also included a ``naive LSO'' 
(defined 
simply by $r_{\rm LSO}^{\rm naive} \equiv 6$ as in the Schwarzschild geometry) 
and an 
energy-LSO (such that ${\cal E}_{\rm real} (t) = {\cal E}_{\rm real}^{\rm LSO} 
(\nu)$). 

\begin{figure}
\centerline{\epsfig{file=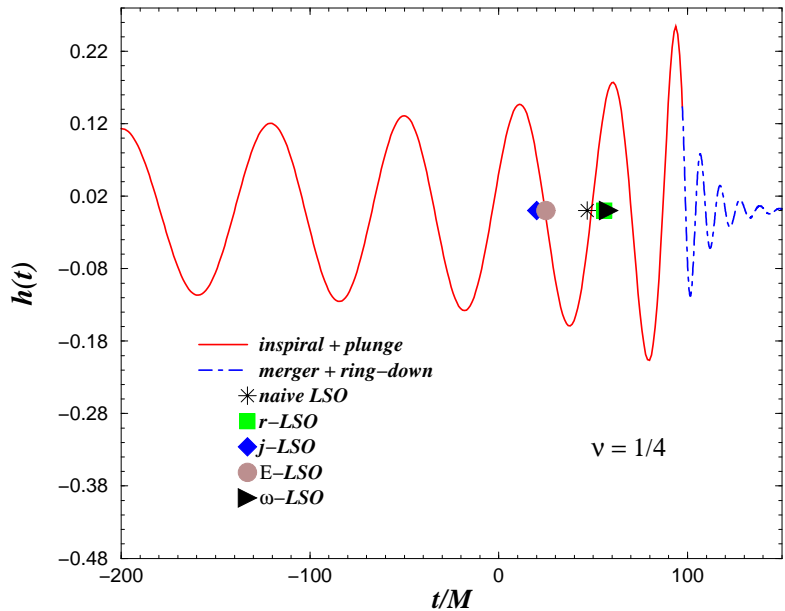,width=0.5\textwidth,angle=-90}}
\caption{\sl Plot of the complete waveform: inspiral and 
plunge followed by merger and ring-down. 
The locations of several possible definitions of LSO crossing are also 
indicated.}
\label{Fig12}
\end{figure}

The corresponding numerical values of the reduced radial coordinate $r$ are:
\bea
\label{5.2}
&& r_{\rm j-LS0} = 6.631\,, \quad \quad 
r^{\rm naive}_{\rm LSO} = 6.000\,, \quad 
\quad r_{\rm {\cal E}-LS0} = 6.534\,,\\
\label{5.3}
&& r_{\rm r-LS0} = 5.718\,, \quad \quad
r_{\rm \omega-LS0} = 5.639\,. 
\eea
As mentioned above, the fact that the various definitions of the LSO differ 
significantly is due to the fact that when $\nu = 1/4$ the GW damping effects 
are 
rather large and blur the transition to the plunge. Note that the number of GW
cycles left after the (exact) $\omega$-LSO (and until the light-ring) is
$N_{\rm GW}^{\rm after} = 
2 N_{\rm orbit}^{\rm after} = 1.2048$ (for $\nu = 1/4$). As said above, this
is smaller than the (physically less relevant) number of cycles left after the
adiabatic LSO (where $\omega \simeq 0.80\, \omega_{\rm LSO}$), which is $\simeq 2.35$.

Even if our estimate of the waveform is admittedly rough, we think that it
can play an important role for defining better filters for the search of
signals in LIGO and VIRGO. In particular, two features of this waveform are
striking: (i) the `plunge' part of the waveform looks like a continuation
of the inspiral part (this is because the orbital motion remains in fact
quasi-circular), and (ii) the adiabatic waveform gets significantly
out of phase with the exact waveform before crossing the LSO. We shall come
back in future work to the consequences of these results for data analysis,
and see how they can be used to improve upon the state-of-the-art filters
constructed in Ref.~\cite{DIS},\cite{DIS2}.

\section{Discussion}
\label{sec7}

In this paper we have extended a methodology introduced in previous papers 
\cite{DIS}, 
\cite{BD99}, and applied it to the study of the transition from inspiral to 
plunge in 
coalescing binary black holes with comparable masses, moving on quasi-circular
orbits. Our philosophy is that it is 
possible to use suitably re-summed versions of post-Newtonian results to write 
an 
explicit (analytical) system of ordinary differential equations describing the 
transition to the plunge. Our explicit proposal is the evolution system 
(\ref{3.28})--(\ref{3.31}) obtained by combining the results of \cite{DIS} for 
the 
re-summation of the gravitational wave damping, and the results of \cite{BD99} 
for the 
re-summation of the conservative part of the dynamics of comparable-mass 
binaries. The 
basic reason why we think the simple evolution system (\ref{3.28})--(\ref{3.31}) 
can 
accurately describe the transition to the plunge is that we have consistently 
checked 
that most of the ``plunge'' motion (at least down to $R \simeq 3 {GM}$) is 
in fact 
very much like a quasi-circular inspiral motion (with $\dot{R}^2 \ll (R \, 
\dot{\varphi})^2$).

In general one needs to numerically integrate the basic evolution system 
(\ref{3.28})--(\ref{3.31}) to get physical results of direct interest. However, 
we have 
shown that one can understand the various physical elements entering this system 
by 
comparing it to several simple approximations: the adiabatic approximation, the 
$\dot 
r$-linearized one, and the universal $\rho$-approximation (valid when $\nu 
\,\laq \,0.05$). In 
particular, the latter approximation allowed us to derive some scaling laws: one 
scaling law (which is very well satisfied, even up to the maximum value $\nu = 
1/4$) 
states that the radial momentum at the Last Stable Orbit (LSO) scales like 
$\nu^{3/5}$, 
while another scaling law (accurately satisfied only for $\nu \ll 1$) states 
that the 
number of cycles left after the LSO scales like $\nu^{-1/5}$.

The two most important consequences of the present approach are: (i) a way to 
compute 
initial dynamical data $(\mbox{\boldmath$q$}_1 , \mbox{\boldmath$q$}_2 , 
\mbox{\boldmath$p$}_1 , \mbox{\boldmath$p$}_2)$ for a comparable-mass binary 
black hole 
system, represented in ADM coordinates, such that only a fraction of an orbit 
needs to 
be further evolved by numerical relativity techniques, and (ii) an estimate of 
the 
complete waveform emitted by a binary black hole coalescence, smoothly combining 
an 
inspiral signal, a plunge signal, a merger signal and a ring-down.

However, much work remains to be done to firm up and complete our approach. We 
checked 
the robustness of our approach by considering an as-well-justified, slightly 
different 
evolution system. But stronger checks are called for. In particular it would be 
quite 
important to extend the present work (which used as input the 2.5PN-accurate 
damping 
and 2PN-accurate dynamics) to higher PN levels, when they become fully 
available. We 
note in this respect the recent work \cite{DJS2} which extended the 
effective-one-body 
approach to the 3PN level. [Note in passing that quasi-static tidal interactions 
between black holes enter only at the 5PN level \cite{D83}.] It is 
quite important to complete our determination of initial {\it dynamical} data 
$(\mbox{\boldmath$q$}_a , \mbox{\boldmath$p$}_a)$ by explicitly constructing the 
initial {\it gravitational} data $(g_{ij} (x) , K_{ij} (x))$ corresponding to 
$(\mbox{\boldmath$q$}_a , \mbox{\boldmath$p$}_a)$ (and containing no free 
incoming 
radiation). When this becomes available it will be possible to further check our 
method 
(by numerically evolving spacetimes starting at various stages of the plunge) 
and to 
provide more accurate estimates of the merger waveform. Though our 
``light-ring-matching'' approach to estimating the complete waveform is 
admittedly rough, we think it can play a useful role for data analysis: it can 
be used to 
test the 
accuracy of present templates (based on the adiabatic approximation) and allow 
one to 
construct more accurate, or at least, more robust, templates. We will come 
back to 
this issue in future work. Finally, let us note that it would be, in principle, 
important to be able to extend our approach to black holes having significant 
intrinsic 
spins. We, however, anticipate that this is a highly non-trivial task.

\acknowledgments
A.B.'s research was supported at Caltech by the Richard C. Tolman Fellowship and 
by NSF Grant AST-9731698 and NASA Grant NAG5-6840.

All the numerical results in the present paper were produced using Mathematica.

\end{document}